\documentclass[12pt]{article}
\usepackage{graphicx}
\usepackage{dsfont}
\usepackage{subcaption}
\usepackage{amsmath,amssymb,amsthm,amsfonts,bbm}
\usepackage{natbib}
\usepackage{xcolor}
\usepackage{setspace} 
\usepackage[margin=1.0in]{geometry}
\usepackage{tikz}
\usepackage[colorlinks = true,
            linkcolor = blue,
            urlcolor  = purple,
            citecolor = blue,
            anchorcolor = blue]{hyperref}
\usepackage{enumitem}
\usepackage{stackengine}
\def\rub{\stackengine{.64ex}{%
  \stackengine{.4ex}{\textbf{\textsf{P}}}{\rule{1ex}{.16ex}\kern.55ex}{O}{r}{F}{F}{L}%
  }{\rule{1ex}{.16ex}\kern.55ex}{O}{r}{F}{F}{L}\kern-.1ex}

\theoremstyle{definition}

\newtheorem{assumption}{Assumption}

\theoremstyle{plain}
\newtheorem{lemma}{Lemma}

\DeclareMathOperator*{\argmin}{arg\,min}

\usepackage{lineno}

\begin{document}

\title{Procurements with Bidder Asymmetry in Cost and Risk-Aversion
}

\author{Gaurab Aryal\thanks{ Department of Economics, Washington University in St. Louis, \href{mailto:aryalg@wustl.edu}{ aryalg@wustl.edu}},
 Hanna Charankevich\thanks{ Biocomplexity Institute,
 University of Virginia, \href{mailto:hc2cc@virginia.edu}{ hc2cc@virginia.edu}}, 
 Seungwon Jeong\thanks{ College of Business, KAIST; School of Economics, University of Bristol, \href{eugene.jeong@gmail.com}{eugene.jeong@gmail.com}}, and 
 Dong-Hyuk Kim\thanks{Corresponding Author: School of Economics, University of Queensland, \href{donghyuk.kim@uq.edu.au}{donghyuk.kim@uq.edu.au}}.
}

\date{\today}

\maketitle

\begin{abstract}
We propose an empirical method to analyze data from first-price procurements where bidders are asymmetric in their risk-aversion (CRRA) coefficients and distributions of private costs. Our Bayesian approach evaluates the likelihood by solving type-symmetric equilibria using the boundary-value method and integrates out unobserved heterogeneity through data augmentation. We study a new dataset from Russian government procurements focusing on the category of printing papers. We find that there is no unobserved heterogeneity (presumably because the job is routine), but bidders are highly asymmetric in their cost and risk-aversion. Our counterfactual study shows that choosing a type-specific cost-minimizing reserve price marginally reduces the procurement cost; however, inviting one more bidder substantially reduces the cost, by at least 5.5\%. Furthermore, incorrectly imposing risk-neutrality would severely mislead inference and policy recommendations, but the bias from imposing homogeneity in risk-aversion is small.
	
\end{abstract}

\section{Introduction}
This paper proposes a Bayesian method to analyze data from first-price (sealed-bid) procurements (FPP) with bidder-asymmetry in cost and risk-aversion.
In particular, we consider a theoretical model of FPP with exogenous entry and type-symmetric Bayesian Nash Equilibrium bidding strategies.
Here, a type of bidder refers to a pair of cost density and constant relative risk-aversion (CRRA) coefficient.
For this setting, \cite{Campo2012} identifies the model primitive (cost densities and CRRA coefficients) and proposes an indirect semiparametric estimation method.
The empirical literature, however, has seldom considered asymmetry in both cost and risk-aversion.
We, therefore, have limited empirical insights on the effects of asymmetric risk-aversion on procurement outcomes.

The main contribution of this paper is to develop a novel empirical procedure that produces reliable inference for asymmetric FPPs by combining and extending several state-of-the-art methods in the literature. 
First, 
the procedure explores the posterior distribution over the space of the model primitive by a Markov chain Monte Carlo (MCMC) algorithm \citep{KimDH_2014_BSL}.
Second, 
we model the (procurement-specific) unobserved heterogeneity as additional latent components that are distributed jointly with the model primitive under the posterior so that we can integrate out the latent components via MCMC.
This strategy to get around the difficulty of handling missing variables is known as \emph{data augmentation}, which \cite{Li_Zheng_2009} and \cite{AryalGrundlKimZhu2017}, among others, use to study auction markets.
Third, 
we extend the boundary-value method of \cite{Fibich_Gavish_2011} to compute the type-symmetric equilibrium strategies for risk-averse bidders in FPPs and use it in our MCMC algorithm.
Note that the literature has often used the backward-shooting method of \cite{MarMeuRichStr_1994} to compute asymmetric bidding strategies, which is, however, shown to be inherently unstable near the boundary of the support; see \cite{Fibich_Gavish_2011}. 
The boundary-value method is reliable everywhere and, therefore, is more suitable for evaluating likelihoods and conducting policy simulations.

In addition, our Bayesian method provides a natural framework for formal decision-making under parameter uncertainty. 
In particular, we consider the policymaker who wishes to choose a reserve price to minimize the (expected) procurement cost when there is uncertainty about the model primitives.
Our decision method computes the procurement cost under a model primitive at a given, \emph{possibly type-specific}, reserve price, which to our knowledge is the first in the empirical auction literature; see \cite{Kotowski2018} for theoretical developments on this topic. 
For this step, we further extend the algorithm of \cite{Fibich_Gavish_2011} to accommodate binding reserve prices and evaluate the 
procurement
cost by simulating equilibrium bids.
The decision method then integrates out the model primitive by the posterior, resulting in 
the (posterior) predictive procurement cost
at  the said reserve price.
Finally, our method selects a reserve price that gives the smallest predictive cost.

This solution is also coherent under the rationality axioms for a decision-maker in \cite{Savage_1954} and \cite{Anscombe_Aumann_1963}. See also \cite{KimDH_2013_IJIO}, \cite{Aryal_Kim_2013}, \cite{KimDH_2014_BSL}, and \cite{AryalGrundlKimZhu2017} for applications of statistical decision theory in empirical auction design problems.
This paper is the first to conduct such detailed counterfactual simulations for FPPs with asymmetric costs and risk-aversion. 

Using our empirical method, we study a new dataset with FPPs from Russian public procurements conducted in 2014. (\cite{Charankevich2020} investigates 
a sample of 
``open auctions"
in the Russian procurements, which is different from the sample of FPPs that we analyze here.) 
In Russia, all
government
units 
purchase a wide range of goods and services through public procurements, which accounts for 7\% of Russian GDP in 2014. Since the economic transition beginning in the 1990s, Russia has been revising the procurement system through several legislative changes and technological innovations to establish transparent competition among the suppliers to reduce the government expenditure. It is, therefore, critical for the policymaker to learn the economic fundamentals of the procurement system and evaluate alternative policy options for further improvements.
We illustrate how the policymaker can achieve the goals applying our method to the data.

The Russian procurement system practices several allocation methods and 
selects one of them, e.g., FPP or ``open auctions," depending on the nature of the project and according to the relevant federal laws. 
For example, small projects (reserve price below \rub 500,000) must use an FPP.
In total, there are 102 different categories in FPPs based on different goods and services.
The categories are separate, each with its own market and a different set of suppliers.
We choose one category to analyze by applying a set of selection criteria mostly concerning the sample size.
We divide bidders into three types: type 1 (frequent) bidders bid in at least 10\% of the procurements and type 3 (fringe) bidders only once, and type 2 are the remainders.
Among the four categories complying with our selection criteria, we choose to analyze the category of ``printing papers," as it has the largest sample with 411 procurements. 
In the ``printing papers" category, 1, 171, and 402 unique bidders are of types 1, 2, and 3, and we observe a total of 58, 625, and 402 bids, respectively.

On average, type 1 bidder bids lower than type 2, and type 2 lower than type 3. 
The differences in bids may arise because of differences in costs or risk-aversion. 
The posterior of the model parameters (model primitive) reveals that the ordering of risk-aversion is the opposite of the observed bid pattern.
In particular, the CRRA coefficient of type 1 (frequent) bidder is roughly 0.2, whereas the coefficients of type 2 and 3 bidders are respectively 0.8 and 0.9. 
But, the predictive cost densities suggest that type 1 bidder tends to draw smaller costs than type 2, and type 2 smaller than type 3. 
Therefore, the difference between the cost densities is substantial enough to explain the bid pattern, despite the reversely ordered risk-aversion.
Our analysis also finds no variation in the unobserved heterogeneity, which is consistent with the job of supplying papers being routine.

Our estimates of the model parameters are coherent with our findings in the counterfactual policy analysis. 
All bidders, except the most frequent bidder, exhibit high risk-aversion. 
So, the policymaker should prefer FPP to second-price procurements (SPP) \citep{Holt1980} and choose a large reserve price in an FPP \citep{HuMatthewsZou2010}. 
In particular, we find that when a single reserve price is used for all bidders, the current mechanism is cost-minimizing.
Recall also that when bidders are asymmetric, choosing a single reserve price would be suboptimal. Although there is no closed-form expression for the reserve prices as in \cite{Myerson_1981}
due to risk-aversion, our method allows evaluating the procurement costs at different type-specific reserve prices.
It recommends lowering the reserve price for type 1 by 4\% from the current reserve price, but leaving other types' prices unchanged. In that case, however, our method predicts a cost reduction only of 0.2\%, suggesting that the current mechanism is effectively cost-minimizing.

In addition, we consider a scenario where the policymaker may invite an additional bidder in the spirit of \cite{bulow_Klemperer_1996}. 
For symmetric FPPs, the article implies that an FPP without a reserve price generates lower expected costs than optimally chosen reserve price but with one less bidder.
We find that this insight holds in our context with bidders who are asymmetric in both cost densities and risk-aversion. 
In particular, adding one bidder of type 1, 2, and 3 reduces
the predictive procurement cost
by 6.2\%, 5.8\%, and 5.5\%, respectively. Thus, even one additional fringe bidder substantially lowers the cost.

Moreover,
we investigate the implications on the procurement cost and efficiency of incorrectly assuming either risk neutrality 
or an identical CRRA parameter for all bidders. 
When one ignores risk-aversion, small bids in the data inflate the left tail of the cost densities, which then \emph{tilts} cost-minimizing reserve prices toward zero. As a result, the misspecified model selects 
a small reserve price,
predicting 14.0\% of cost-reduction. But, this result is misleading because the model with asymmetric CRRAs predicts a 15.2\% of \emph{increase} in the cost at that price. The misspecified model predicts that the efficient bidder wins with 33.2\% of probability, whereas the model with asymmetric CRRAs predicts that the probability would be 6.0\%. 
We find that the model with a common CRRA overall approximates our analysis with asymmetric CRRAs, except for overestimating type 1 bidder's CRRA coefficient. 

Nevertheless, one should not conclude that the model with identical
risk-aversion always approximates the model with 
asymmetric risk-aversion because the approximation quality may depend on the model primitive, \emph{a priori} unknown to the researcher. 
Our method with asymmetric CRRAs is not computationally more expensive than the model with a common CRRA. 
Therefore, there is no reason to impose homogeneity in CRRA coefficients. 
Finally, our analysis is robust to alternative definitions of bidder types, the prior over the parameter that indexes the cost densities, and the density specification of unobserved heterogeneity.

The paper proceeds as follows. 
Section \ref{section:model} describes our model and its identification.  
Sections \ref{section:data} and \ref{sec:method} present our data and econometric method, respectively. 
Section \ref{section:results} discusses the empirical results and counterfactual analysis. 
Section \ref{section:conclusion} concludes
with feasible extensions.
The detailed information about data, computational detail, and
additional results are in Supplementary Appendix \citep{ACJK2021S}.

\section{Model and Identification \label{section:model}}
Consider a procurement that allocates a project to one of the bidders in the set $\mathcal{I}$ with $|\mathcal{I}|\geq2$.
Bidders submit their bids simultaneously, and the one with the lowest bid wins the project at a price equal to her bid, which we refer to as the first-price procurement (FPP). 
Let bidder $i$'s cost be $c_i$, which follows the distribution $F_{i}(\cdot)$ and is independent of other bidders' costs.
We make the following assumptions on the cost distributions. 
\begin{assumption}\label{assF}
Bidder $i$'s cost distribution $F_{i}(\cdot)$ has a density $f_{i}(\cdot)>0$ on the support 
$[\underline{c},\overline{c}]\subset\mathbb{R}_{+}$, 
and for two different bidders $i\neq j$
it can be that $F_i(c) \neq F_j(c)$ for some $c$.
\end{assumption}
In addition, 
bidder $i$ exhibits constant relative risk-aversion (CRRA) with coefficient, $\eta_i$.
\begin{assumption}\label{assU}
Bidder $i$'s utility function is given by  $U_i(\xi)=\xi^{1-\eta_i}$ for consumption $\xi\geq0$ with the parameter $\eta_i\in[0,1)$, and it can be that $\eta_i \neq \eta_j$ if $i\neq j$. 
\end{assumption}
The CRRA specification has been widely used due to its convenient functional form in the auction literature, e.g., \cite{Bajari_Hortacsu_2005}, \cite{Lu_Perrigne_2008}, \cite{Campo_Guerre_Perrigne_Vuong_2011}, and \cite{AryalGrundlKimZhu2017}. Such convenience also allows 
the boundary-value method of
\cite{Fibich_Gavish_2011} to accommodate bidder-specific risk-aversion in this paper, and one may compare the coefficients with previous estimates due to its prevalence. 

If bidder $i$ wins the procurement at price $b_i$, her utility is $(b_i-c_i)^{1-\eta_i}$ under Assumption \ref{assU}.
While the realizations of costs are bidders' private information, cost distributions and risk-aversion parameters, $\{F_{i}(\cdot), \eta_{i}: i\in \mathcal{I}\}$, are assumed to be common knowledge.
Bidder $i$ with cost $c_i$ chooses $b_{i}$ to maximize her expected utility given everyone else's bidding strategy. 
Suppose all bidders  other than bidder $i$ use strictly increasing 
equilibrium bidding strategies $\{\beta_{j}(\cdot|\mathcal{I}): j\neq i, j\in \mathcal{I}\}$. 
Then bidder $i$ solves
\[
\max_{\tilde{b}_i\in\mathbb{R}_{+}}\Big\{
(\tilde{b}_i - c_i)^{1-\eta_i} 
\prod_{j\in\mathcal{I}\setminus\{i\}}
(1-F_{j}(\beta_{j}^{-1}(\tilde{b}_{i}|\mathcal{I})))\Big\}.
\]
Define $\phi_j(b|\mathcal{I}) := \beta_{j}^{-1}(b|\mathcal{I})$,  the inverse bidding strategy of 
bidder $j$.
Then, the optimal bid $b_i$ must satisfy the condition, 
\begin{eqnarray}
1-\eta_i =({b}_i - c_i)
\sum_{j\in\mathcal{I}\setminus\{i\}}
\frac{f_j(\phi_j(b_i|\mathcal{I}))}{1-F_j(\phi_j(b_i|\mathcal{I}))}\frac{1}{\phi_j'(\phi_j(b_i|\mathcal{I})|\mathcal{I})},\label{eq:foc}
\end{eqnarray}
for all $i\in\mathcal{I}$, implying a system of differential equations, which can be numerically solved 
with the boundary conditions $\phi_i(\overline{c}|\mathcal{I})=\overline{c}$ and
$\phi_i(\underline{b}_{\mathcal{I}}|\mathcal{I})=\underline{c}$ for all $i \in \mathcal{I}$.

Since 
$\underline{b}_{\mathcal{I}} = \beta_i(\underline{c}|\mathcal{I})$
is unknown, 
the standard algorithm to solve \eqref{eq:foc}, known as \emph{backward-shooting}, 
starts with a guess of $\underline{b}_{\mathcal{I}}$ and adjusts its guess at each iteration \citep{MarMeuRichStr_1994}. 
\cite{Fibich_Gavish_2011}, however, show that the backward-shooting algorithm is inherently unstable near the boundary and propose the boundary-value method to overcome the problem.
We use their boundary-value method in our empirical method, where $\{\phi_j(\cdot|\mathcal{I})\}$ have to be evaluated at data points, including the ones near the boundary.

\paragraph{\emph{Identification.} }
\cite{Campo2012} uses the exogenous variation in the bidder configuration $\mathcal{I
}$ to identify the parameters of interest.
For completeness, we present the core intuition of the identification strategy in \cite{Campo2012}.
Since $\{\phi_i: i\in \mathcal{I}\}$ are strictly increasing, 
the bid distribution of bidder $i\in\mathcal{I}$ is $G_i(b|\mathcal{I}) = F_i(\phi_i(b|\mathcal{I}))$.
Then, we can rewrite (\ref{eq:foc}) as
\begin{align}
c_i=b_i - \frac{1-\eta_i}{\sum_{j\in\mathcal{I}\setminus\{i\}} \frac{g_j(b_i|\mathcal{I})}{1-G_j(b_i|\mathcal{I})}},
\label{eq:cibi}
\end{align}
where 
$g_i(\cdot|\mathcal{I})$ is the density of $G_i(\cdot|\mathcal{I})$.
Note that 
\eqref{eq:cibi} in a procurement with $\mathcal{I}=\{1,2\}$ 
gives
\[
\underline{b}_{\mathcal{I}} -
\frac{1-\eta_2}{g_1(\underline{b}_{\mathcal{I}}|\mathcal{I})}
= \underline{c} =
\underline{b}_{\mathcal{I}} -
\frac{1-\eta_1}{g_2(\underline{b}_{\mathcal{I}}|\mathcal{I})},
\]
which gives 
$
g_1(\underline{b}_{\mathcal{I}}|\mathcal{I})\eta_1-g_2(\underline{b}_{\mathcal{I}}|\mathcal{I}) \eta_2
=
g_1(\underline{b}_{\mathcal{I}}|\mathcal{I})-g_2(\underline{b}_{\mathcal{I}}|\mathcal{I}).
$
Then, the exogenous variation in $\mathcal{I}$ is sufficient  for identification.
For example, 
if we observe FPPs with
$\mathcal{I}_1 = \{1,2\}$, $\mathcal{I}_2 = \{1,3\}$, and $\mathcal{I}_3 = \{2,3\}$, 
we identify the CRRA coefficients as 
\[
\left(
\begin{array}{c}
\eta_1 \\
\eta_2 \\
\eta_3 
\end{array}
\right)
=
\left(
\begin{array}{ccc}
g_1(\underline{b}_{\mathcal{I}_1}|\mathcal{I}_1) & -g_2(\underline{b}_{\mathcal{I}_1}|\mathcal{I}_1) & 0 \\
g_1(\underline{b}_{\mathcal{I}_2}|\mathcal{I}_2) & 0 & -g_3(\underline{b}_{\mathcal{I}_2}|\mathcal{I}_2) \\
0 & g_2(\underline{b}_{\mathcal{I}_3}|\mathcal{I}_3) & -g_3(\underline{b}_{\mathcal{I}_3}|\mathcal{I}_3) 
\end{array}
\right)^{-1}
\left(
\begin{array}{c}
g_1(\underline{b}_{\mathcal{I}_1}|\mathcal{I}_1)-g_2(\underline{b}_{\mathcal{I}_1}|\mathcal{I}_1) \\
g_1(\underline{b}_{\mathcal{I}_2}|\mathcal{I}_2)-g_3(\underline{b}_{\mathcal{I}_2}|\mathcal{I}_2) \\
g_2(\underline{b}_{\mathcal{I}_3}|\mathcal{I}_3)-g_3(\underline{b}_{\mathcal{I}_3}|\mathcal{I}_3)
\end{array}
\right),
\]
where the right-hand side depends only on the bid densities that are directly identified from the data.
Then, by substituting $\eta_\tau$  in \eqref{eq:cibi}, we identify the cost distributions.

\paragraph{\emph{Unobserved Heterogeneity.}} 
Bidders may observe some aspects of the project that affect their costs (and hence their bids), which the researcher does not observe.  
Let $u\in\mathbb{R}_+$ denote such unobserved characteristics.  

\begin{assumption}\label{as:UH}
\begin{enumerate}
\item $u_t\stackrel{i.i.d}{\sim}F_{u}$ with density $f_u>0$ on the support $[\underline{u},\overline{u}]\in\mathbb{R}_+$ with a location normalization, e.g., $\overline{u}$ is known.
\item In procurement $t$, $u_t$ is independent of bidder's private cost, i.e., $u_t\perp c_{it}$, for all bidders. 
\item The final cost for bidder $i$ with $c_{it}$ in procurement $t$ is given by $c_{it}^o:= u_t\times c_{it}$. 
\end{enumerate}
\end{assumption}
Let $F_i^o$ be the distribution of $c_i^o$ for bidder $i\in \mathcal{I}$ and $\{\beta_i^o: i\in \mathcal{I} \}$ be the associated 
bidding strategies with the unobserved heterogeneity, $u_t$. 
For $c_{it}\in[\underline{c},\overline{c}]$ and $u_t\in [\underline{u}, \overline{u}], \beta_i^o(u_t c_{it}|\mathcal{I})= u_t \times \beta_i(c_{it}|\mathcal{I})$, where $\beta_i(c|\mathcal{I})$ is bidder $i$'s equilibrium bidding strategy when $u_t=1$; see \cite{Liu_Luo_2017}.

The identification argument combines \cite{Campo2012}  with 
\cite{Kotlarski1966} as 
in \cite{LiVuong1998} and \cite{Krasnokutskaya2011}. 
We formalize this result below; see 
its proof in section S2 of the Supplemental Appendix (S for the appendix, hereafter). 

\begin{lemma}
Under Assumptions \ref{assF}--\ref{as:UH}, $\{F_{i},\eta_{i}:i\in \mathcal{I}\}$ and $F_u$ are identified
by the data of all submitted bids $\{(b_{it})_{i\in\mathcal{I}_t}\}$ and bidder configurations $\{\mathcal{I}_t\}$
if every bidder $i$ exogenously enters the procurement with a strictly positive probability (so, $\mathcal{I}_t$ varies exogenously).
\label{identificationUH}
\end{lemma}

\section{Russian Government Procurement\label{section:data}}
This section describes the institutional background of the government procurements in Russia, presents the dataset we analyze, and discusses the implications of the reserve price on our analysis 
in sections \ref{sec:method} and \ref{section:results}.
In particular, the background description here identifies a few cases where observed bids might not be competitive and justifies the data we study as equilibrium outcomes after excluding those suspicious cases. 
It also motivates us to take cost-minimization as the primary policy objective in counterfactual analysis.

\subsection{Institutional Background}\label{sec:background}
All government bodies and public units in Russia purchase goods and services through government procurements. Examples
of potential buyers include federal public authorities, regional governments, city councils, public hospitals, and schools.
Hundreds of goods and services, e.g.,  car tires, hardcover textbooks, road maintenance, and printing papers, are traded via the official platform, ``Unified Information System (UIS: \href{https://zakupki.gov.ru/epz/main/public/home.html#statAnchor}{zakupki.gov.ru})."
(We use quotation marks to indicate field terms in their closest English translation. Moreover, procurement here is a general term referring to procuring something unless it comes with a technical qualifier, e.g., first-price procurement.) 
The public procurements are economically significant; according to the UIS, concluded contracts in 2014 add up to 
5.47 trillion RUB ($\approx$7\% of Russian GDP, \href{https://www.statista.com/statistics/1055726/russia-nominal-gdp/}{statistical.com}).

Since the economic transition beginning in the 1990s, Russia has been revising the procurement system through several changes in legislation and modernizing it via technical innovations. Those reforms aim to reduce government spending and improve outcomes by creating a competitive environment for suppliers; especially by encouraging suppliers' participation as well as eliminating corruption between government agents and suppliers. In particular, Federal Law No. 94-FZ (21 July 2005) laid a foundation for the current form of the system. The Law introduced the concept of ``maximum (or initial contract) prices" and prohibited ``closed procurements," i.e., 
a negotiation inviting a single supplier, except for special cases such as projects involving state secrets.

Before the Law (94-FZ), the government agent overseeing the procurements 
had considerable freedom to choose a supplier and set a price. Therefore,  
the agent (buyer) could set a high price and select an ``insider"  (a supplier in collusion) to carry out the project at a cost lower than the price and share the margin.
The legislation mandates that
a ``maximum price" must be chosen in such a way that 
participation of general suppliers is encouraged for ``healthy competition,"  
and the price can still be justifiable given the nature of the project, market conditions, and historical data.
For example, a buyer should be able to purchase comparable goods and services at the ``maximum price" outside the procurement system.
(The UIS provides protocols and official methods for setting a ``maximum price" to handle different situations.)
Section \ref{sec:reserve}  below discusses the implications of ``maximum price" for our analysis.

For projects with a ``maximum price" above \rub 100,000 ($\approx$\$2,600), the procurer must select a supplier through a competitive procedure.
The ``maximum price" is then publicly announced, and any legal entity can participate with no entrance fee. 
The announcement should be placed at least four business days before the closing date if the ``maximum price" is below \rub 250,000
(and seven days if above)
to prevent buyers from selecting an ``insider" by 
setting a tight deadline.

In addition, to improve transparency, by Federal Law No. 44-FZ (1 January 2014), the UIS publicly announces forthcoming procurements and maintains all procurement data, e.g., participants' identities, their offers, and the winner for every procurement, and the UIS provides 
a
platform to run a procurement. 
The system practices several allocation methods, including negotiation and contest, and one of them is ``sealed-bid auction" in the field term. 
When the latter is implemented, bidders submit their initial documents in a sealed envelope (or online, but rarely in 2014). 
In the presence of all participants, 
then,
each bidder makes a final decision on her bid
and, finally,
the procurer opens all the bids  (Federal Law No. 44-FZ, article 78 of paragraph 3).
Note that the procurer, here, 
is a government agency running procurements on behalf of buyers, other government agents. 
The lowest bidder 
wins at a price equal to her bid, if not higher than the ``maximum price."
The mechanism is, therefore, the first-price procurement (FPP), where bidders know whom they oppose, and the ``maximum price" plays the role of a reserve price in FPPs.

FPPs are used for cases with a  ``maximum price" below \rub 500,000, i.e., small projects. 
But, the penalty for a supplier in corruption, e.g., ``insider," is \rub 500,000 plus a full reimbursement of expenses. The supplier is also publicly marked as ``unreliable" for two years. In addition, the fine for the 
government agent in corruption is \rub50,000 with 
a three-year job suspension.

Despite all the legal devices to eliminate corruption, a buyer can still invite an ``insider" only. 
For example, a buyer could deliberately make a typographical error, e.g., replacing a Cyrillic letter with a similar Latin letter in keywords. Then, only the insider can easily search for it in the system.
Therefore, the cases with only one bidder can be suspicious. 
Even when multiple bidders appear in a procurement, there can be corruption.  
For example, the agent could invite shill bidders who submit bids with no chance of winning, e.g., bidding higher than the reserve price. 
Alternatively, the agent may manipulate submitted bids to increase the winning probability of the insider.
\cite{Charankevich2020} reports evidence of bid manipulation by exploring procurement outcomes and non-reported (or missing) bids in ``open auctions," which can be viewed as an oral-descending procurement.
Considering all these, in our analysis, we exclude observations with only one bidder, bids above reserve prices, or missing bids to avoid those suspicious cases.

\subsection{Data: category ``printing papers"}\label{sec:printing}
The procurements with FPPs provide an ideal setting to study asymmetry in both cost densities and risk-aversion for the following reasons. 
First, bid data from FPPs allow us to identify the risk-aversion parameters. 
Second, FPPs are less prone to collusion among the bidders than oral-descending procurements because bidders do not observe other bidders' behavior \citep{Robinson85}.
Third, since FPPs are used for projects with small budgets,
they may attract small firms that are likely to be risk-averse \citep{Herranz_et_al2015}.
Indeed, the test of \cite{JunZincenko2022} rejects risk-neutrality in favor of risk-aversion (p-value $<0.0001$) for each symmetric FPPs, i.e., type
2 bidders only and type 3 bidders only with types defined below.

Finally, projects with small budgets are homogeneous and 
frequent,
i.e., the bid preparation cost would be minimal, and the jobs are routine. 
To this end, we fail to reject the independence of any two randomly selected bids in the same FPP (p-value $>0.1$) via the test used by \cite{Krasnokutskaya2011} with 1,000 bootstrapped samples. 
For these reasons, the procurement-specific unobserved heterogeneity might not be substantial, which section \ref{section:results} shall confirm from the data. 

The UIS provides data on 42,828 FPPs in 2014 across 102 categories of different goods and services; see section S3.1 for the list of the categories and some statistics.
We consider each category as an independent set of procurements because they are separated by industries with different sets of suppliers.
Now, we explain how we select a category to analyze; see also section S3.1.
In each category, first, we rule out all procurements with one bidder because they are vulnerable to corruption (section \ref{sec:background}) and are not even bidding competition.
We then exclude procurements with missing bids, which may arise due to bid manipulation \citep{Charankevich2020}. 
We also discard the procurements with bids larger than reserve prices because those bids could be shill bids as discussed or could signal that the reserve prices are set too low, violating the UIS protocols to select a reserve price.

After excluding 
these three cases,
we group bidders 
into
three types: in each category, type 1 bidders bid in at least ten percent of the FPPs, type 3 bidders bid once, and type 2 bidders are all the others. 
We define the bidder types by the participation rate because it is the only exogenous bidder-specific attribute 
in the data, besides
their identities. 
(Section \ref{sec:sensitivity} considers alternative definitions.) 
Since the identification strategy relies on the variation in the bid for each type and bidder configuration, we sort out categories with at least 50 bids for each type to estimate the type-specific parameters. 
Four categories satisfy this condition. 
Among them, we analyze the one with the largest sample, 
which is the category of ``printing papers" with $T=411$ procurements.
(In the previous version of this paper, \cite{Kim2020} study the four categories separately, where the other ones have 228, 235, and 305 procurements.)

The category ``printing papers" refers to white A4 paper for printing, copying, and faxing. A typical order specifies A4 white papers in packs of 500 sheets. The paper can vary from ``regular white" of 92-94\% (ISO) to ``premium white" above 98\% (ISO).
A standard contract includes the delivery term that can range from 5 to 30 days and provision for the fulfillment of incomplete orders and replacement or exchange of damaged goods within 3 to 10 days after delivery. 
As mentioned above, any government unit can purchase the products, and any legal entity that can supply papers may bid in this category. 

\begin{figure}[t!]
\begin{center}
\caption{Bid Data, `printing paper'\label{fig:data}}
    \includegraphics
[width=0.8\textwidth]   
{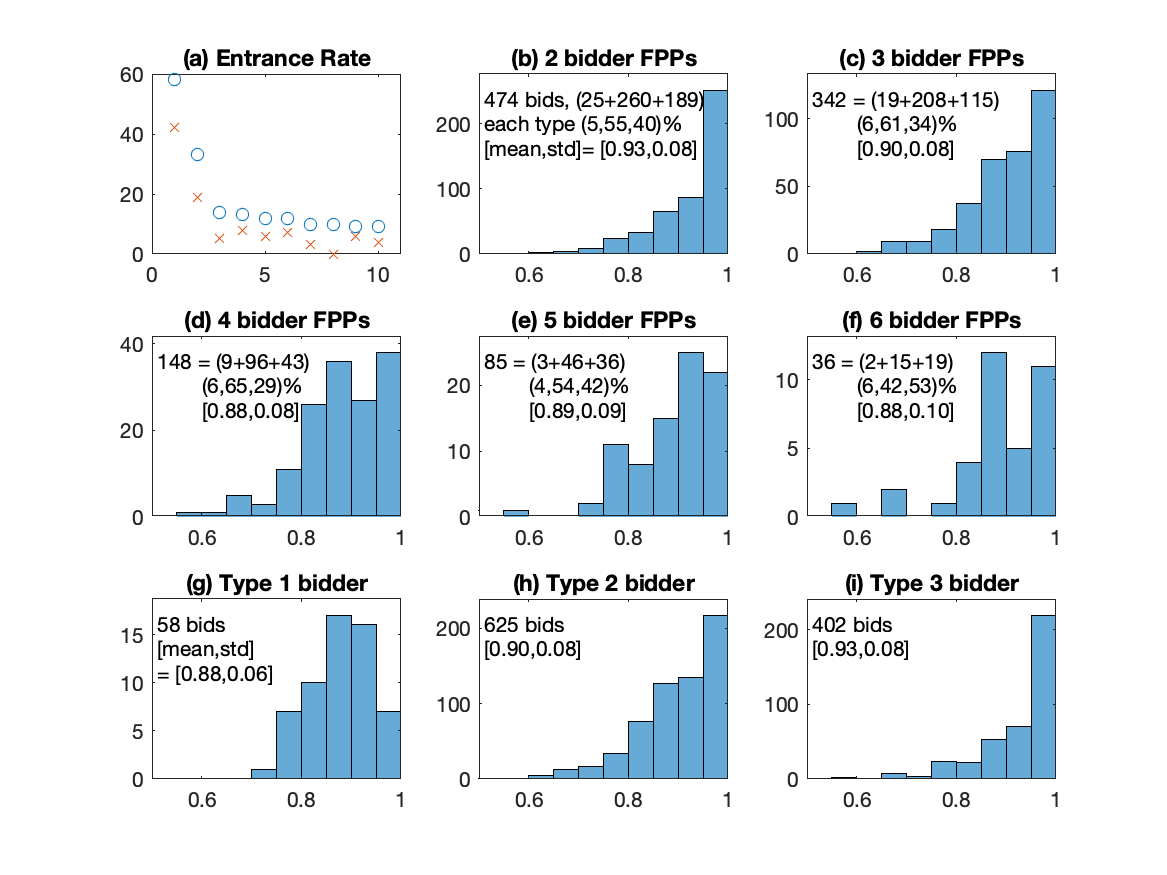}
\caption*{\footnotesize 
Panel (a) shows the number of entrants $(\circ)$ for the ten 
frequent bidders and their numbers of wins $(\times)$. 
Panels (b) to (f) present bid histograms for different numbers of bidders 
and (g) to (i) for different types of bidders. 
All bids are divided by reserve prices.
Panels (b) to (i) show the number and  [average, standard deviation] of bids, and (b) to (f) counts bids with the proportion for each type in ( ). 
At the 1\% level, the KS test rejects the homogeneity in bid distributions 
between (b) and any of (c) $\sim$ (f), 
between (c) and (d), and for all pairs in \{(g),(h),(i)\}.
}
\end{center}
\end{figure}

Now, we document some descriptive statistics of the data from the ``printing papers" category.
This category initially has 536 FPPs.
Among them, we exclude 114 for having only one bid, 
three for missing bids, and 
eight for bids above the reserve price.
Among the eight bids, five (three) were submitted by bidders who bid once (twice).
No bidder repeats bidding above the reserve price, and no one wins with such a bid.
In the sample of 411 $(=536-(114+3+8))$  procurements,
we have one type 1 bidder who bids 58 times, 
171 type 2 bidders with 625 bids, 
and 402 type 3 bidders.  
The second and third frequent bidders appear 33 and 14 times in the data.
To
assess
how the type definition affects our analysis, 
in section \ref{sec:sensitivity} we change the definition of type 1 to include the second frequent bidder and then to also include the third frequent bidder.
The entrance rate dramatically drops only for the first a few bidders;
see the circles in Figure \ref{fig:data}(a), suggesting that those 
bidders may 
differ
from other bidders, i.e.,
asymmetry in model primitives. 
Section \ref{sec:sensitivity} also considers alternative type definitions based on how often each bidder wins; see the crosses in Figure \ref{fig:data}(a).

Figure \ref{fig:data}(b) shows the histogram of bids from procurements with two bidders.
It has 474 bids and $(25, 260, 189)$ of them, i.e., $(5,55,40)$\%, are submitted by type $(1,2,3)$ bidders, respectively.
The sample mean and standard deviation of the bids are 0.93 and 0.08. 
Panels (c)$\sim$(f) similarly show the bid data for procurements with 3 to 6 bidders.

Recall that bidders know 
their competitors
when bidding; see section \ref{sec:background}. 
In the data, bidders appear to use the information on the competition. 
The bid distributions vary with the number of bidders, and there is a tendency that the average bid decreases in the number of bidders.
We conduct the Kolmogorov-Smirnov (KS) test against the hypothesis that two marginal bid distributions are identical. 
The $p$ values are close to zero for all pairs involving the two bidder case (b) and for the pair of three bidder case (c) and four bidder case (d), 
rejecting the hypothesis of identical distributions. 
That is, we have some evidence that bidders appear to behave differently according to the competition level; section S3.2 
gives more 
evidence.
As the number of bidders grows, however, the bid distributions get harder to distinguish statistically. 
That might be because the bid weakly converges to the cost with the competition level. The number of bidders also gets noisier in measuring the competition 
because the bidder configuration becomes more variable when bidders are asymmetric.

On average, type 3 bidders bid higher than type 2 and type 2 higher than type 1 (panels (g)$\sim$(i)). (For any pair of two types, the $p$-value of the KS test is close to zero.) 
This pattern arises if type $\tau$ is either more efficient or more risk-averse than type $\tau + 1$. But, the cause of the pattern cannot be detected by reduced form analysis, motivating a structural approach. 

\begin{figure}[t!]
\begin{center}
\caption{Bid Data, Symmetric Procurements, `printing paper'\label{fig:data_sym}}
    \includegraphics
[width=0.8\textwidth]   
{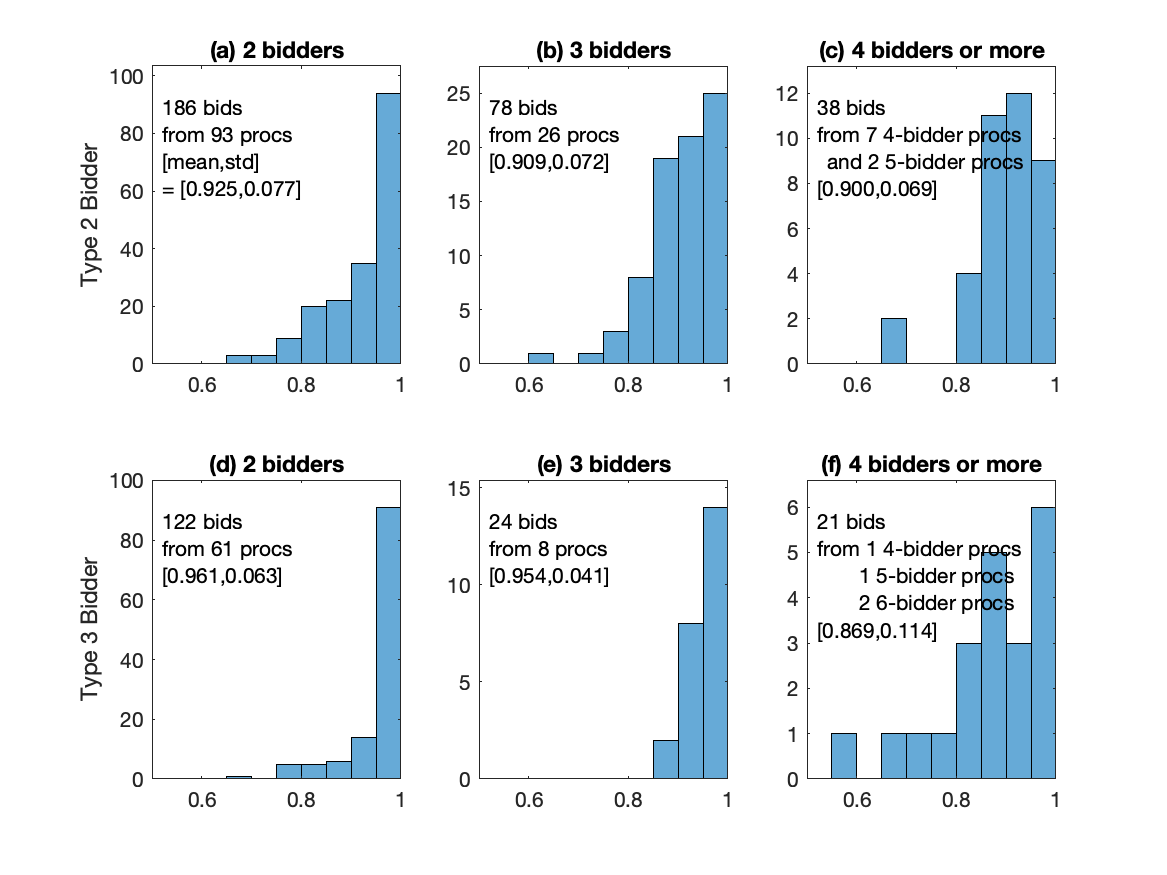}
\caption*{\footnotesize 
For types 2 (upper) and 3 (lower), the left/middle/right panel shows the histogram of bid data from \emph{symmetric} procurements with 2/3/4+ bidders.
All bids are divided by reserve prices.
Each panel shows the number of bids and their [average, standard deviation]. 
\emph{For each type} [For each number of bidders], the KS test rejects the distributional homogeneity against the alternative that the bid \emph{with the larger number of} [of type 2] bidders is first-order stochastically dominated at the 5\% level except for the pair of \emph{(b) and (c)} [(c) and (f)].
}
\end{center}
\end{figure}

Figure \ref{fig:data_sym} shows the bid histograms for symmetric procurements with only type 2 (3) bidders 
in the upper (lower) panels. Note that there is no symmetric procurements with type 1 bidder, as there is one type 1 bidder.
The left and middle panels are for 2 and 3 bidder procurements and the right ones for the rest in the data.
As discussed there, we conduct the KS test and get additional evidence that bidders' behavior depends on the competition level, especially for the cases with sufficiently large bid samples. 

Although we have excluded the procurements with bids above reserve prices, the high density near the reserve price may still seemingly suggest the presence of shill bids (no winning chance). 
According to the data, however, the bid of 0.99 gives  a 14\% chance of winning. Hence, the high density does not indicate shill bids. It is, instead, plausible that bidders' optimal bids are actually near the reserve price because the reserve price has to be justifiable as a market price as discussed 
in section \ref{sec:background}; 
see also section \ref{sec:reserve}.

Finally, a group of suppliers may collude (not necessarily involving the government agent) to submit noncompetitive bids. 
Empirical methods to detect a bidding ring require bidders to be risk-neutral and a suspicious group of bidders to bid together in many procurements; see \cite{Schurter2020} and references therein.
It is infeasible to study a bidding ring using our data because even if the currently available methods may extend for risk-aversion, 
the dropping entrance rate in Figure \ref{fig:data}(a) does not leave us any group of bidders jointly appearing in sufficiently many procurements.
In particular, among 45 pairs of top ten frequent bidders, only eight have ever bid in the same procurements, but rarely: type 1 bidder met the $(2, 4, 6, 7, 9)^{\rm th}$ frequent bidders $(5,1,1,7,4)$ times, respectively, and the $2^{\rm nd}$ and $6^{\rm th}$ bidders met twice, the $3^{\rm rd}$ and $5^{\rm th}$ bidders three times, and the $5^{\rm th}$ and $10^{\rm th}$
bidders 
five times.

\subsection{Reserve Price}\label{sec:reserve}
Recall that the reserve price is set sufficiently high to encourage suppliers' participation and yet still justifiable as a market price to purchase comparable goods and services outside the procurement system.
This description has three important implications on our analysis in the following sections. 
First, the procurement system does not use the reserve price to deter any bidder from entering,
which alone 
might
validate the reserve price as non-binding.
Second, the description of the reserve price also implies that if a supplier incurred a cost higher than the reserve price, the cost is too high for the supplier to operate in the market, even outside the procurements.
Hence, we do not consider such an inefficient supplier as a potential bidder in the procurement, and section \ref{sec:method} specifies the cost densities to have their support below the current reserve price, 
i.e., the latter is nonbinding.
Third, in our policy simulations, 
when no bidder can bid below the counterfactual reserve price, 
we assume that the procurer carries out the job at the current reserve price;  see section \ref{sec:CF}.

\section{Inference Method}\label{sec:method}
A set $\mathcal{I}_t$ of bidders who can beat the reserve price
is exogenously given for each procurement $t \in \{1,\ldots,T\}$.
Bidder $i \in \mathcal{I}_t$ with type $\tau(i) \in \{1,\ldots,\overline{\tau}\}$ has the CRRA coefficient $\eta_{\tau(i)}$ and 
draws her cost $c_{it}$ from the cost distribution $F_{\tau(i)}(\cdot)$ with density $f_{\tau(i)}(\cdot)$ independently of other bidders.
The cost density is strictly positive on $[0,1]$, corresponding to the bid data normalized by the reserve price.
Let $f:= \{f_{\tau}(\cdot):\tau\in\{1,\ldots,\overline{\tau}\}\}$, $\eta:= \{\eta_{\tau}:\tau\in\{1,\ldots,\overline{\tau}\}\}$,  and $\theta:= (f,\eta)$.
Let $\beta_{\tau(i)} (\cdot|\theta, \mathcal{I}_t)$
be the bidding strategy when $u_t = 1$.
That is,
bidder $i$  bids 
    $
    b_{it} = \beta_{\tau(i)} (c_{it}|\theta,\mathcal{I}_t) \in [\underline{b}(\theta,\mathcal{I}_t),1],
    $
    where $\underline{b}(\theta,\mathcal{I}_t) :=\beta_{\tau(i)} (0|\theta,\mathcal{I}_t)$ for all $i \in \mathcal{I}_t$.
     When $u_t \neq 1$, bidder $i$ observes her cost $u_t c_{it} \in u_t\times[0,1] = [0,u_t]$ and bid 
    $
    b_{it} = u_t   \beta_{\tau(i)} (c_{it}|\theta,\mathcal{I}_t)
    $,
    which lies in $[u_t\underline{b}(\theta,\mathcal{I}_t),u_t]$. 
    We model 
    $
    u_1,\ldots,u_t \stackrel{iid}{\sim} f_u(\cdot)  
    $
    with the support of $[\underline{u},1]$ and $\underline{u}>0.$    
Note that $u_t\leq1$ reflects the institutional feature that the reserve price does not exclude any bidder.

For estimation, we specify the cost density as
    \begin{align}
        f_\tau(c) := f(c|\psi_\tau) 
        := 
        \left\{0.01 + 
        0.99 \times \frac
        {\exp\left[ \phi(c)'\psi_\tau\right]}
        {\int_0^1\exp\left[ \phi(\tilde{c})'\psi_\tau\right]d\tilde{c}} 
        \right\}
        \times \mathbbm{1}(c \in [0,1]),
        \label{ftau}
    \end{align}
    where 
    $\psi_{\tau}\in \mathbb{R}^k$ is the vector of parameters and 
    $\phi(\cdot)$ is the vector of the $k$ subsequent Legendre polynomials, defined on $[0,1]$.
	Specifically, the $j$-th entry in $\phi(\cdot)$ is given as 
    $
    \phi_j(c):=\sqrt{2j+1}\times\tilde{\phi}_j(2u-1),    
    $
    where $\tilde{\phi}_j(x):=\frac{d^j}{dx^j}(x^2-1)^j/(2^jj!)$. The uniform component in (\ref{ftau}) with the small weight $(0.01)$ ensures 
    the density is strictly bounded away from zero.

    Note that $\phi_j$ has $j-1$ extrema; 
    see section S4
    for graphs of $\phi_j(\cdot)$ for some $j$s. 
    As $k$ increases, the density of $f_{\tau}(c)$ defined in 
    (\ref{ftau}) can approximate more complicated  densities, i.e., the ones with many inflection points.
Believing that the true cost densities are smooth, we put a smaller prior probability on larger $j$ by the prior $\pi_{\psi_\tau}(\psi_\tau)=\prod_{j=1}^k \pi_j(\psi_{j,\tau})$ with $\psi_{j,\tau}\sim \mathcal{N}(0, (2^{-j})^2)$ for all $j\in \{1,\ldots,k\}$. 
Since 
    the prior variance of $\psi_j$ decreases in $j$, 
    $\psi_j$ gets close to zero as $j$ increases, squeezing out the contribution of higher order components. 
    Hence, 
    cost densities with oscillations are less likely under our prior.
    Note that \eqref{ftau} is the uniform density at the prior mean of $\psi_\tau$.

For the unobserved heterogeneity, we specify $f_u(\cdot)$ 
such that it implies 
    \begin{align}
        u_t \stackrel{i.i.d}{\sim} \mathcal{N}(1,\sigma_u^2) \times \mathbbm{1}(u_t \in [\underline{u}(\sigma_u),1]),\label{ut}
    \end{align}
where the lower bound $\underline{u}(\sigma_u):=1-c_u\times\sigma_u$ with $c_u$ such that $\Pr(|\mathcal{N}(0,1)| \leq c_u)=0.99$, and the upper bound is set to 1     
because the reserve price is non-binding, larger than the upper bound of the cost.
To ensure that $u_t$ is sufficiently larger than zero, we restrict $\underline{u}(\sigma_u)\geq 0.1$. 
This restriction implies the upper bound of $\sigma_u$, i.e., $\sigma_u \leq \overline{\sigma}_u:=(1-0.1)/c_u$.
We then use the uniform prior $\pi_{\sigma_u}(\sigma_u) \propto \mathbbm{1}(0,\overline{\sigma}_u).$
For the CRRA parameter, $\eta_\tau$, we use a uniform prior $\pi_\eta(\eta_\tau) \propto \mathbbm{1}(0,0.9)$, which excludes values near 1
to ensure that our computation does not fail due to the flat utility function, i.e., $u(x) = x^{1-\eta} \rightarrow 1$ as $\eta \rightarrow 1$.
We collect the priors by 
    $
    \pi(\theta,\sigma_u):= \prod_{\tau=1}^{\overline{\tau}} 
    \pi_{\psi_\tau}(\psi_\tau) \pi_{\sigma_u}(\sigma_u)\pi_\eta(\eta_\tau),
    $
where $\theta := \{\theta_\tau\}_{\tau=1}^{\overline{\tau}}$ with  $\theta_\tau:=(\psi_\tau,\eta_\tau)$ and $\overline{\tau}=3$.

Conditional on $(u_t,\theta)$, the bid $b_{it}$ in procurement $t$  has the density 
    \begin{align}
        g_{\tau(i)}(b_{it}|u_t,\theta,\mathcal{I}_t) 
        := \frac{1}{u_t}\times 
        \frac{f[\beta_{\tau(i)}^{-1}(b_{it}/u_t|\theta,\mathcal{I}_t)|\psi_{\tau(i)}]}
        {\beta_{\tau(i)}'[\beta_{\tau(i)}^{-1}(b_{it}/u_t|\theta,\mathcal{I}_t)|\theta,\mathcal{I}_t]}
        \times \mathbbm{1} 
        (
        b_{it}/u_t \in [\underline{b}(\theta,\mathcal{I}_t),1]
        ),\label{biddensity}
    \end{align}
    for which we need to evaluate $\beta_{\tau(i)}^{-1}$.
    To do so, section S5 modifies the boundary-value method to accommodate bidders with CRRA utility in FPPs, which \cite{Fibich_Gavish_2011} originally propose for risk-neutral bidders in high-bid auctions.

Then, the posterior density of the latent variables is given as
    \begin{align*}
        \pi(\theta,\sigma_u,(u_t)_{t=1}^T|z) \propto 
        \pi(\theta,\sigma_u) \prod_{t=1}^T \left\{f_u(u_t|\sigma_u) 
        \prod_{i\in\mathcal{I}_t} g_{\tau(i)}(b_{it}|u_t,\theta,\mathcal{I}_t) \right\},
    \end{align*}
where $z:= ((b_{it})_{i\in\mathcal{I}_t}, \mathcal{I}_t)_{t=1}^T $, including bids and bidder configurations in the dataset. 
Once we have the posterior of the structural parameter $\theta$, 
we are mostly interested in the posterior moments of important functions of $\theta$.
For a measurable function $h(\theta)$, its 
$r^{\rm th}$ posterior moment is 
$
E[h(\theta)^r|z] :=
\int h(\theta)^r \pi(\theta|{z})
d\theta$.
The posterior mean ($r=1$) is often presented along with some uncertainty notions such as the posterior standard deviation, for which $r=2$ is also used.
To evaluate the moments we first draws $\theta^{(1)},\ldots,\theta^{(M)} \sim \pi_\theta(\theta|z)$ by a standard Markov chain Monte Carlo (MCMC) algorithm; see section S6.
Then, we evaluate the moments by the MCMC draws, 
i.e., 
$M^{-1} \sum_{m=1}^M h(\theta^{(m)})^r \stackrel{a.s}{\longrightarrow}E[h(\theta)^r|z]$. 

We conclude this section with a summary of
section S7 
that evaluates the performance of our method using simulated data. 
We consider three types of bidders, each with a different cost density and CRRA parameter, in FPPs with a substantial variation of bidder configurations: $A\in\{(11), (111), (22), (222), (33), (333), (12), (13), (23),(123)\}$. 
We consider two cases: one with a substantial variation in the unobserved heterogeneity, $\sigma_u > 0$, and the other with no variation 
$\sigma_u = 0$. 
For each case, we consider two different sample sizes, $\tilde{T}_A\in\{20,100\}$ for all $A$s.
Using those bid data we find that the MCMC traces are stable and converge quickly, and parameters are accurately estimated.
As desired, the posterior of quantities of interests, such as $f_\tau(\cdot|\theta)$ and $f_u(\cdot|\sigma_u^2)$, becomes more precise around the true values, whether $\sigma_u > 0$ or $\sigma_u = 0$, as the sample size increases.

\section{Inference and Counterfactual Results\label{section:results}}
This section first summarizes the posterior distribution 
of the parameters of the model with 
asymmetric
risk-aversion, and discusses the
bias from two model misspecifications:
imposing symmetric risk-aversion or ignoring risk-aversion altogether.
Then, the section predicts procurement costs under counterfactual scenarios, and also quantifies the impacts of model misspecification in terms of procurement costs.
Finally, the section concludes with some sensitivity analysis.

\subsection{Posterior Inference}\label{section:estimate}
\paragraph{\emph{Asymmetric
Risk-Aversion.}}
We sample $(\theta,\sigma_u)$ from the posterior using the data discussed in section \ref{section:data}, allowing for bidder asymmetry in cost density and risk-aversion.
The top block of Table \ref{tab:estimates} shows the posterior mean, standard deviation, and a 95\% credible interval (2.5 and 97.5 percentiles) for each CRRA coefficient $\eta_\tau$ for $\tau\in\{1,2,3\}$ and $\sigma_u$. 

The estimates suggest that risk-aversion varies across the three types.
Type 1 (most frequent) bidder is the least risk-averse, and type 3 (one-time) bidders are the most risk-averse.
The posterior of $\eta_1$ has a considerable variation, and its mean is substantially smaller than the ones of $\eta_2$ and $\eta_3$. 
The distribution of $\eta_2$ also differs from the one of $\eta_3$, 
where the latter is more precise and (slightly) larger.
Formally, the $p$ values of KS test are all close to zero, rejecting the hypothesis that $\eta_\tau$ and $\eta_{\tilde{\tau}}$ follow 
the identical marginal posterior distributions for $\tau\neq\tilde{\tau}$. Section S8 discusses the use of KS  test in Bayesian analysis.

In addition, there is no unobserved heterogeneity, $\sigma_u\approx 0$, which is consistent with supplying A4 papers being routine. 
If present, any measurable variation in the unobserved heterogeneity should have led to a non-degenerate posterior of $u_t$, even if the parametric assumption in \eqref{ut} 
was incorrect. 
Section \ref{sec:sensitivity} also obtains a degenerate posterior of $u_t$ with a more flexible density of $u_t$. 
\begin{table}[t!]  
 \begin{center}
 \caption{Posterior Distributions of CRRA Coefficients and UH Parameters\label{tab:estimates}}
  \scalebox{0.85}{\begin{tabular}{r|cccccc}
\hline
\hline
& Posterior & Posterior & Posterior  &\multicolumn{3}{c}{$p$-value of KS test }\\
& Mean   & St. Dev. & 95\% Cred. Int.  &$\eta_1$&$\eta_2$&$\eta_3$\\
&(1)&(2)&(3)&(4)&(5)&(6)\\
\hline
Heterogeneous CRRA  $\eta_{1}$ 	 & 0.196 	 & 0.165 	 & [0.005, 0.596] 	 & 	 & 0.000  	 & 0.000 \\
 $\eta_{2}$ 	 & 0.828 	 & 0.043 	 & [0.729, 0.893] 	 & 	 & 	 & 0.000 \\
 $\eta_{3}$ 	 & 0.891 	 & 0.009 	 & [0.867, 0.900] 	 & 	 & 	 & \\
 $\sigma_u$ 	 & 0.000 	 & 0.000 	 & [0.000, 0.000] 	 & 	 & 	 & \\
&&&&&&\\
Homogeneous CRRA  $\eta$ 	 & 0.889 	 & 0.009 	 & [0.865, 0.900] 	 & 0.000 	 & 0.000 	 & 0.000 \\
 $\sigma_u$ 	 & 0.000 	 & 0.000 	 & [0.000, 0.000] 	 & 	 & 	 & \\
&&&&&&\\
No CRRA  $\sigma_u$ 	 & 0.066 	 & 0.001 	 & [0.064,0.068] 	 & 	 & 	 & \\
\hline
\end{tabular}}
 \caption*{\footnotesize The table shows the posterior distributions of CRRA parameters and the parameter of the distribution of the unobserved heterogeneity by their posterior means (1),  standard deviations (2), and posterior 95\% credible intervals (3).
 The columns (4) to (6) report the $p$-values of the KS test against the hypothesis that the two CRRA coefficients are drawn from the same (posterior) distribution.
 }
 \end{center}
\end{table}  

\begin{figure}[t!]
\begin{center}
\caption{Posterior Predictive Cost Densities\label{fig:densities}}
    \includegraphics
[width=0.8\textwidth]   
{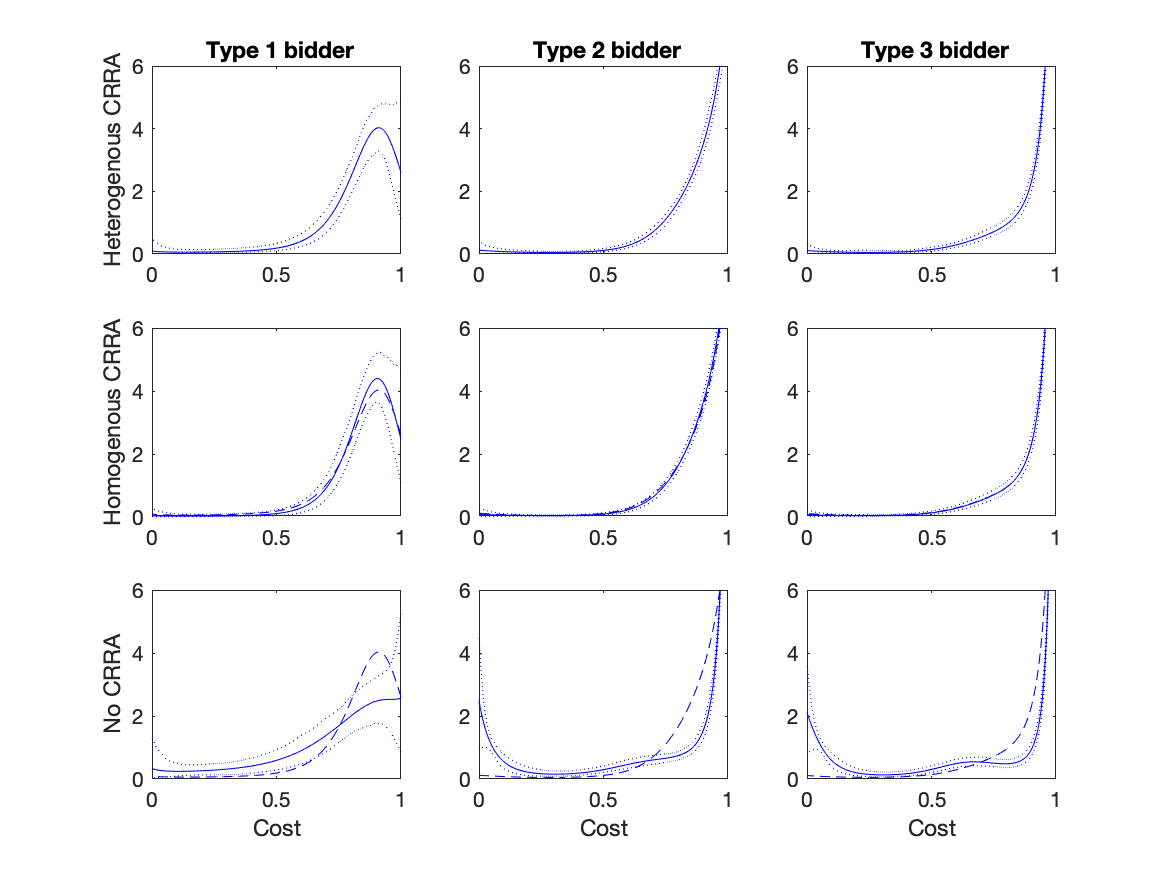}
\caption*{\footnotesize 
The upper block shows the posterior predictive cost density (solid) and 95\% credible band (dotted) for each type. 
The middle and lower repeat the exercise by imposing homogeneity in risk-aversion and imposing no risk-aversion, respectively. The dashed lines are the predictive densities pasted from the upper block. 
}
\end{center}
\end{figure}

The top panels in Figure \ref{fig:densities} show,
for each type $\tau \in \{1,2,3\}$,
the posterior mean of the cost density \eqref{ftau} at every point $c \in [0,1]$ by a solid line and a 95\% credible band around the predictive density by dotted lines. 
Bidders are asymmetric in the cost densities.
Type 1 (frequent) bidder is more efficient in supplying papers than the other bidders, with type 3 (one-time) bidders being the least efficient. 
This prediction, especially for types 2 and 3, is precise, as indicated by the tight credible bands. 
Moreover, by the KS test, we reject the hypothesis that 
the posterior predictive cost distributions
are identical for each pair of types; the relevant $p$-values are all close to zero. Note that we apply the KS test to cost samples drawn from the predictive cost densities; see section S8.

\paragraph{\emph{Homogeneous Risk-Aversion.}}
We analyze the data again but imposing $\eta = \eta_\tau$ for all $\tau \in \{1,2,3\}$. 
The middle block of Table \ref{tab:estimates} summarizes the posterior distribution of $\eta$, which precisely predicts that all bidders are highly risk-averse. 
Note that we reject the hypothesis that the posterior distribution of the constrained $\eta$ is identical to the 
posterior
of $\eta_\tau$ from the upper block (heterogeneous CRRA) for each $\tau\in \{1,2,3\}$.
For this constrained model, there is no unobserved heterogeneity $\sigma_u \approx 0$ as with asymmetric risk-aversion above. 
The middle panels of Figure \ref{fig:densities} show the posterior predictive densities (solid) and 95\% credible bands (dotted), where the dashed lines copy the predictive densities from the upper block for comparison. 
Imposing homogeneity in risk-aversion generates a slightly different predictive cost density for type 1 bidder, but the other types remain the same.
The KS test  with the 5\% level rejects 
the hypothesis that the predictive cost distribution remains the same under the constraint for type 1 bidders.

\paragraph{\emph{Risk-Neutrality.}}
We repeat the exercise but restricting $\eta_\tau = 0$ for all $\tau$.
Table \ref{tab:estimates} and Figure \ref{fig:densities} (bottom) suggest that 
this misspecification induces an overestimation of unobserved heterogeneity and substantial bias in predicting cost densities; the more risk-averse, the larger the bias.
Once we impose risk-neutrality, smaller bids in our sample must be justified by other model components, causing the bias pattern as we observe.
For all $\tau\in\{1,2,3\}$, the KS test rejects, by $p$-value $\approx0$, the hypothesis that the cost under $\eta_\tau = 0$ follows the same predictive distribution as the cost 
without the constraint.

\subsection{Counterfactual Analysis}\label{sec:CF}
Russia has constantly been updating the procurement system mainly to reduce government spending, as we mentioned in section \ref{sec:background}.
We, therefore, evaluate counterfactual scenarios by predictive procurement costs and investigate implications of misspecified risk-aversion. 
We also compute predictive costs and efficiency at other relevant policy options for comparison.

\paragraph{\emph{Decision Theoretic Approach.}}
Since we study the policymaker's \emph{decision} problem, we use a statistical decision-theoretic approach; see \cite{Berger_1985} for a survey. 
Note that \cite{KimDH_2013_IJIO} introduces the approach for empirical auction design and \cite{Aryal_Kim_2013}, \cite{KimDH_2014_BSL}, and  \cite{AryalGrundlKimZhu2017} use or extend for different contexts.

If the policymaker knows $\theta$, 
he can choose 
an action $\rho\in\mathcal{A}$, e.g., a reserve price, 
to minimize the (expected) procurement cost $\Lambda(\rho,\theta)$.
Let $\rho^*(\theta):=\argmin_{\rho\in\mathcal{A}} \Lambda(\rho,\theta)$.
Choosing $\rho^*(\theta)$ is infeasible, however, as $\theta$ is uncertain.
The posterior $\pi(\theta|z)$ represents the policymaker's uncertainty about $\theta$,
combining his prior 
and the sample 
$z$. 
Thus, he should choose
\begin{align}
\rho_B(z) := \argmin_{\rho\in\mathcal{A}} \left\{\int \Lambda(\rho,\theta) \pi(\theta|z) d\theta = E[ \Lambda(\rho,\theta)|z]\right\}.\label{eq:Baction}
\end{align}
This is the idea of the Bayesian decision theory and it is similar to the usual expected utility theory.
Choosing $\rho_B(z)$, known as a \emph{Bayes action}, is rational under the axioms of \cite{Savage_1954} and \cite{Anscombe_Aumann_1963}.
A \emph{decision rule} that maps every data $z$ to $\rho_B(z)$ is optimal under a frequentist perspective (Bayes risk principle).

This approach formally considers the structure of the procurement cost and uncertainty and, therefore, it
may incur smaller procurement costs than a `plug-in' method.
Since 
$\theta$
is unknown, the policymaker would choose some $\rho \neq \rho^*(\theta)$ in practice.
Consider a cost function that drops sharply before $\rho^*(\theta)$ and slowly increases after 
$\rho^*(\theta)$.
For this cost structure, the policymaker must prefer  $\rho_{\rm large}>\rho^*(\theta)$ to $\rho_{\rm small}<\rho^*(\theta)$ for the same error, i.e., $\rho_{\rm large}-\rho^*(\theta) = \rho^*(\theta) - \rho_{\rm small}$.
Especially, if the cost  is (almost) flat after $\rho^*(\theta)$,
then,  $\rho_{\rm max} =\max\mathcal{A}$ is (almost) equivalent to $\rho^*(\theta)$. 
The extent to which the policymaker prefers a large action must depend on the cost structure and amount of uncertainty. 
For example, if there is no uncertainty about $\theta$, he would pick $\rho^*(\theta)$ regardless of the cost structure. 
Alternatively, if the cost is flat after $\rho^*(\theta)$ for all $\theta$ as in the case with a large number of bidders, he would choose $\rho_{\rm max}$.
Solution \eqref{eq:Baction} formalizes the idea of making a decision considering the cost structure and uncertainty. 
On the other hand, the plug-in approach, 
which is first popularized in empirical auctions by \cite{Paarsch_1997}, chooses $\rho^*(\hat{\theta}(z))$. 
That is, the plug-in approach regards the estimate $\hat{\theta}(z)$ as the \emph{true} parameter in the decision problem, meaning that it ignores parameter uncertainty and the shape of the cost function (other than the fact that $\rho^*(\theta)$ is a minimizer of $\Lambda(\rho,\theta)$.)

After formalizing uncertainty by the data $z$ and maintained assumptions (model and prior), $\rho_B(z)$ is \emph{certainly} the best action. 
So, no uncertainty notion comes with $\rho_B(z)$. 
Specifically, a credible interval represents uncertainty associated with $\theta$, but $\rho_B(z)$ is the action after integrating out $\theta$ by $\pi(\theta|z)$.
We instead provide a credible interval of 
procurement cost
at $\rho_B(z)$ or other notable actions, which is natural because the cost is the outcome of interest and is still uncertain at $\rho_B(z)$.
This is, however, different from the convention to construct a confidence interval around the plug-in \emph{estimate} $\rho^*(\hat{\theta}(z))$ to consider the variation in random data $z$ in repeated sampling. 
The confidence interval is designed for testing a hypothesis like $\rho^* = 0$, which is a decision problem different from the policymaker's problem.

We explain in section S9 our algorithm to evaluate the predictive procurement cost, $E[\Lambda(\rho,\theta)|z]$, which integrates out $\theta$ and $\mathcal{I}$, respectively, by the posterior and the empirical distribution of $\mathcal{I}$, considering (stochastic) refinements of $\mathcal{I}$ due to binding reserve prices.   

\begin{figure}[t!]
\begin{center}
\caption{Counterfactual Analysis, Common Reserve Price\label{fig:common}}
    \includegraphics
[width=0.8\textwidth]   
{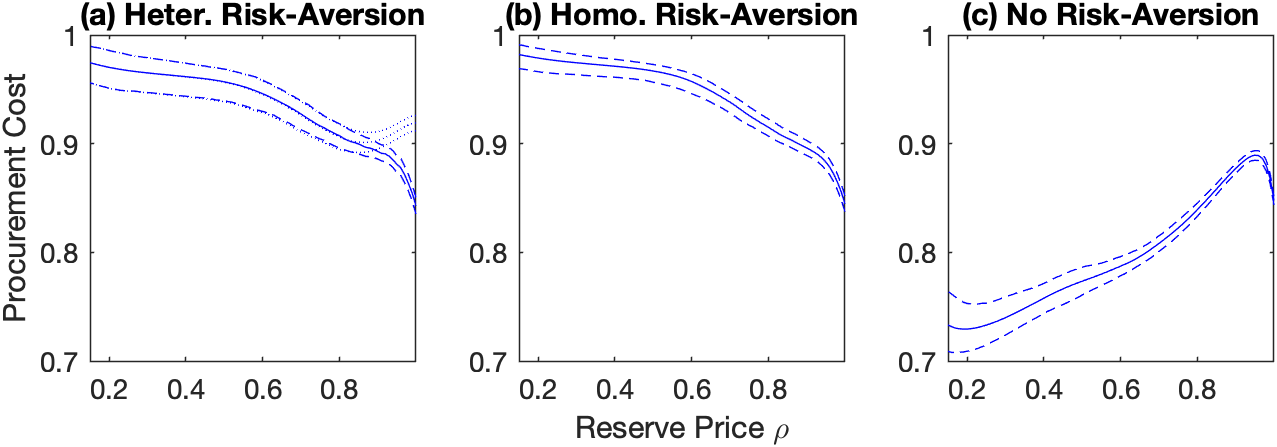}
\caption*{\footnotesize 
Each panel shows 
the predictive procurement cost
at each reserve price $\rho$ (solid line) along with its 95\% credible band (dashed) for the models with heterogeneous risk-aversion, homogeneous risk-aversion, and no risk-aversion. 
Panel (a) also shows the predictive cost under the second-price procurement (dotted). 
}
\end{center}
\end{figure}

\paragraph{\emph{Common Reserve Prices.}} 
We consider a situation where the policymaker wishes to choose one reserve price and apply it to all bidders regardless of their types, i.e., $\rho_1 = \rho_2 = \rho_3$.
Figure \ref{fig:common}(a) shows $E[\Lambda(\rho,\theta)|z]$ as a function of $\rho$ (solid line) and its 95 percent credible band (dashed lines), i.e., the 2.5 and 97.5 percentiles of $\Lambda(\rho,\theta)$ under the posterior at every $\rho$ in the figure.
The three dotted lines show the posterior predictive cost and its credible band if the policymaker implements the second-price procurements (SPP).
When bidders are risk-averse, bidders bid  more aggressively in an FPP than in an SPP.
In particular, the figure shows that the FPP with $\rho_c=(1,1,1)$ results in lower costs than the SPP with the cost-minimizing reserve price. 
Panels (b) and (c) show the predictive costs under the models with homogeneous risk-aversion and no risk-aversion, respectively.
When bidders are modeled to be risk-averse, the predictive cost is minimized at $\rho_c$ whether or not risk-aversion is
type-specific.
When risk-aversion is ignored, however, the method recommends a much smaller reserve price. 
This result follows from the fact that
the cost densities in Figure \ref{fig:densities} (bottom) falsely predict a large probability of small costs, especially for type 2 and 3 bidders, while these bidders would draw high costs more likely, Figure \ref{fig:densities} (top).

The first block (common $\rho$) in Table \ref{tab:CF}  documents that 
the predictive procurement cost
is 0.843 with a 95\% credible interval of $[0.835, 0.850]$ at the cost-minimizing reserve price $\rho_B^{C}(z) := \rho_c$, where the superscript $C$ indicates that the Bayes action is selected under the restriction of common reserve price.
The table also shows that the efficient bidder wins the procurement with a 99.2\% chance at $\rho_B^{C}(z)$. 
This prediction is similar even when risk-aversion is restricted to be homogeneous. 
Table \ref{tab:CF} also shows that the model with  risk-neutrality selects $\rho_B^{C,RN}(z):=(0.19,0.19,0.19)$ and predicts 
that the procurement cost
would be 0.729 at $\rho_B^{ C,RN}(z)$, 
where the superscript $RN$ is for risk-neutrality.
That is, this misspecified model predicts $14.0\% \approx |0.729-0.848| \div 0.848$ of cost-reduction at $\rho_B^{ C,RN}(z)$, where 0.848 is the predictive cost at $\rho_c$; see the third block (for comparison) in  Table \ref{tab:CF}.
However, this prediction is misleading: the model with asymmetric CRRAs predicts the 
procurement
cost of 0.971 at $\rho_B^{C,RN}(z)$, increasing the cost by  $15.2\% \approx |0.971-0.843|\div 0.843$.
At $\rho_B^{C,RN}(z)$, moreover, the model under risk-neutrality also predicts that the efficient bidder would win 
with a
33.2\% of chance, but the chance of allocation is only 3.6\% under the model with asymmetric CRRAs.
That is, if one ignores risk-aversion, the proposed policy will substantially increase the
procurement 
cost, and most procurements will fail to find a supplier.
\begin{table}[t]  
 \begin{center}
 \caption{Counterfactual Analysis\label{tab:CF}}
  \scalebox{0.85}{\begin{tabular}{r|cccc}
\hline
\hline
 	&  Cost  Min.	& \multicolumn{1}{c}{Predictive}	& \multicolumn{1}{c}{Prob. that} 	& \multicolumn{1}{c}{Prob. of}		\\
 	& Reserve Price			& \multicolumn{1}{c}{Procurement Cost} 		& \multicolumn{1}{c}{Lowest Wins} 	& \multicolumn{1}{c}{Transaction, if $\neq1$}	\\
 	&(1)&(2)&(3)&(4)\\
\hline 
\multicolumn{1}{l|}{\textbf{Common $\rho$}} &&&&\\
 Heterogeneous CRRA	 & (1.00, 1.00, 1.00)	 & 0.843 [0.835, 0.850] 	 & 0.992 [0.989, 0.996] 	 & 
 \\
 Homogenous CRRA 	 & (1.00, 1.00, 1.00)	 & 0.845 [0.837, 0.852] 	 & 0.994 [0.993, 0.996] 	 & 
 \\
 No Risk-Averson 	 & (0.19, 0.19, 0.19)	 & 0.729 [0.708, 0.754] 	 & 0.332 [0.301, 0.358] 	 & 0.333 [0.303, 0.359] \\
 &&&&\\
\multicolumn{1}{l|}{\textbf{Type-Specific $\rho$}} &&&&\\
 Heterogeneous CRRA	 & (0.96, 1.00, 1.00)	 & 0.841 [0.834, 0.849] 	 & 0.991 [0.989, 0.995] 	 & 
 \\
 Homogenous CRRA 	 & (1.00, 1.00, 1.00)	 
 &&&\\
 No Risk-Averson 	 & (0.75, 0.20, 0.20)	 & 0.722 [0.703, 0.747] 	 & 0.368 [0.335, 0.392] 	 & 0.376 [0.344, 0.401] \\
 &&&&\\
\multicolumn{1}{l|}{\textbf{For Comparison}} &&&&\\
Heterogeneous CRRA	 & (0.19, 0.19, 0.19)	 & 0.971 [0.952, 0.987] 	 & 0.036 [0.016, 0.059] 	 & 0.036 [0.016, 0.059] \\
Heterogeneous CRRA	&  (0.75, 0.20, 0.20)	 & 0.965 [0.944, 0.980] 	 & 0.060 [0.037, 0.087] 	 & 0.063 [0.040, 0.091] \\
No Risk Aversion  	&  (1.00, 1.00, 1.00)	 & 0.848 [0.843, 0.851] 	 & 0.985 [0.982, 0.988] 	 & 
\\
 &&&&\\
\multicolumn{1}{l|}{\textbf{Additional Bidder }} &&&&\\
 Type 1 Bidder &	 & 0.791 [0.775, 0.804] 	 & 0.988 [0.981, 0.996] \\
 Type 2 Bidder &	 & 0.794 [0.782, 0.805] 	 & 0.996 [0.994, 0.998] \\
 Type 3 Bidder &	 & 0.797 [0.784, 0.811] 	 & 0.996 [0.993, 0.997] \\
\hline 
\end{tabular}}
 \caption*{\footnotesize 
 Column (1) shows the Bayes action, i.e., cost-minimizing reserve price, 
 and columns (2) to (4) show  predictive outcome variables along with 95\% credible intervals: the
 procurement
 cost, the probability that the bidder with the lowest cost wins, and the probability that at least one bidder has a cost below the (or her own) reserve price. 
Column (4) shows the predictive probabilities only when the prediction is different from one.
 }
 \end{center}
\end{table}  

\paragraph{\emph{Type-Specific Reserve Prices.}} 
Now, we consider the policymaker who can select type-specific reserve prices.
The second block (type-specific $\rho$) in Table \ref{tab:CF} shows that the model with asymmetric risk-aversion recommends $\rho_B^{T}(z)=(0.96,1.00,1.00)$, which may screen out type 1 (most frequent) bidder and this solution, $\rho_B^{T}(z)$, results in the predictive cost of 0.841, where the superscript $T$ indicates
that the Bayes action is 
type-specific. 
But, the cost saving is marginal relative to the current cost of 0.843 at $\rho_c = (1,1,1)$, suggesting that the predictive cost is flat in $\rho_1$ around 0.96 near one for $(\rho_2,\rho_3)=(1,1)$ and, therefore, any
$\rho_1$ near one
would be practically cost-equivalent.

When risk-aversion is restricted to be homogeneous, 
the method still chooses $\rho_c$. 
However, ignoring risk-aversion introduces a large bias: the method selects $\rho_B^{T,RN}(z) = (0.75,0.20,0.20)$ at which it predicts the 
procurement
cost of 0.722 ($14.9\%\approx |0.722 - 0.848|\div0.848$ of cost-saving), but the model with heterogeneous risk-aversion predicts the cost of 0.965 at $\rho_B^{T,RN}(z)$ ($14.5\% \approx |0.965 - 0.843|\div0.843$ of cost increase). 
Although efficiency is by assumption not of the first order interest 
for the policymaker, 
the efficiency measures predicted with risk-neutrality are also severely biased: 
the model with heterogeneous risk-aversion predicts that roughly 94\% of procurements would fail to find a supplier through the mechanism.

\paragraph{\emph{Cost Reduction from Inviting an Additional Bidder.}} 
Finally, we consider the case in which the buyer can invite one additional (genuine) bidder. 
\cite{bulow_Klemperer_1996} show that 
inviting an additional bidder would improve the seller's revenue more than choosing a revenue maximizing reserve price in a standard symmetric auction with risk-neutral bidders.
But, it is unclear whether this finding would hold in our asymmetric model with risk-averse bidders.      

We evaluate the predictive procurement cost at $\rho_c$ using the posterior with the model where bidders are heterogeneous in risk-aversion. 
When the buyer invites a type 1 bidder, i.e., type 1 bidder is added to all bidder configurations 
$\mathcal{I}$,
our method predicts the cost of 0.791 with a 95\% credible interval of $[0.775,0.804]$. 
The predictive procurement cost
in this case is 6.2\% $(\approx  |0.791 - 0.843| \div 0.843)$ smaller than 
the predictive cost at $\rho_c$, the first row in Table \ref{tab:CF}, with non-overlapping credible intervals.
We have similar results when the procurer invites a bidder from other types. 
When the procurer invites one additional type 2 (3) bidder, the predictive cost would be 0.794 (0.797), which is a 5.8 (5.5)\% of cost reduction, with a 95\% credible interval of $[0.782,0.805]$ (of $[0.784,0.811]$).

Note that even if a fringe bidder (type 3) is invited, this bidder will lower the 
procurement
cost more than selecting the 
cost-minimizing reserve price.
Hence, we find that the insight of \cite{bulow_Klemperer_1996} holds for the ``printing papers" category of Russian procurements where bidders are asymmetric in both cost density and risk-aversion.

\subsection{Sensitivity Analysis}\label{sec:sensitivity}
This subsection examines how sensitive our empirical findings are to the definition of bidder types and prior specifications.
In this
subsection, the \emph{main} specification refers to the one with heterogeneous risk-aversion outlined in section \ref{sec:method}, 
which gives the estimates in the top block of Table \ref{tab:estimates}.
We then consider six alternative specifications.
The first (second) specification classifies the two (three) most frequent bidders as 
a type 1 bidder.
Recall that the most (second-most) frequent bidder appears in the data 58 (33) times, but   
the frequency does not change much starting from the third frequent bidder, who appears 14 times; see Figure \ref{fig:data}(a).

The type definitions here suggest that bidders' entry depends on model primitives such as risk-aversion and cost density. To avoid resorting to any entry model, however, one might want to alternatively define the types, e.g., by how often they win. In our data, the two most frequent entrants are also the most frequent (42 and 19 times) winners. Thus, the analysis remains the same even if one defines type 1 bidder based on the winning rate for the first two bidders. However, the fourth entrant is the third (8 times) winner. The third specification, \emph{3 Type 1 Bidders (win)}, defines type 1 bidders as the three most frequent winners. The winning rate does not drop after that; see Figure \ref{fig:data}(a).

The prior variance of $\psi$ of the fourth (fifth) specification is four times smaller (larger) than the main specification; see 
the cost density \eqref{ftau}
where we introduce $\psi$. 
The last one adopts a more flexible density for 
the unobserved heterogeneity $\{u_t\}$. 
Extending \eqref{ut}, we specify $u_t \stackrel{i.i.d}{\sim} \mathcal{N}(\mu_u,\sigma_u^2) \times \mathbbm{1}(u_t \in [\underline{u},1]),$
where $\sigma_u>0$, $\underline{u} > 0$, and $\mu_u\in [\underline{u},1)$ so that the distribution of unobserved heterogeneity is indexed by a three-dimensional parameter vector $(\mu_u,\sigma_u,\underline{u})$.
Recall that the main specification indexes $f_u$ by a one-dimensional parameter $\sigma_u$ with  
the restriction of $(\mu_u,\underline{u})=(1,1-c_u\times\sigma_u)$;
see \eqref{ut}. 
Note that we use a flat prior for all the other parameters such as $\sigma_u$ and $(\eta_1,\eta_2,\eta_3)$ 
in the main specification.

All the six alternative specifications produce predictive cost densities close to the ones under the main specification; see section S10.1. 
We test the hypothesis that the predictive distribution of type 1 bidder's cost remains the same when we change the definition of type 1 bidder to include the second frequent bidder: 
Table \ref{tab:sens_par_p} reports that the associated $p$-value is 0.425, and we fail to reject the hypothesis at any conventional level.
Similarly, we conduct the hypothesis testing for other types.
Then, we repeat it
for the other specifications. 
In all cases, we fail to reject the hypothesis that the predictive cost distribution remains the same. 

\begin{table}[t]  
 \begin{center}
 \caption{Sensitivity Analysis on Latent Variables \label{tab:sens_par_p}}
  \scalebox{0.85}{
  \begin{tabular}{l|ccc|ccc}
\hline
\hline
& \multicolumn{3}{c|}{Cost $\{c_\tau\}$, type $\tau=1,2,3$} & \multicolumn{2}{c}{Unobserved Heterogeneity $\{u_t\}$}  \\
& \multicolumn{3}{c|}{KS test, $p$ values} & KS test & Predictive Distribution  \\
Specifications & Type 1 & Type 2 & Type 3 & $p$ value &  mean (stdev) [95\% CI] \\
&(1)&(2)&(3)&(4)&(5)\\
\hline 
(0) Main && NA & & NA & 1.000 (0.000) [0.999,1.000] \\
(1) 2 Type 1 Bidders	& 0.425 	& 1.000 	& 1.000 	& 0.000 	& 1.000 (0.000) [0.999,1.000] \\
(2) 3 Type 1 Bidders	& 0.258 	& 0.999 	& 0.999 	& 0.000 	& 1.000 (0.000) [1.000,1.000] \\
(3) 3 Type 1 Bidders (wins)	& 0.194 	& 0.679 	& 0.952 	& 0.000 	& 1.000 (0.000) [1.000,1.000] \\
(4) Small Prior $V(\psi)$	& 0.105 	& 0.789 	& 0.716 	& 0.000 	& 1.000 (0.000) [0.999,1.000] \\
(5) Large Prior $V(\psi)$	& 0.307 	& 0.999 	& 0.952 	& 0.000 	& 1.000 (0.000) [0.999,1.000] \\
(6) Alternative $f_u(\cdot)$	& 0.988 	& 1.000 	& 0.988 	& 0.000 	& 1.000 (0.000) [1.000,1.000] \\
\hline 
\end{tabular}}
 \caption*{\footnotesize 
 Columns (1) to (3) show the $p$-values of the KS test against the hypothesis that the predictive cost distribution remains the same under the 
 alternative specifications (within each type). 
 Column (4) does similarly for the distribution of the unobserved heterogeneity. 
 Column (5) summarizes the posterior of $u_t$.
}
\end{center}
\end{table}  

We also consider the predictive distribution of the unobserved heterogeneity $\{u_t\}$. 
The KS test strongly rejects, with $p$ values close to zero, the hypothesis that the predictive distribution of $\{u\}$ under the main specification is the same as the distribution under the alternative specification for each of the 
six
cases; see column (4) of Table \ref{tab:sens_par_p}.
The specifications, however, \emph{unanimously} predict that $\{u_t\}$ is practically degenerate at one: its mean and standard deviation are approximately one and zero; see column (5) of Table \ref{tab:sens_par_p}.
This is an example 
to show that a \emph{statistically} significant difference can be \emph{economically} meaningless.

Table \ref{tab:sens_eta} summarizes the posterior of the type-specific CRRA coefficients, $\eta_\tau$ for $\tau \in \{1,2,3\}$. 
Including the second frequent bidder in type 1 does not change the prediction on  $\eta_1$, but the third bidder, when classified as type 1, inflates the prediction on $\eta_1$.
However, that should be natural if the third bidder is similar to type 2 bidders as suggested by the entrance rate (Figure \ref{fig:data}(a)) because type 2 bidders are highly risk-averse. 
Similarly, the specification defining the three most winning bidders as type 1 bidder predicts a high $\eta_1$, which should also be natural because type 1 bidder includes the fourth frequent entrant, who is a type 2 bidder with a high risk aversion under the main specification.

The stronger prior on $\psi$ shrinks the prediction on $\eta_1$ toward zero, but the weaker prior on $\{u_t\}$ does not substantially change the prediction. 
The prediction on  $(\eta_2,\eta_3)$ is more robust than $\eta_1$ because the bid samples of those types are 7 to 10 times larger than type 1. 
Overall, all the specifications give qualitatively the same prediction on $(\eta_1,\eta_2,\eta_3)$:  
type 1 bidders are the least risk-averse, and the other bidders are highly risk-averse with $\eta_2<\eta_3$.
Note that the KS test rejects, at any conventional level, the hypothesis that the posterior of $\eta_\tau$ under the main specification 
equals
the posterior of $\eta_\tau$ under the alternative specification for each of the 
six
cases.
\begin{table}[t]  
 \begin{center}
 \caption{Sensitivity Analysis on Posterior of CRRA Coefficients \label{tab:sens_eta}}
  \scalebox{0.85}{
  \begin{tabular}{l|ccc}
\hline
\hline
 & \multicolumn{3}{c}{Posterior mean (standard deviation) [95\% credible interval]} \\
Specifications & $\eta_1$ & $\eta_2$ & $\eta_3$     \\
\hline 
Main	& 0.196 (0.165) [0.005,0.596] 	& 0.828 (0.043) [0.729,0.893] 	& 0.891 (0.009) [0.867,0.900] \\
2 Type 1 Bidders	& 0.196 (0.152) [0.007,0.548] 	& 0.819 (0.042) [0.736,0.888] 	& 0.891 (0.008) [0.872,0.900] \\
3 Type 1 Bidders	& 0.350 (0.176) [0.029,0.692] 	& 0.817 (0.042) [0.725,0.886] 	& 0.892 (0.008) [0.870,0.900] \\
3 Type 1 Bidders (wins)	& 0.597 (0.127) [0.331,0.823] 	& 0.865 (0.032) [0.779,0.899] 	& 0.888 (0.012) [0.856,0.900] \\
Small Prior $V(\psi)$	& 0.141 (0.118) [0.005,0.443] 	& 0.798 (0.043) [0.704,0.869] 	& 0.890 (0.010) [0.863,0.900] \\
Large Prior $V(\psi)$	& 0.242 (0.197) [0.006,0.720] 	& 0.842 (0.043) [0.740,0.898] 	& 0.891 (0.008) [0.870,0.900] \\
Alternative $f_u(\cdot)$	& 0.202 (0.162) [0.008,0.585] 	& 0.819 (0.041) [0.727,0.887] 	& 0.890 (0.010) [0.858,0.900] \\
\hline 
\end{tabular}}
\caption*{\footnotesize Each column shows the posterior mean of $\eta_\tau$, standard deviation in ( ), and a 95\% credible interval in [ ].  }
\end{center}
\end{table}  

However, the \emph{statistically} significant differences in the posterior distributions of $\eta_\tau$ between the specifications would not be large enough to induce an \emph{economically} significant impact on the policymaker's decision problem. 
When the policymaker applies a common reserve price to all bidders, our decision method selects the current reserve price as the cost-minimizing price under all the specifications, giving similar predictions on the procurement cost and the likelihood of the lowest bidder winning the procurement; see the upper block of Table \ref{tab:sen_rho}.
When the policymaker can choose bidder-specific reserve prices, our method selects different reserve prices for type 1 depending on the specification. 
As we discussed in the previous subsection, the predictive cost with the main specification is practically the same for $\rho_1$ near one at $(\rho_2,\rho_3) = (1,1)$. 
All the alternative specifications produce similar predictive costs as a function of $\rho_1$ given  $(\rho_2,\rho_3) = (1,1)$. That is, the method could select any price $\rho_1$ near one, especially when the cost functions are evaluated by Monte Carlo, but all giving similar predictions on the outcome variables of interest; see the lower block of Table \ref{tab:sen_rho}.
All the specifications predict that the current mechanism is effectively cost-minimizing.
Finally, we repeat the counterfactual analysis of inviting one additional bidder for each alternative specification and obtain predictions on the outcome variables of interest that are similar to the prediction under the main specification; see section S10.2.
\begin{table}[t]  
 \begin{center}
 \caption{Sensitivity Analysis on Counterfactual Studies\label{tab:sen_rho}}
  \scalebox{0.85}{\begin{tabular}{l|ccc}
\hline
\hline
 	&  Cost  Min.	& Predictive	& Probability that 	\\
 	& Reserve Price			& Procurement Cost 		& Lowest Cost Bidder Wins 	\\
 	&(1)&(2)&(3)\\
\hline 
\multicolumn{1}{l|}{\textbf{Common $\rho$}:} &&&\\
Main Spec	 & (1.00,1.00,1.00) 	 & 0.843 (0.004) [0.835, 0.850] 	 & 0.992 (0.002) [0.989, 0.996] \\
2 Type 1 Bidders	 & (1.00,1.00,1.00) 	 & 0.843 (0.004) [0.836, 0.850] 	 & 0.991 (0.002) [0.988, 0.995] \\
3 Type 1 Bidders	 & (1.00,1.00,1.00) 	 & 0.843 (0.004) [0.836, 0.850] 	 & 0.992 (0.002) [0.989, 0.996] \\
3 Type 1 Bidders (wins)	 & (1.00,1.00,1.00) 	 & 0.842 (0.004) [0.834, 0.849] 	 & 0.994 (0.002) [0.991, 0.997] \\
Small Prior $V(\psi)$	 & (1.00,1.00,1.00) 	 & 0.840 (0.004) [0.832, 0.848] 	 & 0.992 (0.001) [0.989, 0.994] \\
Large Prior $V(\psi)$	 & (1.00,1.00,1.00) 	 & 0.844 (0.004) [0.837, 0.850] 	 & 0.992 (0.002) [0.989, 0.996] \\
Alternative $f_u(\cdot)$	 & (1.00,1.00,1.00) 	 & 0.843 (0.004) [0.836, 0.851] 	 & 0.992 (0.002) [0.990, 0.996] \\
&&&\\
\multicolumn{1}{l|}{\textbf{Type Specific $\rho$}:} &&&\\
Main Spec	 & (0.96,1.00,1.00)	 & 0.841 (0.004)  [0.834,0.849] 	 & 0.991 (0.001)  [0.989,0.995] \\
2 Type 1 Bidders	 & (0.93,1.00,1.00)	 & 0.841 (0.004)  [0.832,0.849] 	 & 0.986 (0.001)  [0.984,0.989] \\
3 Type 1 Bidders	 & (0.93,1.00,1.00)	 & 0.843 (0.005)  [0.833,0.851] 	 & 0.986 (0.002)  [0.983,0.989] \\
3 Type 1 Bidders (wins)	 & (1.00,1.00,1.00)	 & 0.842 (0.004)  [0.834,0.849] 	 & 0.994 (0.002)  [0.991,0.997] \\
Small Prior $V(\psi)$	 & (0.95,1.00,1.00)	 & 0.838 (0.004)  [0.830,0.846] 	 & 0.991 (0.001)  [0.988,0.993] \\
Large Prior $V(\psi)$	 & (0.95,1.00,1.00)	 & 0.843 (0.004)  [0.835,0.850] 	 & 0.991 (0.002)  [0.988,0.994] \\
Alternative $f_u(\cdot)$	 & (0.95,1.00,1.00)	 & 0.842 (0.004)  [0.833,0.849] 	 & 0.991 (0.001)  [0.989,0.994] \\
\hline 
\end{tabular}}
 \caption*{\footnotesize  Column (1) shows cost-minimizing type-specific reserve prices (Bayes actions), and (2) and (3) show the posterior predictive cost and efficiency, posterior standard deviation in ( ), and a 95\% credible interval in [ ].}
\end{center}
\end{table}  

\section{Concluding Remarks}\label{section:conclusion}
We conclude this paper with a discussion about some possible extensions to our method. 
First, one may consider a non-separable unobserved heterogeneity instead of the separable one as introduced in Assumption \ref{as:UH}-3. 
To be specific, if $u$ is discrete with a finite support and if
$\underline{b}_{\mathcal{I}}(u) = \beta_i(\underline{c}|u,\mathcal{I})$  is strictly increasing in $u$  for all $i\in\mathcal{I}$, 
the bid distribution for $\mathcal{I}$ with $|\mathcal{I}|\geq3$
identifies the conditional bid distribution given $u$ for the given $\mathcal{I}$, 
following \cite{Hu_McAdams_Shum_2013}.
Those identified conditional bid distributions at any $u$ in its finite support then identify $\{F_i,\eta_i\}$ with variation in $\mathcal{I}$ following \cite{Campo2012}.
In our data, 237 procurements out of $T=411$, approximately 58\%, have only  $|\mathcal{I}| = 2$ bidders, and we do not consider this specification for estimation.

Second, one may treat bidders' type as an additional parameter to estimate instead of fixing a type for each bidder. 
For example, \cite{An2017} proposes a method to estimate bidder types.
Our sample, however, does not meet its requirement. In particular, every bidder must appear at least in three different procurements. But, this is not the case in our data, where 84\% of bidders enter once or twice.
In our Bayesian setting, alternatively, we may model each bidder's membership to a type as a random variable following the Dirichlet prior. This approach is standard with a range of applications, e.g., Dirichlet process mixture; see \cite{Ferguson_1973} and \cite{Escobar_West_1995} among others. 
If applied here, however, it would further complicate our method and substantially increase computing time.

Finally, a dataset may contain procurement-specific covariates, $x_t$. 
We can also adapt the specification \eqref{ftau} to accommodate $x_t$.
In particular, let us use $h(\cdot|\psi_\tau)$ to denote \eqref{ftau} and $f$ for the cost density with $x_t$.
Let $\widetilde{f}(c|x_t,\gamma_\tau)$ be a low-dimensional parametric density indexed by $(x_t,\gamma_\tau)$, where $\gamma_\tau$ is a vector of parameters.
For example, $\widetilde{f}(c|x_t,\gamma_\tau)$ can be the exponential density with the mean $\exp(x_t'\gamma_\tau)$. 
Then, we may specify the CDF of the cost by  
$
F(c|x_t,\gamma_\tau,\psi_\tau) 
:= H(\widetilde{F}(c|x_t,\gamma_\tau)|\psi_\tau),
$
where $H$ and $\widetilde{F}$ are the CDFs of $h$ and $\widetilde{f}$.
To understand this specification, consider its log density, $\log f(c|x_t,\gamma_\tau,\psi_\tau) 
\approx
\log \widetilde{f}(c|x_t,\gamma_\tau) 
+ \psi_{1,\tau} \phi_1(\widetilde{x}_{\tau,t}) + \psi_{2,\tau} \phi_2(\widetilde{x}_{\tau,t}) + \cdots,$
where $\widetilde{x}_{\tau,t}:=\widetilde{F}(c|x_t,\gamma_\tau)$.
That is, this specification
first approximates the cost density by  the parametric family and reduces the error by the additional terms.
Therefore, if the parametric family offers a good fit, $\psi_\tau$ need not be high dimensional for a given approximation quality. 
So, the specification is a parsimonious and yet flexible representation of the cost density with covariates.  
This approach has been used before, e.g., see 
\cite{KimDH_2013_IJIO}, 
\cite{Aryal_Kim_2013},
\cite{KimDH_2014_BSL}, 
and 
\cite{AryalGrundlKimZhu2017}.

{
\singlespacing
\bibliographystyle{econometrica}
\bibliography{bibliography}
}

\clearpage
\newpage
\setcounter{section}{0}
\setcounter{figure}{0}
\setcounter{table}{0}
\setcounter{page}{0}
\renewcommand{\thepage}{S\arabic{page}} 
\renewcommand{\thesection}{S\arabic{section}}  
\renewcommand{\thetable}{S\arabic{table}}  
\renewcommand{\thefigure}{S\arabic{figure}}
\renewcommand{\theequation}{S\arabic{equation}} 
\begin{center}
\Large{\bf Supplementary Appendix}
\end{center}

\section{Roadmap}\label{sec:road}
This supplementary appendix provides additional materials.
The order of the contents in this appendix closely follows the order in the main paper, \cite{ACJK2021}.

Section \ref{sec:lemma2} proves Lemma 1 in section 2 in the main paper.
Section \ref{sec:data} provides a complete list of 102 categories, for each of which the section presents some additional statistics. The section then explains the selection criteria by which we choose the ``printing papers" category. Then, 
the section
documents further evidence that 
bidders bid depending on the level of competition for the category.

The next five sections in this appendix discuss details about the computational algorithms used in our estimation method; see sections 4 and 5 in the main paper. 
In particular, section \ref{section:lgndr} plots graphs of several Legendre polynomial basis functions that we use 
to specify 
the cost densities, and 
section \ref{section:gavish} provides computational details on solving the asymmetric equilibria and evaluating the likelihood in a Markov chain Monte Carlo (MCMC) method.
Section \ref{section:posteriorcomputation} explains the MCMC algorithm that we use, and
section \ref{section:montecarlo} presents 
the estimation results with 
simulated data. 
Section 4 in the main paper has a summary of the findings from this section.
Section \ref{sec:KStest} discusses 
how the Kolmogorov-Smirnov (KS) test are used to examine if multiple MCMC outcomes are drawn from the same (posterior) distribution.
 
Section \ref{sec:algorithmPolicy} discusses the algorithm for counterfactual simulations in section 5.2. Finally, section \ref{sec:sen} provides additional results of sensitivity analysis in section 5.3.

\section{Proof of Lemma 1}\label{sec:lemma2}
Consider any bidder configuration $\mathcal{I}$ with $|\mathcal{I}|\geq2$, and let $G^0(\cdots|\mathcal{I})$ be the joint distribution of bids submitted by bidders in  $\mathcal{I}$.
Note that $G^0(\cdots|\mathcal{I})$ is directly identified from the bid data generated from procurements with $\mathcal{I}$.
By Theorem 1 in \cite{Krasnokutskaya2011}, then, we can identify the marginal bid distribution of bidder $i$ when there is no unobserved heterogeneity, i.e., $G_i(\cdot|1, \mathcal{I})$, for all $i\in\mathcal{I}$ and the distribution of the unobserved heterogeneity, $F_u(\cdot)$.
By Lemma 2 in \cite{Campo2012}, then, we can identify the risk-aversion coefficient and the cost distribution for bidder $i$, $\{\eta_i,F_i(\cdot)\}$  
using $\{G_i(\cdot|1, \mathcal{I})\}_{i\in\mathcal{I}}$ when $\mathcal{I}$ exogenously varies.

\section{Data}\label{sec:data}
This section explains the selection criteria that we use to choose a category to analyze. Then, we provide additional statistical evidence that bidders in the category ``printing papers" 
bid depending on the level of competition.
\subsection{Selecting a Category}\label{sec:selcat}
The Russian government uses the first-price procurement (FPP) 
(along with several other allocation methods)
to buy goods and services from private bidders.
Tables \ref{Table:AllCat1} to \ref{Table:AllCat4} list all the 102 categories 
for which FPPs were used in 2014.
Some categories have names that are too long to fit in the table, in which case their full names appear below the table along with their identification numbers (ID). 
It is evident that the categories are in different industries with separate markets.
The tables show the total numbers of procurements, bids, and bidders in each category in columns (1), (2), and (3), respectively.
For example, the second job category, ``educational service," has 763 procurements in total, where 1,079 bidders submit a total of 1,607 bids. 
(As in the paper, we use double quotation marks to indicate field terms in their closest English translation.)
Columns (4), (5), and (6)
show the numbers of procurements with only one bidder, unrecorded bids, and bids larger than the reserve price, respectively. 
In the second category, ``educational service," 292 procurements have only one bidder, two procurements have missing bids, and 14 procurements have at least one bid larger than the reserve price.

\begin{table}[t!]
\centering
\caption{All Categories, Number of Procurements, Bids, Bidders, etc}
\label{Table:AllCat1}
\scalebox{0.83}{\begin{tabular}{rl|rrr|rrr|rr|rrr}
\hline\hline
 &                      & \multicolumn{3}{l|}{Number (\#) of } & \multicolumn{3}{l|}{\# of procs } & \multicolumn{2}{l|}{\# of }  & \multicolumn{3}{l}{\# of bids }
 \\
 &                      &  \multicolumn{3}{l|}{procurements, }& \multicolumn{3}{l|}{1 bid only, } & \multicolumn{2}{l|}{remaining }  & \multicolumn{3}{l}{submitted by }
 \\
 & & \multicolumn{3}{l|}{bids, }  & \multicolumn{3}{l|}{missing bids, }  & \multicolumn{2}{l|}{procs }  & \multicolumn{3}{l}{type (1,2,3) }
 \\
\multicolumn{2}{l|}{Category}& \multicolumn{3}{l|}{bidders } & \multicolumn{3}{l|}{bids $> \rho$}  & \multicolumn{2}{l|}{(\%) } & \multicolumn{3}{l}{bidders }\\
ID & Names & 
(1) & (2) & (3) & (4) & (5) & (6) & (7) & (8) & (9) & (10) & (11) \\
\hline 
 1 & car tires & 272 & 609 & 448 & 83 & 0 & 12 & 177 &(65\%)  & 0 & 199 & 293 \\
 2 & educational services* & 763 & 1607 & 1079 & 292 & 2 & 14 & 458 &(60\%)  & 0 & 629 & 647 \\
 3 & training courses & 211 & 461 & 325 & 81 & 0 & 7 & 123 &(58\%)  & 0 & 162 & 198 \\
 4 & construction: other building r & 1259 & 3471 & 2285 & 257 & 1 & 36 & 975 &(77\%)  & 0 & 1601 & 1525 \\
 5 & metrology services & 443 & 671 & 372 & 280 & 1 & 27 & 154 &(35\%)  & 0 & 150 & 222 \\
 6 & IT and computer services* & 467 & 897 & 716 & 153 & 0 & 22 & 293 &(63\%)  & 0 & 201 & 493 \\
 7 & wooden office furniture & 209 & 510 & 412 & 47 & 1 & 7 & 154 &(74\%)  & 0 & 141 & 298 \\
 8 & M\&R cars & 290 & 626 & 518 & 65 & 0 & 19 & 206 &(71\%)  & 0 & 146 & 362 \\
 9 & garbage collection & 278 & 550 & 431 & 100 & 0 & 17 & 166 &(60\%)  & 0 & 145 & 277 \\
 10 & system maintenance & 1430 & 2947 & 1703 & 298 & 1 & 80 & 1053 &(74\%)  & 0 & 1463 & 996 \\
 11 & advanced professional training & 719 & 1363 & 652 & 366 & 2 & 12 & 339 &(47\%)  & 0 & 599 & 360 \\
 12 & construction: admin and busine & 590 & 1519 & 1128 & 129 & 2 & 25 & 439 &(74\%)  & 0 & 553 & 774 \\
 13 & Internet and regional network  & 770 & 1414 & 534 & 315 & 1 & 31 & 427 &(55\%)  & 208 & 526 & 295 \\
 14 & other drugs & 329 & 596 & 218 & 140 & 0 & 8 & 181 &(55\%)  & 55 & 277 & 105 \\
 15 & engineering evaluation and res & 701 & 1775 & 1204 & 197 & 2 & 21 & 482 &(69\%)  & 0 & 730 & 781 \\
 16 & other design and engineering s & 324 & 907 & 675 & 58 & 1 & 14 & 252 &(78\%)  & 0 & 293 & 504 \\
 17 & milk & 289 & 585 & 442 & 79 & 0 & 9 & 202 &(70\%)  & 0 & 172 & 314 \\
 18 & other personal services & 685 & 1500 & 1256 & 198 & 2 & 21 & 466 &(68\%)  & 0 & 308 & 934 \\
 19 & inventorying and certification & 396 & 1124 & 558 & 82 & 0 & 12 & 305 &(77\%)  & 95 & 565 & 359 \\
 20 & other workshops and training c & 217 & 432 & 315 & 106 & 1 & 4 & 106 &(49\%)  & 0 & 109 & 206 \\
 21 & general medical checkup and ex & 828 & 1498 & 874 & 446 & 0 & 20 & 366 &(44\%)  & 0 & 620 & 386 \\
 22 & other entertaining services & 327 & 722 & 545 & 74 & 0 & 16 & 238 &(73\%)  & 0 & 243 & 364 \\
 23 & sanitation improvement* & 336 & 814 & 702 & 70 & 0 & 25 & 246 &(73\%)  & 0 & 144 & 528 \\
 24 & M\&R lifting and conveying mac & 541 & 1056 & 585 & 163 & 2 & 29 & 349 &(65\%)  & 0 & 518 & 297 \\
\hline
\end{tabular}}
\caption*{\footnotesize  
* The name is appended with `(other).'  
The truncated product names are as below.
}
\scalebox{0.83}{\begin{tabular}{rl}
 4 & construction: other building repairing \\
 11 & advanced professional training for employees with higher education \\
 12 & construction: admin and business buildings, bus and railway stations, airports \\
 13 & Internet and regional network services \\
 15 & engineering evaluation and research* \\
 16 & other design and engineering services \\
 19 & inventorying and certification of non-housing stock \\
 20 & other workshops and training courses \\
 21 & general medical checkup and examination services \\
 24 & M\&R lifting and conveying machines \\
\end{tabular}}
\end{table}
\begin{table}[t!]
\centering
\caption{All Categories, Number of Procurements, Bids, Bidders, etc. Table continued.}
\label{Table:AllCat2}
\scalebox{0.83}{\begin{tabular}{rl|rrr|rrr|rr|rrr}
\hline\hline
 &                      & \multicolumn{3}{l|}{Number (\#) of } & \multicolumn{3}{l|}{\# of procs } & \multicolumn{2}{l|}{\# of }  & \multicolumn{3}{l}{\# of bids }
 \\
 &                      &  \multicolumn{3}{l|}{procurements, }& \multicolumn{3}{l|}{1 bid only, } & \multicolumn{2}{l|}{remaining }  & \multicolumn{3}{l}{submitted by }
 \\
 & & \multicolumn{3}{l|}{bids, }  & \multicolumn{3}{l|}{missing bids, }  & \multicolumn{2}{l|}{procs }  & \multicolumn{3}{l}{type (1,2,3) }
 \\
\multicolumn{2}{l|}{Category}& \multicolumn{3}{l|}{bidders } & \multicolumn{3}{l|}{bids $> \rho$}  & \multicolumn{2}{l|}{(\%) } & \multicolumn{3}{l}{bidders }\\
ID & Names & 
(1) & (2) & (3) & (4) & (5) & (6) & (7) & (8) & (9) & (10) & (11) \\
\hline 
 25 & motor fuel & 626 & 1015 & 514 & 294 & 0 & 16 & 319 &(51\%)  & 0 & 399 & 291 \\
 26 & out-of-schedule public transpo & 398 & 681 & 537 & 167 & 0 & 16 & 220 &(55\%)  & 0 & 186 & 297 \\
 27 & office and school plastic supp & 221 & 564 & 401 & 41 & 0 & 5 & 176 &(80\%)  & 0 & 224 & 289 \\
 28 & medical lab services* & 279 & 438 & 283 & 168 & 0 & 6 & 109 &(39\%)  & 0 & 121 & 144 \\
 29 & copier P\&A & 581 & 1623 & 1030 & 93 & 3 & 16 & 471 &(81\%)  & 0 & 756 & 719 \\
 30 & computer P\&A* & 357 & 862 & 644 & 95 & 3 & 10 & 250 &(70\%)  & 0 & 277 & 452 \\
 31 & office and school plastic supp & 340 & 932 & 564 & 51 & 2 & 8 & 279 &(82\%)  & 33 & 437 & 377 \\
 32 & individual car M\&R & 212 & 435 & 371 & 58 & 0 & 11 & 143 &(67\%)  & 0 & 95 & 251 \\
 33 & M\&R IT equipment & 418 & 1019 & 839 & 87 & 1 & 14 & 318 &(76\%)  & 0 & 262 & 629 \\
 34 & A/C equipment & 390 & 1270 & 911 & 48 & 2 & 18 & 322 &(83\%)  & 0 & 475 & 653 \\
 35 & ground-based vehicles insuranc & 389 & 1125 & 85 & 74 & 1 & 46 & 273 &(70\%)  & 671 & 170 & 40 \\
 36 & off-the-shelf software and use & 316 & 624 & 473 & 103 & 0 & 11 & 202 &(64\%)  & 0 & 174 & 319 \\
 37 & lead-free gasoline, sulfur con & 314 & 517 & 311 & 147 & 1 & 8 & 158 &(50\%)  & 0 & 178 & 173 \\
 38 & advanced professional training & 380 & 575 & 308 & 274 & 1 & 4 & 101 &(27\%)  & 12 & 110 & 163 \\
 39 & security services & 1378 & 3448 & 1921 & 336 & 3 & 59 & 988 &(72\%)  & 0 & 1841 & 1096 \\
 40 & subscription for newspapers, m & 356 & 671 & 130 & 114 & 0 & 15 & 228 &(64\%)  & 229 & 230 & 64 \\
 41 & subscription for domestic news & 371 & 703 & 121 & 112 & 2 & 20 & 243 &(65\%)  & 254 & 251 & 46 \\
 42 & property and assets evaluation & 591 & 1695 & 773 & 133 & 3 & 9 & 446 &(75\%)  & 0 & 1078 & 445 \\
 43 & architectural design & 211 & 610 & 411 & 50 & 0 & 5 & 156 &(74\%)  & 0 & 279 & 270 \\
 44 & construction completion* & 1379 & 3492 & 2579 & 327 & 6 & 64 & 994 &(72\%)  & 0 & 1191 & 1799 \\
 45 & drivers' medical examination & 751 & 1138 & 867 & 482 & 1 & 23 & 247 &(33\%)  & 0 & 245 & 353 \\
 46 & vehicle P\&A* & 446 & 969 & 671 & 139 & 1 & 22 & 287 &(64\%)  & 0 & 329 & 436 \\
 47 & printing paper* & 536 & 1237 & 641 & 114 & 3 & 8 & 411 &(77\%)  & 58 & 625 & 402 \\
 48 & software & 1292 & 2756 & 1686 & 370 & 0 & 37 & 888 &(69\%)  & 0 & 1213 & 1081 \\
 49 & desktop computers & 337 & 676 & 527 & 105 & 1 & 9 & 225 &(67\%)  & 0 & 188 & 363 \\
 50 & snow cleaning & 545 & 1118 & 796 & 192 & 0 & 26 & 332 &(61\%)  & 0 & 388 & 482 \\
 51 & M\&R office equipment & 959 & 2521 & 1748 & 171 & 1 & 31 & 758 &(79\%)  & 0 & 1049 & 1212 \\
 52 & construction management & 202 & 465 & 285 & 49 & 1 & 13 & 142 &(70\%)  & 0 & 216 & 176 \\
\hline
\end{tabular}}
\caption*{\footnotesize  
* The name is appended with `(other).'  
The truncated product names are as below.
}
\scalebox{0.83}{\begin{tabular}{rl}
 26 & out-of-schedule public transportation* \\
 27 & office and school plastic supplies: binders, briefcases, folders, book covers \\
 31 & office and school plastic supplies: paper press, paper cutters, blotting-pads, pen cases, bookmarks, etc. \\
 35 & ground-based vehicles insurance \\
 36 & off-the-shelf software and user licenses \\
 37 & lead-free gasoline, sulfur content < 150 mg/kg \\
 38 & advanced professional training for employees with vocational education \\
 40 & subscription for newspapers, magazines and other periodical literature \\
 41 & subscription for domestic newspapers, magazines, etc. \\
 42 & property and assets evaluation services \\
\end{tabular}}
\end{table}
\begin{table}[t!]
\centering
\caption{All Categories, Number of Procurements, Bids, Bidders, etc. Table continued.}
\label{Table:AllCat3}
\scalebox{0.83}{\begin{tabular}{rl|rrr|rrr|rr|rrr}
\hline\hline
 &                      & \multicolumn{3}{l|}{Number (\#) of } & \multicolumn{3}{l|}{\# of procs } & \multicolumn{2}{l|}{\# of }  & \multicolumn{3}{l}{\# of bids }
 \\
 &                      &  \multicolumn{3}{l|}{procurements, }& \multicolumn{3}{l|}{1 bid only, } & \multicolumn{2}{l|}{remaining }  & \multicolumn{3}{l}{submitted by }
 \\
 & & \multicolumn{3}{l|}{bids, }  & \multicolumn{3}{l|}{missing bids, }  & \multicolumn{2}{l|}{procs }  & \multicolumn{3}{l}{type (1,2,3) }
 \\
\multicolumn{2}{l|}{Category}& \multicolumn{3}{l|}{bidders } & \multicolumn{3}{l|}{bids $> \rho$}  & \multicolumn{2}{l|}{(\%) } & \multicolumn{3}{l}{bidders }\\
ID & Names & 
(1) & (2) & (3) & (4) & (5) & (6) & (7) & (8) & (9) & (10) & (11) \\
\hline 
 53 & M\&R alarm systems & 275 & 848 & 578 & 58 & 0 & 6 & 211 &(77\%)  & 23 & 322 & 430 \\
 54 & gasoline & 662 & 1043 & 576 & 346 & 1 & 23 & 296 &(45\%)  & 0 & 307 & 339 \\
 55 & area cleanup with special equi & 233 & 466 & 400 & 78 & 0 & 13 & 147 &(63\%)  & 0 & 97 & 273 \\
 56 & land management & 367 & 994 & 612 & 82 & 0 & 7 & 278 &(76\%)  & 40 & 487 & 365 \\
 57 & road surfacing & 211 & 391 & 308 & 70 & 1 & 11 & 136 &(64\%)  & 0 & 106 & 205 \\
 58 & special construction works & 235 & 522 & 376 & 75 & 2 & 12 & 152 &(65\%)  & 0 & 152 & 271 \\
 59 & recreational and entertaining  & 396 & 855 & 632 & 75 & 0 & 17 & 306 &(77\%)  & 0 & 298 & 442 \\
 60 & road construction & 233 & 425 & 289 & 81 & 0 & 13 & 144 &(62\%)  & 0 & 146 & 180 \\
 61 & wiring* & 293 & 764 & 615 & 80 & 0 & 19 & 199 &(68\%)  & 0 & 208 & 432 \\
 62 & construction: multi-apartment  & 249 & 609 & 422 & 80 & 0 & 7 & 164 &(66\%)  & 0 & 219 & 297 \\
 63 & door and window installation & 239 & 680 & 539 & 34 & 1 & 5 & 199 &(83\%)  & 0 & 192 & 433 \\
 64 & road construction & 210 & 354 & 282 & 84 & 0 & 12 & 121 &(58\%)  & 0 & 84 & 173 \\
 65 & sanitary and epidemiological s & 233 & 339 & 187 & 156 & 1 & 14 & 73 &(31\%)  & 26 & 44 & 104 \\
 66 & database services* & 548 & 1078 & 670 & 117 & 0 & 47 & 385 &(70\%)  & 0 & 464 & 396 \\
 67 & entrepreneurial services* & 506 & 1096 & 921 & 175 & 3 & 24 & 311 &(61\%)  & 0 & 224 & 638 \\
 68 & property inspection & 249 & 974 & 546 & 43 & 2 & 10 & 195 &(78\%)  & 21 & 500 & 353 \\
 69 & M\&R general use equipment and & 230 & 528 & 458 & 57 & 0 & 14 & 161 &(70\%)  & 0 & 93 & 342 \\
 70 & office and school supplies ret & 211 & 496 & 366 & 58 & 0 & 9 & 144 &(68\%)  & 15 & 118 & 269 \\
 71 & fire alarm system installation & 310 & 786 & 540 & 86 & 0 & 13 & 212 &(68\%)  & 0 & 283 & 381 \\
 72 & peacekeeping and public securi & 344 & 813 & 603 & 86 & 2 & 26 & 237 &(69\%)  & 0 & 240 & 425 \\
 73 & other safety services & 218 & 632 & 524 & 48 & 0 & 6 & 164 &(75\%)  & 0 & 158 & 406 \\
 74 & firewood & 201 & 337 & 277 & 86 & 0 & 15 & 100 &(50\%)  & 0 & 70 & 147 \\
 75 & white paper & 204 & 462 & 323 & 50 & 0 & 4 & 150 &(74\%)  & 17 & 152 & 231 \\
 76 & exhibition and trade show orga & 220 & 488 & 348 & 39 & 0 & 8 & 173 &(79\%)  & 0 & 174 & 256 \\
 77 & anti-cancer drugs & 209 & 346 & 73 & 105 & 0 & 3 & 101 &(48\%)  & 107 & 104 & 24 \\
 78 & construction completion works* & 257 & 629 & 534 & 42 & 0 & 14 & 203 &(79\%)  & 0 & 132 & 419 \\
\hline
\end{tabular}}
\caption*{\footnotesize  
* The name is appended with `(other).'  
The truncated product names are as below.
}
\scalebox{0.83}{\begin{tabular}{rl}
 55 & area cleanup with special equipment \\
 59 & recreational and entertaining services* \\
 62 & construction: multi-apartment buildings \\
 65 & sanitary and epidemiological services \\
 69 & M\&R general use equipment and machinery* \\
 70 & office and school supplies retail services \\
 72 & peacekeeping and public security services* \\
 76 & exhibition and trade show organization \\
\end{tabular}}
\end{table}
\begin{table}[t!]
\centering
\caption{All Categories, Number of Procurements, Bids, Bidders, etc. Table continued.}
\label{Table:AllCat4}
\scalebox{0.83}{\begin{tabular}{rl|rrr|rrr|rr|rrr}
\hline\hline
 &                      & \multicolumn{3}{l|}{Number (\#) of } & \multicolumn{3}{l|}{\# of procs } & \multicolumn{2}{l|}{\# of }  & \multicolumn{3}{l}{\# of bids }
 \\
 &                      &  \multicolumn{3}{l|}{procurements, }& \multicolumn{3}{l|}{1 bid only, } & \multicolumn{2}{l|}{remaining }  & \multicolumn{3}{l}{submitted by }
 \\
 & & \multicolumn{3}{l|}{bids, }  & \multicolumn{3}{l|}{missing bids, }  & \multicolumn{2}{l|}{procs }  & \multicolumn{3}{l}{type (1,2,3) }
 \\
\multicolumn{2}{l|}{Category}& \multicolumn{3}{l|}{bidders } & \multicolumn{3}{l|}{bids $> \rho$}  & \multicolumn{2}{l|}{(\%) } & \multicolumn{3}{l}{bidders }\\
ID & Names & 
(1) & (2) & (3) & (4) & (5) & (6) & (7) & (8) & (9) & (10) & (11) \\
\hline 
 79 & databases & 216 & 430 & 305 & 39 & 0 & 29 & 152 &(70\%)  & 0 & 146 & 188 \\
 80 & fire alarm system maintenance & 494 & 1676 & 1115 & 65 & 0 & 21 & 410 &(83\%)  & 0 & 732 & 786 \\
 81 & sewage utility services & 1188 & 2281 & 1345 & 453 & 5 & 69 & 677 &(57\%)  & 0 & 1033 & 653 \\
 82 & M\&R electric equipment* & 224 & 590 & 492 & 40 & 0 & 11 & 175 &(78\%)  & 0 & 146 & 382 \\
 83 & test of integral mechanical an & 242 & 775 & 601 & 55 & 0 & 7 & 183 &(76\%)  & 0 & 256 & 451 \\
 84 & general cleaning* & 444 & 1141 & 807 & 99 & 1 & 15 & 331 &(75\%)  & 0 & 434 & 570 \\
 85 & other healthcare services & 340 & 603 & 450 & 185 & 1 & 16 & 144 &(42\%)  & 0 & 155 & 231 \\
 86 & other engineering and technica & 261 & 675 & 495 & 53 & 5 & 5 & 199 &(76\%)  & 0 & 242 & 348 \\
 87 & eggs & 339 & 652 & 497 & 123 & 1 & 9 & 208 &(61\%)  & 0 & 171 & 337 \\
 88 & special care services & 261 & 453 & 352 & 133 & 0 & 8 & 123 &(47\%)  & 0 & 111 & 196 \\
 89 & housing rental & 269 & 323 & 275 & 217 & 0 & 2 & 51 &(19\%)  & 0 & 54 & 50 \\
 90 & demolition & 246 & 761 & 479 & 38 & 5 & 9 & 195 &(79\%)  & 0 & 351 & 324 \\
 91 & lead-free gasoline, sulfur con & 230 & 376 & 238 & 103 & 0 & 8 & 121 &(53\%)  & 0 & 127 & 131 \\
 92 & M\&R other medical equipment & 504 & 936 & 561 & 156 & 0 & 13 & 337 &(67\%)  & 0 & 408 & 344 \\
 93 & special safety and security se & 214 & 548 & 435 & 46 & 0 & 15 & 155 &(72\%)  & 0 & 144 & 319 \\
 94 & M\&R special-use equipment* & 299 & 801 & 650 & 56 & 0 & 15 & 229 &(77\%)  & 0 & 223 & 485 \\
 95 & car rental with driver & 293 & 581 & 431 & 85 & 0 & 15 & 199 &(68\%)  & 0 & 198 & 276 \\
 96 & M\&R heat meters & 251 & 644 & 416 & 49 & 0 & 5 & 197 &(78\%)  & 0 & 303 & 278 \\
 97 & hardcover textbooks & 299 & 660 & 221 & 56 & 0 & 9 & 235 &(79\%)  & 72 & 402 & 110 \\
 98 & M\&R surgical equipment & 480 & 940 & 515 & 139 & 0 & 11 & 330 &(69\%)  & 0 & 483 & 291 \\
 99 & anti-viral drugs & 219 & 372 & 157 & 111 & 0 & 8 & 102 &(47\%)  & 32 & 134 & 81 \\
 100 & other tools, equipment and dev & 220 & 457 & 349 & 58 & 0 & 3 & 159 &(72\%)  & 0 & 129 & 262 \\
 101 & oxygen & 231 & 350 & 171 & 123 & 1 & 12 & 97 &(42\%)  & 10 & 107 & 88 \\
 102 & other medical tools and device & 272 & 591 & 413 & 63 & 0 & 6 & 203 &(75\%)  & 0 & 227 & 281 \\
\hline
\end{tabular}}
\caption*{\footnotesize  
* The name is appended with `(other).'  
The truncated product names are as below.
}
\scalebox{0.83}{\begin{tabular}{rl}
 83 & test of integral mechanical and electrical systems \\
 86 & other engineering and technical services \\
 91 & lead-free gasoline, sulfur content $<$ 1000 mg/kg \\
 93 & special safety and security services \\
 100 & other tools, equipment and devices \\
 102 & other medical tools and devices \\
\end{tabular}}
\end{table}

For each category, we exclude all the procurements with either one bidder, missing bids, or bids larger than the reserve prices.
Note that some procurements may meet multiple conditions for exclusion, e.g., when a procurement has only one bidder, this bidder bids above the reserve price, or the bid is not recorded. 
Columns (7) and (8) show the numbers of remaining procurements and their proportion in parentheses, respectively. 
For example, the ``educational service" category has 458 procurements after dropping the three cases, and the proportion of the remaining procurements is roughly 60 percent ($\approx 458 \div 763$).
The tables show that the main reason for us to discard procurements is that there are procurements with only one bidder.

Our first criterion for selecting a category for analysis is that the number of procurements is at least 200 after excluding all procurements with either only one bidder, missing bids, or bids above the reserve price.
That is, we sort out categories with a reasonably large sample because the sample size is critical for precise estimation. 
The second criterion is that at least 60 percent of procurements are not excluded for those three reasons, i.e., column (8) is larger than or equal to 60 percent. 
Note that section 3.1 in the main paper explains that those three cases for exclusion, say columns (4), (5), and (6), are vulnerable to corruption. 
If a high proportion of procurements are excluded for those three reasons, it may suggest that the remaining procurements might not represent the competition in the category.

Within each category, we group bidders into three types 
for the remaining procurements as counted in column (7).
Type 1 bidders bid in at least 10\% of the 
procurements, type 3 bidders bid once, and type 2 bidders are all the others. 
Columns (9), (10), and (11) show the number of bids submitted by each type in the remaining procurements. 
The ``educational service" category, for example, does not have any bid submitted by type 1 bidder, but type 2 and 3 bidders bid 629 and 647 times, respectively.
It is noticeable that a bidder hardly appears in more than 10\% of the procurements, i.e., it is rare that a bidder frequently enters the procurements.
Therefore, if a bidder does so, it suggests that the bidder can be inherently different from other bidders, motivating an asymmetric model. 

The number of bids for each type is important because the bid variation identifies the model primitives along with the variation in bidder configuration. 
Hence, we collect categories where the number of bids is at least 50 for each type, which is our third criterion to select a category.
Out of all the job categories, 
only four categories meet the three criteria:
categories with IDs 19, 40, 47, and 97.
Among them, we analyze 
category 47, ``printing papers,"  in the paper, as it has the largest number of procurements $(T=411)$. 

\subsection{Additional Details, ``Printing Papers"}
In the category ``printing papers," the bidder configuration has a rich variation with 32 distinct observed configurations. 
Table \ref{Table:bids4eachI} lists the configurations that appear at least ten times in the data.
For each of those frequent configurations, Table \ref{Table:bids4eachI} shows the sample average of the normalized bids (bids divided by the reserve price) and associated standard deviation in parentheses. 
The numbers in brackets in column (A) for all types are the number of procurements, 
and the numbers in brackets in columns (1), (2), and (3) are the number of bids for each type of bidders, respectively.
For example, there are 40 procurements with two type 2 bidders and one type 3 bidder (223). 
For this configuration, we have 80 (40) bids submitted by type 2 (3) bidders.
Table \ref{Table:bids4eachI} also shows that the occurrence rate drops quickly. For example, the most frequent configuration (22) appears 93 times, but the second most (33) 61 times, the fifth (222) 26 times, and the ninth (2223) 11 times. 
\begin{table}[t!]
\centering
\caption{Descriptive Statistics for Frequent Bidder Configurations in ``printing papers" }
\label{Table:bids4eachI}
\scalebox{0.83}{\begin{tabular}{r|llll}
\hline\hline
& 
\multicolumn{4}{|c}{Average (Standard Deviation) [Number] of Procurements or Bids } 
\\
Bidder & 
\multicolumn{1}{|c}{All Types} 
& 
\multicolumn{1}{c}{Type 1}
& 
\multicolumn{1}{c}{Type 2} 
& 
\multicolumn{1}{c}{Type 3}
\\
Configuration & 
\multicolumn{1}{|c}{(A)} 
&
\multicolumn{1}{c}{(1)} 
&
\multicolumn{1}{c}{(2)} 
&
\multicolumn{1}{c}{(3)} 
\\
\hline 
 (22)	 	& 0.925 (0.077) [93] 	& 	& 0.925 (0.077) [186] 	&  \\
 (222)	 	& 0.909 (0.072) [26] 	& 	& 0.909 (0.072) [78] 	&  \\
 (33)	 	& 0.961 (0.063) [61] 	& 	& 	& 0.961 (0.063) [122]  \\
 (223)	 	& 0.893 (0.088) [40] 	& 	& 0.882 (0.090) [80] 	& 0.916 (0.081) [40]  \\
 (23)	 	& 0.911 (0.084) [58] 	& 	& 0.893 (0.087) [58] 	& 0.929 (0.077) [58]  \\
 (12)	 	& 0.922 (0.060) [16] 	& 0.907 (0.066) [16] 	& 0.937 (0.051) [16] 	&  \\
 (233)	 	& 0.880 (0.098) [21] 	& 	& 0.835 (0.089) [21] 	& 0.902 (0.096) [42]  \\
 (122)	 	& 0.898 (0.065) [12] 	& 0.866 (0.070) [12] 	& 0.915 (0.057) [24] 	&  \\
 (2223)	 	& 0.870 (0.087) [11] 	& 	& 0.865 (0.082) [33] 	& 0.884 (0.105) [11]  \\
\hline
\end{tabular}}
\caption*{\footnotesize  
Each cell shows the sample mean (standard deviation) of the bids for the given bidder configuration.
Column (A) shows the number of procurements in the brackets [  ], 
and columns (1), (2), and (3) show the number of bids in the brackets [ ] for each type of bidders. 
}
\end{table}

We provide additional evidence that bidders bid differently across different configurations, suggesting that they 
take into account
the level of competition when bidding.
Table \ref{Table:KStest4AB} conducts the KS test against the null hypothesis that the \emph{type-specific} bid distributions are identical in procurements with different bidder configurations. 
We consider the nine frequent bidder configurations in Table \ref{Table:bids4eachI}. 
The first row of Table \ref{Table:KStest4AB} considers the bid distributions of type 1 bidder. 
Since type 1 bidder appears only in configurations (122) and (12) among the ones listed in Table \ref{Table:bids4eachI}, there is only one pair to consider. 
The second block (type 2) considers the distributions of bids submitted by type 2 bidders across 
all the pairs of two bidder configurations, both involving type 2 bidders. 
(Both configurations, say $A$ and $B$, must have type 2 bidders to compare the bid distribution of type 2 in  configuration $A$ and the bid distribution of type 2 in configuration $B$.)
Similarly, the third block repeats this exercise for type 3 bidders.

\begin{table}[t!]
\centering
\caption{$p$ values for KS test; Homogeneity in Bid Distributions}
\label{Table:KStest4AB}
\scalebox{0.83}{\begin{tabular}{rrrrrrrccc}
\hline\hline
	& \multicolumn{2}{c}{Bidder} 
	& \multicolumn{2}{c}{Number of} 
	& \multicolumn{2}{c}{KS test $p$ values}  
	& \multicolumn{3}{c}{Fail to reject $H_0$?}  
	\\ 
	& \multicolumn{2}{c}{Configurations} 
	& \multicolumn{2}{c}{Bids in}
	& \multicolumn{2}{c}{with Alt. Hypothesis}  
	& Small & No & Fringe 
	\\ 
 	&  $A$ & $B$ &   Config $A$ &   Config $B$ & $H_1:A \neq B$ & $H_1:A < B$
	& Sample & Theory & Bidders
	\\
	&
	(1) &
	(2) &
	(3) &
	(4) &
	(5) &
	(6) &
	(7) &
	(8) &
	(9) \\
\hline
 Type 1 & (122) & (12) & 12 & 16 & 0.357 & 0.179 & Yes & &  \\
 \\
 Type 2 & (222) & (22) & 78 & 186 & **0.013 &  ***0.006 & & &  \\
& (223) & (22) & 80 & 186 & ***0.000 &  ***0.000 & & &  \\
& (23) & (22) & 58 & 186 & ***0.003 & & & &  \\
& (12) & (22) & 16 & 186 & 0.627 & & Yes & Yes &  \\
& (233) & (22) & 21 & 186 & ***0.000 & & & &  \\
& (122) & (22) & 24 & 186 & **0.044 &  **0.022 & & &  \\
& (2223) & (22) & 33 & 186 & ***0.000 &  ***0.000 & & &  \\
& (223) & (222) & 80 & 78 & 0.124 & & & Yes & Yes  \\
& (222) & (23) & 78 & 58 & 0.846 & & & Yes & Yes  \\
& (222) & (12) & 78 & 16 & 0.315 & & Yes & Yes &  \\
& (233) & (222) & 21 & 78 & ***0.000 & & & &  \\
& (122) & (222) & 24 & 78 & 0.904 & & Yes & Yes &  \\
& (2223) & (222) & 33 & 78 & ***0.002 &  ***0.001 & & &  \\
& (223) & (23) & 80 & 58 & 0.338 & 0.170 & & & Yes  \\
& (223) & (12) & 80 & 16 & *0.074 & & & &  \\
& (233) & (223) & 21 & 80 & **0.019 & & & &  \\
& (122) & (223) & 24 & 80 & 0.216 & & Yes & Yes & Yes  \\
& (2223) & (223) & 33 & 80 & 0.195 &  *0.098 & & &  \\
& (12) & (23) & 16 & 58 & 0.179 & & Yes & Yes & Yes  \\
& (233) & (23) & 21 & 58 & ***0.001 &  ***0.001 & & &  \\
& (122) & (23) & 24 & 58 & 0.773 & & Yes & Yes & Yes  \\
& (2223) & (23) & 33 & 58 & **0.017 &  ***0.009 & & &  \\
& (233) & (12) & 21 & 16 & ***0.001 & & & &  \\
& (122) & (12) & 24 & 16 & 0.419 & 0.212 & Yes & &  \\
& (2223) & (12) & 33 & 16 & ***0.004 & & & &  \\
& (122) & (233) & 24 & 21 & ***0.001 & & & &  \\
& (2223) & (233) & 33 & 21 & 0.480 & & Yes & Yes & Yes  \\
& (2223) & (122) & 33 & 24 & **0.017 & & & &  \\
 \\
 Type 3 & (223) & (33) & 40 & 122 & ***0.000 & & & &  \\
& (23) & (33) & 58 & 122 & ***0.001 & & & &  \\
& (233) & (33) & 42 & 122 & ***0.000 &  ***0.000 & & &  \\
& (2223) & (33) & 11 & 122 & **0.014 & & & &  \\
& (223) & (23) & 40 & 58 & 0.647 & 0.336 & & & Yes  \\
& (233) & (223) & 42 & 40 & 0.782 & & & Yes & Yes  \\
& (2223) & (223) & 11 & 40 & 0.279 & 0.140 & Yes & & Yes  \\
& (233) & (23) & 42 & 58 & 0.212 & 0.106 & & & Yes  \\
& (2223) & (23) & 11 & 58 & 0.116 &  *0.058 & & &  \\
& (2223) & (233) & 11 & 42 & 0.777 & & Yes & Yes & Yes  \\
\hline
\end{tabular}}
\caption*{\footnotesize  
(*,**,***) for the (10, 5, 1) percent significance.
}
\end{table}

For each block, i.e., for type $\tau$ bidder, columns (1) and (2) show the pair of bidder configurations under consideration, say configurations $A$ and $B$, and columns (3) and (4) show the number of  bids submitted by type $\tau$ bidders in procurements with configuration $A$ and the number of  bids submitted by type $\tau$ bidders in procurements with configuration $B$, respectively.
We conduct the KS test against the null hypothesis that 
the bid distribution of type $\tau$ bidders in procurements with configuration $A$ is identical to the bid distribution of type $\tau$ bidders in procurements with configuration $B$.
There are two columns, (5) and (6), of $p$ values of the KS test.
Column (5) is associated with the alternative hypothesis that the distributions are not equal (two-sided test). Column (6) is associated with the alternative hypothesis that the bid in configuration $A$ is smaller than in configuration $B$ (one-sided test).
We consider the second alternative hypothesis only when configuration $B$ is nested in configuration $A$ and, therefore, type $\tau$ bidders would bid lower in configuration $A$ than in $B$, as stated in the alternative hypothesis. 
(No configuration $A$ is nested in configuration $B$.)

For many cases, $p$ values are sufficiently small to reject the null hypothesis at a conventional significance level.
When we fail to reject the null,
Table \ref{Table:KStest4AB} provides possible explanations in 
columns (7), (8), and (9).
For example, the reason that we do not reject the null 
for the pair of configurations (12) and (22) for type 2 bidder 
might be because the number of bids from type 2 in configuration (12) is only 16, i.e., the sample is too small. 
The table indicates if the sample size is small, i.e., with less than 30 bids, for any of the two configurations.
Moreover, for the pair of   (12) and (22), economic theory does not predict in which configuration type 2 bidders would bid lower. 
Due to asymmetric risk-aversion, 
that statement is true even if type 1 bidder is on average more efficient (smaller costs) than type 2 bidders.

We have three cases for which either of the first two justifications (small sample or no theory) does not explain the failure of rejecting the null hypothesis: type 2 $\{(223),(23)\}$, type 3 $\{(223),(23)\}$, and type 3 $\{(233),(23)\}$. 
Note that type 3 (fringe) bidders are in all those configurations.  
Recall that we group all bidders bidding only once in type 3 because there is no bidder-specific information other than how many times they enter.  
However, the bidders might have observed bidder-heterogeneity even for  those one-time bidders and, therefore, bid accordingly. This may explain why we fail to reject the null more often for the pairs with fringe bidders, including the three cases for which the first two explanations do not apply.

Therefore, the comparison of bid distributions for various pairs of two different bidder configurations mostly reveals that bidders bid 
depending on whom they oppose, 
and they bid more aggressively in more competitive procurements.

\section{Graphs of Legendre Polynomials\label{section:lgndr}}
\begin{figure}[t!]
\caption{Legendre Polynomials}\label{fig:lgdr}
\begin{center}
\begin{tabular}{c}
\includegraphics[width=4in]{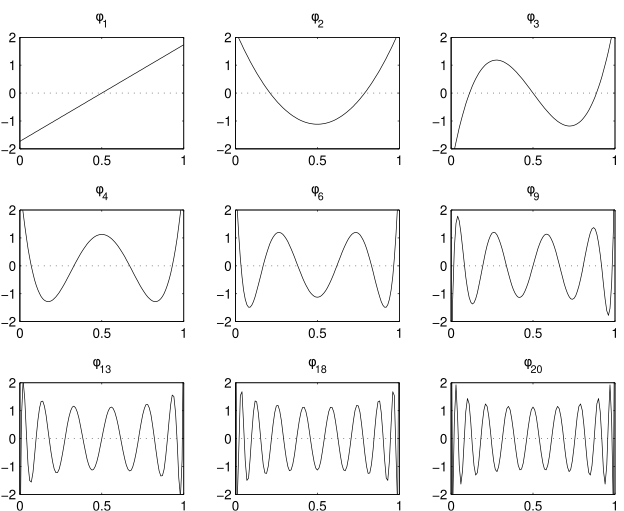}
\end{tabular}
\caption*{\footnotesize 
Notes. Each panel shows a basis function of the Legendre polynomials 
$\phi_j$ for some $j\in\{1,2,\ldots,20\}$ that are re-centered at 0.5 with support $[0,1]$. The function $\phi_j$ has $j-1$ extrema with support $[0,1]$,
i.e.,  as $j$ increases $\phi_j$ becomes more ``wavy."}
\end{center}
\end{figure}

In this section, we present some basis functions for our specification of the cost densities. 
We construct the basis functions by shifting and re-scaling the Legendre polynomials so that they are defined on the unit interval $[0,1]$.
	Specifically, the $j$-th entry in $\phi(\cdot)$ is given as 
    \begin{align}
    \phi_j(c):=\sqrt{2j+1}\times\tilde{\phi}_j(2u-1),    
    \end{align}
    where $\tilde{\phi}_j(x):=\frac{d^j}{dx^j}(x^2-1)^j/(2^jj!)$. 
  Figure \ref{fig:lgdr} shows $\phi_j(\cdot)$ for some $j$ to give an idea about the shape of the basis functions. 
    Observe that $\phi_j$ has $j-1$ extrema in the interior.
    So, as the number  $k$ of basis functions in the cost density increases, the density function specified by Legendre polynomials can approximate more complicated (i.e., with many inflection points) densities. 
    An additional advantage of Legendre polynomials is that they form an orthonormal and orthogonal polynomial system that is complete and has a unique representation of the cost density. They also have a smaller variance than any non-orthonormal polynomial; see \cite{Szego1975}. 
 
\section{Computation of Equilibrium Strategies\label{section:gavish}}
\subsection{Adaptation of the Boundary-Value Method}
We explain how to adapt the method from \cite{Fibich_Gavish_2011}, which is designed  for  asymmetric  high-bid auctions to asymmetric low-bid auctions, i.e., procurements. 
Then, we extend it to incorporate risk-aversion (CRRA) in the algorithm.
For a given FPP, we develop an associated first-price auction model with valuation distributions $\{\tilde{F}_i(\cdot|\psi)\}_{i\in\{1,\ldots,n\}}$.
For the auction model,  we obtain the equilibrium bidding strategy $\{\tilde{\beta}_i\}_{i\in\{1,\ldots,n\}}$ using the algorithm of \cite{Fibich_Gavish_2011}.
Then, we convert the outcome back to the original procurement.

In this section, we fix $\mathcal{I} = \{1,\ldots,n\}$ and allow every bidder to be of her own type, which nests the case where some bidders are of the same type, as a special case. 
In addition, we suppress the dependence of the cost distributions and the (inverse) bidding strategies on the model parameters when it does not cause a confusion.

\subsubsection*{Risk-Neutrality}
In an FPP, bidder $i$ solves 
\begin{align}
\max_b (b-c_i) \prod_{j\neq i} \left\{ 1 - F_j\left[\beta_j^{-1}(b)\right]\right\}\label{procure}
\end{align}
where $c_i \sim F_i(\cdot)$ with a compact support of $[\underline{c},\overline{c}]$ and 
$\beta_i$ is the equilibrium bidding strategy for the game induced by the FPP. 
Now consider a (first-price) auction where bidder $i$'s valuation is given as 
\begin{align}
v_i := \frac{\overline{c} - c_i}{\overline{c}-\underline{c}}.\label{conv}
\end{align}
Then, $v_i \sim \tilde{F}_i(\cdot)$ where $\tilde{F}_i(x) := 1 - F_i[\overline{c}-x(\overline{c}-\underline{c})]$ for all $x \in [0,1]$ by change of variables.
Bidder $i$ in this \emph{associated} auction solves 
\begin{align}
\max_{\tilde{b}} (v_i-\tilde{b}) \prod_{j\neq i} \tilde{F}_j\left[\tilde{\beta}_j^{-1}(\tilde{b})\right]\label{auction}
\end{align}
where $\tilde{\beta}_i$ is the equilibrium bidding strategy for the game induced by the auction.
As in the paper, define the inverse bidding function $\tilde{\phi}_i  = \tilde{\beta}_i^{-1}$.
Then, the algorithm of \cite{Fibich_Gavish_2011} solves for $\tilde{\phi}_i$ in the system of differential equations (FOCs), 
\begin{align}
\tilde{\phi}_i'(b) = \frac{\tilde{F}_i(\tilde{\phi}_i(b))}{\tilde{f}_i(\tilde{\phi}_i(b))}
\left[
\left(
\frac{1}{n-1} \sum_{j=1}^n \frac{1}{\tilde{\phi}_j(b)-b}
\right)
-\frac{1}{\tilde{\phi}_i(b)-b}
\right],\label{procFOC}
\end{align}
for all $i \in \mathcal{I}$.
Once we obtain $\{\tilde{\phi}_i\}$,  
using the definitions of $v_i$ and $\tilde{F}_i$ in \eqref{conv}, we can construct the bidding strategies for the original problem, \eqref{procure}.
Note that (\ref{auction}) can be rewritten as
\begin{align*}
\max_{b} \left[\frac{\overline{c}-c_i}{\overline{c}-\underline{c}}-\frac{\overline{c}-b}{\overline{c}-\underline{c}}\right] \prod_{j\neq i} \left\{ 1 - F_j\left[\overline{c}-\tilde{\beta}_j^{-1}\left(\frac{\overline{c}-b}{\overline{c}-\underline{c}}\right)(\overline{c}-\underline{c})\right]\right\},
\end{align*}
which is equivalent to (\ref{procure}) if and only if, for all $i$,
\begin{align}
\beta_i^{-1}(b) = \overline{c} - \tilde{\beta}_i^{-1}\left(\frac{\overline{c}-b}{\overline{c}-\underline{c}}\right)(\overline{c}-\underline{c}). \label{equiv}
\end{align}
Hence, once we solve \eqref{procFOC}, we can construct the equilibrium bidding strategies for \eqref{procure}.

\subsubsection*{Constant Relative Risk Aversion}
Now, we consider risk-aversion. Suppose bidders are risk-averse with bidder-specific CRRA coefficients, $\{\eta_i\}$. Then, we modify the objective function \eqref{procure} as  
\begin{align}
\max_b (b-c_i)^{1-\eta_i} \prod_{j\neq i} \left\{ 1 - F_j\left[\beta_j^{-1}(b)\right]\right\}\label{procure1}
\end{align}
Following the same logic as above for the case of risk-neutrality, we construct the solution to \eqref{procure1}, which is equivalent to
\begin{align}
\arg\max_{\tilde{b}} (v_i-\tilde{b})^{1-\eta_i} \prod_{j\neq i} \tilde{F}_j\left[\tilde{\beta}_j^{-1}(\tilde{b})\right]
=
\arg\max_{\tilde{b}} (v_i-\tilde{b}) \prod_{j\neq i} \tilde{F}_j\left[\tilde{\beta}_j^{-1}(\tilde{b})\right]^{\frac{1}{1-\eta_i}}.\label{ok11}
\end{align}
Then, we can apply the method of \cite{Fibich_Gavish_2011} to the distribution functions $\tilde{F}_j(\cdot)^{\frac{1}{1-\eta_i}}$.
Alternatively, we can write
the FOC for \eqref{ok11} as below, which
gives a system of differential equations,
\begin{align}
\tilde{\phi}_i'(b) = (1-\eta_i) \left(\frac{\tilde{F}_i(\tilde{\phi}_i(b))}{\tilde{f}_i(\tilde{\phi}_i(b))}\right)
\left[
\left(
\frac{1}{n-1} \sum_{j=1}^n \frac{1}{\tilde{\phi}_j(b)-b}
\right)
-\frac{1}{\tilde{\phi}_i(b)-b}
\right]\label{procFOC1}
\end{align}
for $i \in \mathcal{I}$,
which is similar to \eqref{procFOC}, but the only difference is that the right-hand side of \eqref{procFOC1} is multiplied by a constant $(1-\eta_i)$.
Thus, by slightly modifying the algorithm of \cite{Fibich_Gavish_2011} we can obtain the equilibrium inverse bidding strategies $\{\tilde{\phi}_i\}$ for \eqref{ok11}.
Then, we construct the bidding strategies for \eqref{procure1} via the relationship \eqref{equiv}, and we evaluate the likelihood using the bidding strategies, as section \ref{eq:likelihood} explains.

\subsection{Evaluation of Likelihoods}\label{eq:likelihood}
Following the notation in section 4 of the main paper, 
let $\theta_i = (\psi_i,\eta_i)$ and $\theta = (\theta_1,\ldots,\theta_n)$.
Since the paper uses the unit interval for the cost support, here we assume that the support is known. 
Observe that 
\begin{align}
\overline{c} - \tilde{\beta}_i^{-1}\left(\frac{\overline{c}-b_{it}}{\overline{c}-\underline{c}}\Big|\theta\right)(\overline{c}-\underline{c}) = 
\beta_i^{-1}(b_{it}|\theta) = c_{it} = \overline{c}-v_{it}(\overline{c}-\underline{c}),\label{eq:okdo}
\end{align}
where 
the first equality comes from (\ref{equiv}),
the second is the definition of $\beta_i$, and 
the third is from (\ref{conv}).
The most left-hand side and the most right-hand side of \eqref{eq:okdo} imply that 
\begin{align}
\tilde{\beta}_i^{-1}\left(\frac{\overline{c}-b_{it}}{\overline{c}-\underline{c}}\Big|\theta\right) = v_{it}\label{vit}
\end{align}
Then the bid density in (5) with $u_t = 1$ can be written, in terms of $\{\tilde{F}_i,\tilde{\beta}_i\}$, as 
\begin{align}
g_i(b_{it}|1,\theta,\mathcal{I})&:= 
\tilde{f}\left[\tilde{\beta}_i^{-1}\left(\frac{\overline{c}-b_{it}}{\overline{c}-\underline{c}}\Big|\theta,\mathcal{I}\right)\Big|\psi_i\right] \times\Big|\frac{\partial}{\partial b}\tilde{\beta}_i^{-1}\left(\frac{\overline{c}-b_{it}}{\overline{c}-\underline{c}}\Big|\theta,\mathcal{I}\right)\Big|\nonumber
\\&=  
\left(\frac{1}{\overline{c}-\underline{c}}\right)
\frac{\tilde{f}\left[\tilde{\beta}_i^{-1}\left(\frac{\overline{c}-b_{it}}{\overline{c}-\underline{c}}\Big|\theta,\mathcal{I}\right)\Big|\psi_i\right]}{\Big|\tilde{\beta}_i'\left[\tilde{\beta}_i^{-1}\left(\frac{\overline{c}-b_{it}}{\overline{c}-\underline{c}}\Big|\theta,\mathcal{I}\right)
\Big|\theta,\mathcal{I}\right]\Big|}\label{gb}
\end{align}
for $b_{it}$ such that 
$\tilde{\beta}_i^{-1}\left(\frac{\overline{c}-b_{it}}{\overline{c}-\underline{c}}\Big|\theta,\mathcal{I}\right)
\in[0,1]$ and 
$g_i(b_{it}|1,\theta,\mathcal{I}) = 0$, otherwise.
For $u_t \neq 1$, by Lemma 1, $b_{it}  = u_t \beta_i(c_{it}|\theta,\mathcal{I}) \Longleftrightarrow b_{it}/u_t =
\beta_i(c_{it}|\theta,\mathcal{I})$, implying that 
\begin{align}
g_i(b_{it}|u_t,\theta,\mathcal{I}) 
=  
\frac{1}{u_t(\overline{c}-\underline{c})}
\frac{\tilde{f}\left[\tilde{\beta}_i^{-1}\left(\frac{\overline{c}-b_{it}/u_t}{\overline{c}-\underline{c}}\Big|\theta,\mathcal{I}\right)\Big|\psi_i\right]}{\Big|\tilde{\beta}_i'\left[\tilde{\beta}_i^{-1}\left(\frac{\overline{c}-b_{it}/u_t}{\overline{c}-\underline{c}}\Big|\theta,\mathcal{I}\right)
\Big|\theta,\mathcal{I}\right]\Big|}\label{gb1} 
\end{align}
for $b_{it}$ such that 
$\tilde{\beta}_i^{-1}\left(\frac{\overline{c}-b_{it}/u_t}{\overline{c}-\underline{c}}\Big|\theta,\mathcal{I}\right)
\in[0,1]$ and 
$g_i(b_{it}|u_t,\theta,\mathcal{I}) = 0$, otherwise.
This implies that the support of $b_{it}$ depends on the latent components, $(\theta,\{u_t\})$.
We can evaluate the bid density because we specify $\tilde{f}_i$ and obtain $\tilde{\beta}_i$ as we explained above. Then, we can evaluate the likelihood. 
Note that it is necessary to evaluate the likelihood ratio in the algorithm to simulate the posterior in the next section.

\section{Posterior Computation\label{section:posteriorcomputation}}
While there are various ways of drawing latent variables from the posterior, we find that the algorithm we describe below, a variation of Metropolis-within-Gibbs, works well for our purpose. 
The MCMC algorithm we use is standard, although there can be many variations. 
Our objective here is to introduce an algorithm that works and document what we do for the posterior inference. 

For our empirical application, we consider 
nine
different cases: 
the main specification described in section 4, 
the specification imposing homogeneity in risk-aversion ($\eta = \eta_\tau,\forall\tau$),
the one imposing risk-neutrality ($\eta_\tau = 0,\forall\tau$),
and 
six
other ones for sensitivity analysis; 
see section 5.3.
Here, we focus on the main specification (the algorithms for other specifications are similar).

 In the  $m^{\rm th}$ iteration, our algorithm consists of these steps:   
    \begin{enumerate}
        \item [Step 1.]
        For $\tau=1$, letting $\theta_1^{(m)}:=\theta_1^{(m-1)}=(\psi_1^{(m-1)},\eta_1^{(m-1)})$, we draw a candidate $\tilde{\theta}_1 \sim \mathcal{N}(\theta_1^{(m-1)},\Omega_1^{(m)})$ where $\Omega_1^{(m)}$ is the tuning parameter (we describe how to choose this parameter, shortly below), and update 
        $\theta_1^{(m)} = \tilde{\theta}_1$ with probability 
        \begin{align}
            \min\left\{
            \frac{\pi(\tilde{\theta}_1,\theta_2^{(m-1)},\theta_3^{(m-1)},\sigma_u^{(m-1)},\{u_t^{(m-1)}\}|z)}{\pi(\theta_1^{(m-1)},\theta_2^{(m-1)},\theta_3^{(m-1)},\sigma_u^{(m-1)},\{u_t^{(m-1)}\}|z)},1
            \right\}.\label{ar:1}
        \end{align}
        
       \item [Step 2.]
        Similarly, for $\tau=2$, we repeat Step (1) by letting $\theta_2^{(m)}:=\theta_2^{(m-1)}=(\psi_2^{(m-1)},\eta_2^{(m-1)})$ and drawing a candidate $\tilde{\theta}_2 \sim \mathcal{N}(\theta_2^{(m-1)},\Omega_2^{(m)})$ and updating 
        $\theta_2^{(m)} = \tilde{\theta}_2$ with probability 
        \begin{align}
            \min\left\{
            \frac{\pi(\theta_1^{(m)},\tilde{\theta}_2,\theta_3^{(m-1)},\sigma_u^{(m-1)},\{u_t^{(m-1)}\}|z)}{\pi(\theta_1^{(m)},\theta_2^{(m-1)},\theta_3^{(m-1)},\sigma_u^{(m-1)},\{u_t^{(m-1)}\}|z)},1
            \right\}.\label{ar:2}
        \end{align}
        \item [Step 3.] Same as before, for $\tau=3$, letting $\theta_3^{(m)}:=\theta_3^{(m-1)}=(\psi_3^{(m-1)},\eta_3^{(m-1)})$ we draw a candidate $\tilde{\theta}_3 \sim \mathcal{N}(\theta_3^{(m-1)},\Omega_3^{(m)})$ and update 
        $\theta_3^{(m)} = \tilde{\theta}_3$ with probability 
        \begin{align}
            \min\left\{
            \frac{\pi(\theta_1^{(m)},\theta_2^{(m)},\tilde{\theta}_3,\sigma_u^{(m-1)},\{u_t^{(m-1)}\}|z)}{\pi(\theta_1^{(m)},\theta_2^{(m)},\theta_3^{(m-1)},\sigma_u^{(m-1)},\{u_t^{(m-1)}\}|z)},1
            \right\}.\label{ar:3}
        \end{align}  
        
        \item [Step 4.] To update $\sigma_u$, we let $\sigma_u^{(m)}:=\sigma_u^{(m-1)}$ and draw a candidate $\tilde{\sigma}_u \sim q_{\sigma_u}^m(\cdot|\sigma_u^{(m-1)})$ and update $\sigma_u^{(m)}:=\tilde{\sigma}_u$ with probability
        \begin{align}
            \min\left\{
            \frac{\pi(\theta_1^{(m)},\theta_2^{(m)},\theta_3^{(m)},\tilde{\sigma}_u,\{u_t^{(m-1)}\}|z)}{\pi(\theta_1^{(m)},\theta_2^{(m)},\theta_3^{(m)},\sigma_u^{(m-1)},\{u_t^{(m-1)}\}|z)} 
            \times 
            \frac
            {q_{\sigma_u}^m(\sigma_u^{(m-1)}|\tilde{\sigma}_u)}
            {q_{\sigma_u}^m(\tilde{\sigma}_u|\sigma_u^{(m-1)})}
            ,1
            \right\},\label{ar:4}
        \end{align} 
                where the proposal density $q_{\sigma_u}^m(\cdot|\cdot)$ will be discussed below. 

        \item [Step 5.]        To update $u_t$ for each $t=1,\ldots,T$, let $u_t^{(m)}:=u_t^{(m-1)}$. 
        Then we draw a candidate $\tilde{u}_t \sim q_u^m(\cdot|u_t^{(m-1)})$ and update $u_t^{(m)}:=\tilde{u}_t$ with probability
        \begin{align}
            \min\left\{
            \frac
            {
            f_u(\tilde{u}_t|\sigma_u^{(m)}) 
            \prod_{i\in\mathcal{I}_t} g_{\tau(i)}(b_{it}|\tilde{u}_t,\theta^{(m)},\mathcal{I}_t)
            }
            {
            f_u(u_t^{(m-1)}|\sigma_u^{(m)}) 
            \prod_{i\in\mathcal{I}_t} g_{\tau(i)}(b_{it}|u_t^{(m-1)},\theta^{(m)},\mathcal{I}_t)
            }
            \times
            \frac
            {q_u^m(u_t^{(m-1)}|\tilde{u}_t)}
            {q_u^m(\tilde{u}_t|u_t^{(m-1)})}
            ,
            1
            \right\}.\label{ar:6}
        \end{align}
    \end{enumerate}
 
        Note also that in Steps (1)-(3), the ratio of the proposal densities is one  because of the symmetry of the Gaussian densities, and it does not appear in any of the acceptance probabilities: \eqref{ar:1}, \eqref{ar:2}, and \eqref{ar:3}. The acceptance probability can be simplified in general by canceling the common components on the numerator and denominator. For example, in Step (4), the posterior odd ratio becomes 
        \begin{align}
            \frac
            {
            \pi(\theta_1^{(m)},\theta_2^{(m)},\theta_3^{(m)},\tilde{\sigma}_u,\{u_t^{(m-1)}\}|z)
            }
            {
            \pi(\theta_1^{(m)},\theta_2^{(m)},\theta_3^{(m)},\sigma_u^{(m-1)},\{u_t^{(m-1)}\}|z)
            } 
            =
            \frac
            {
            \pi_{\sigma_u}(\tilde{\sigma}_u)
            }
            {
            \pi_{\sigma_u}(\sigma_u^{(m-1)})
            }
            \prod_{t=1}^T
            \frac
            {
            f_u(u_t^{(m-1)}|\tilde{\sigma}_u)
            }
            {
            f_u(u_t^{(m-1)}|\sigma_u^{(m-1)})
            }.\label{ar:5}
        \end{align}  
        We iterate $M_1=10^5$ times, but we only keep every $\texttt{thin} = 50^{\rm th}$ draw, so we have 2,000 draws. Then, to minimize the influence of the initial point we discard the first $\texttt{burn-in} =1,000$ draws and use the second $M_2=1,000$ draws for the posterior inference. 
        
\begin{figure}[t!]
\caption{MCMC traces; The Main Specification}\label{fig:mcmc}
\begin{center}
\begin{tabular}{c}
\includegraphics[width=5in]{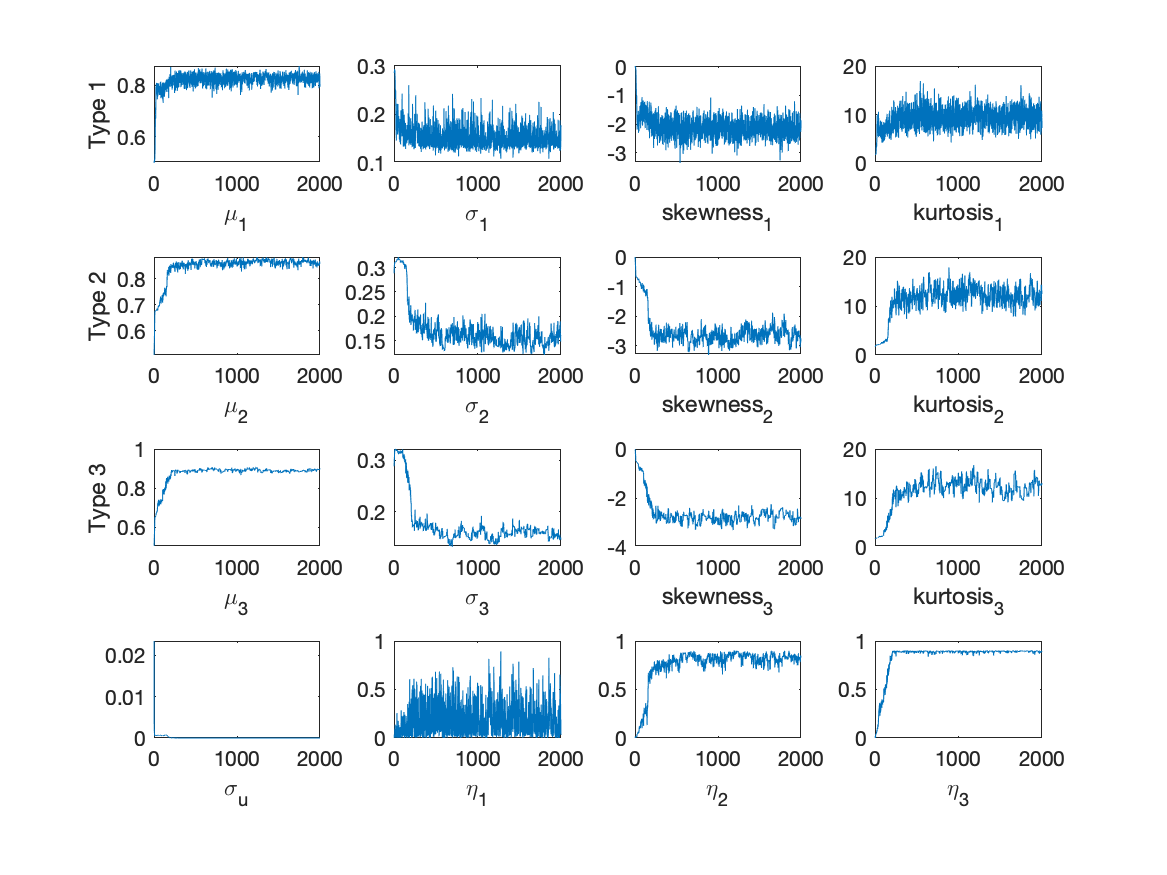}
\end{tabular}
\caption*{\footnotesize
The $(1^{\rm st},2^{\rm nd},3^{\rm rd})$ row shows the  shows MCMC traces of the mean, standard deviation, skewness, and kurtosis of the cost density of type (1,2,3) bidders, respectively.
The fourth row shows the MCMC traces of $\sigma_u$ and $\eta_\tau$ for $\tau = 1,2,3$. 
}
\end{center}
\end{figure}

\begin{figure}[t!]
\caption{MCMC traces; The Specification with Homogeneous CRRA}\label{fig:mcmc_HoCRRA}
\begin{center}
\begin{tabular}{c}
\includegraphics[width=5in]{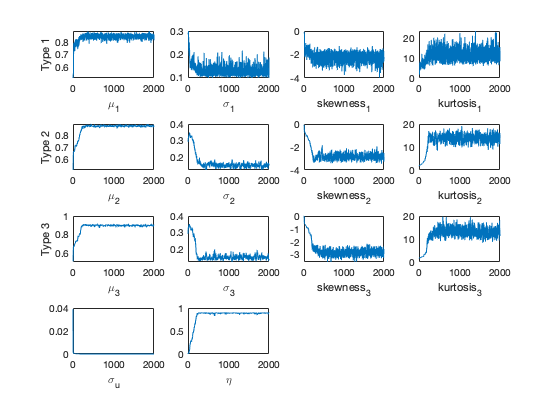}
\end{tabular}
\caption*{\footnotesize
The $(1^{\rm st},2^{\rm nd},3^{\rm rd})$ row shows the  shows MCMC traces of the mean, standard deviation, skewness, and kurtosis of the cost density of type (1,2,3) bidders, respectively.
The fourth row shows the MCMC traces of $\sigma_u$ and $\eta_\tau$ for $\tau = 1,2,3$. 
}
\end{center}
\end{figure}

\begin{figure}[t!]
\caption{MCMC traces; The Specification with Risk Neutrality}\label{fig:mcmc_NoCRRA}
\begin{center}
\begin{tabular}{c}
\includegraphics[width=5in]{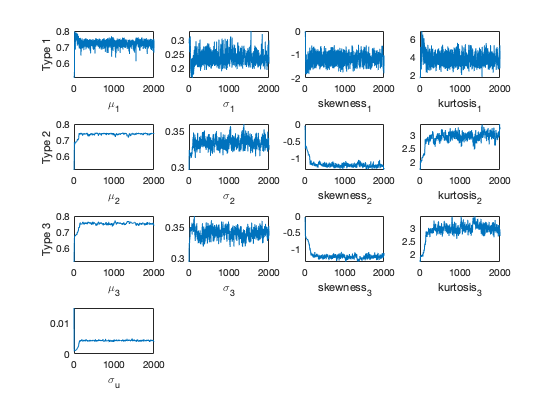}
\end{tabular}
\caption*{\footnotesize
The $(1^{\rm st},2^{\rm nd},3^{\rm rd})$ row shows the  shows MCMC traces of the mean, standard deviation, skewness, and kurtosis of the cost density of type (1,2,3) bidders, respectively.
The fourth row shows the MCMC traces of $\sigma_u$ and $\eta_\tau$ for $\tau = 1,2,3$. 
}
\end{center}
\end{figure}

\begin{figure}[t!]
\caption{MCMC traces; The Specification with Alternative Type 1 (top 2)}\label{fig:mcmc_2Big}
\begin{center}
\begin{tabular}{c}
\includegraphics[width=5in]{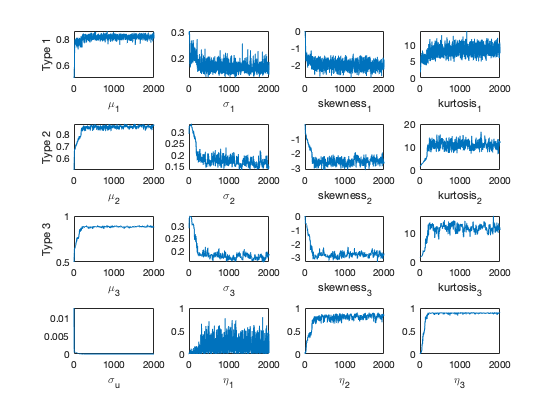}
\end{tabular}
\caption*{\footnotesize
The $(1^{\rm st},2^{\rm nd},3^{\rm rd})$ row shows the  shows MCMC traces of the mean, standard deviation, skewness, and kurtosis of the cost density of type (1,2,3) bidders, respectively.
The fourth row shows the MCMC traces of $\sigma_u$ and $\eta_\tau$ for $\tau = 1,2,3$. 
}
\end{center}
\end{figure}

\begin{figure}[t!]
\caption{MCMC traces; The Specification with Alternative Type 1 (top 3)}\label{fig:mcmc_3Big}
\begin{center}
\begin{tabular}{c}
\includegraphics[width=5in]{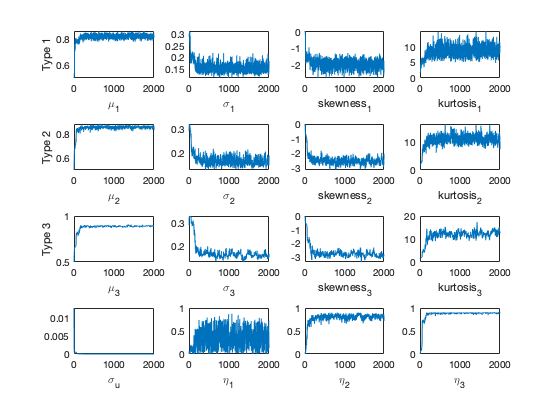}
\end{tabular}
\caption*{\footnotesize
The $(1^{\rm st},2^{\rm nd},3^{\rm rd})$ row shows the  shows MCMC traces of the mean, standard deviation, skewness, and kurtosis of the cost density of type (1,2,3) bidders, respectively.
The fourth row shows the MCMC traces of $\sigma_u$ and $\eta_\tau$ for $\tau = 1,2,3$. 
}
\end{center}
\end{figure}

\begin{figure}[t!]
\caption{MCMC traces; The Specification with Alternative Prior, Small Variance of $\psi$}\label{fig:mcmc_small}
\begin{center}
\begin{tabular}{c}
\includegraphics[width=5in]{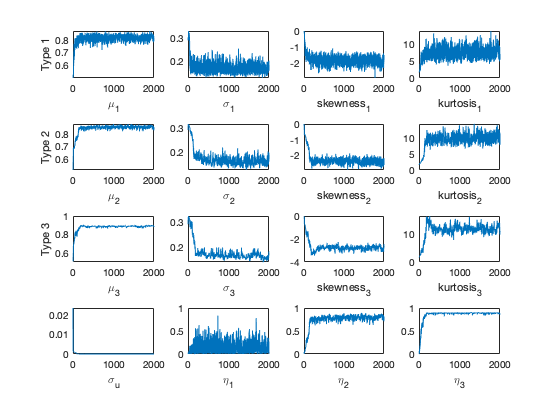}
\end{tabular}
\caption*{\footnotesize
The $(1^{\rm st},2^{\rm nd},3^{\rm rd})$ row shows the  shows MCMC traces of the mean, standard deviation, skewness, and kurtosis of the cost density of type (1,2,3) bidders, respectively.
The fourth row shows the MCMC traces of $\sigma_u$ and $\eta_\tau$ for $\tau = 1,2,3$. 
}
\end{center}
\end{figure}

\begin{figure}[t!]
\caption{MCMC traces; The Specification with Alternative Prior, Large Variance of $\psi$}\label{fig:mcmc_large}
\begin{center}
\begin{tabular}{c}
\includegraphics[width=5in]{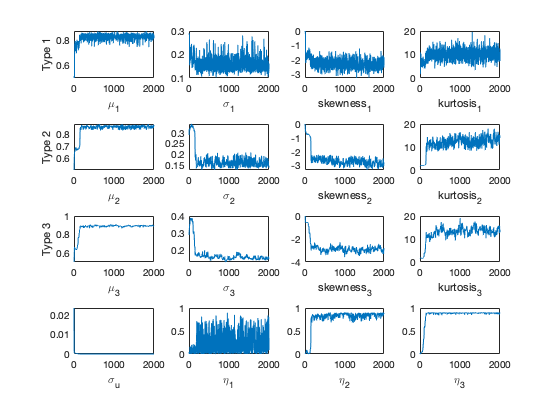}
\end{tabular}
\caption*{\footnotesize
The $(1^{\rm st},2^{\rm nd},3^{\rm rd})$ row shows the  shows MCMC traces of the mean, standard deviation, skewness, and kurtosis of the cost density of type (1,2,3) bidders, respectively.
The fourth row shows the MCMC traces of $\sigma_u$ and $\eta_\tau$ for $\tau = 1,2,3$. 
}
\end{center}
\end{figure}

\begin{figure}[t!]
\caption{MCMC traces; The Specification with Alternative $f_u(u_t)$}\label{fig:mcmc_fUH}
\begin{center}
\begin{tabular}{c}
\includegraphics[width=5in]{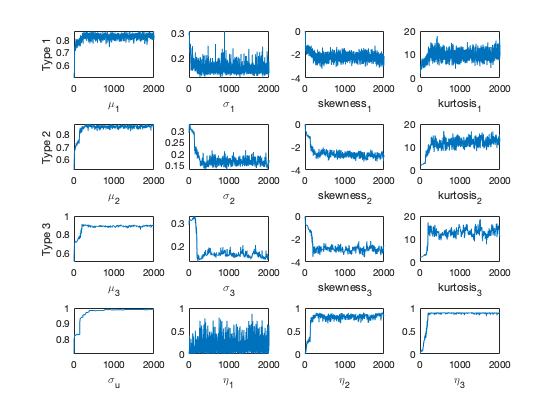}
\end{tabular}
\caption*{\footnotesize
The $(1^{\rm st},2^{\rm nd},3^{\rm rd})$ row shows the  shows MCMC traces of the mean, standard deviation, skewness, and kurtosis of the cost density of type (1,2,3) bidders, respectively.
The fourth row shows the MCMC traces of $\sigma_u$ and $\eta_\tau$ for $\tau = 1,2,3$. 
}
\end{center}
\end{figure}

Figure \ref{fig:mcmc} shows MCMC traces of the mean, standard deviation, skewness, and kurtosis of the cost density of type (1,2,3) bidders in the $(1^{\rm st},2^{\rm nd},3^{\rm rd})$ row, respectively.
The fourth row shows the MCMC traces of $\sigma_u$ and $\eta_\tau$ for $\tau = 1,2,3$, respectively. 
Overall, the algorithm converges before the $\texttt{burn-in}$ iterations.
Figures \ref{fig:mcmc_HoCRRA} to \ref{fig:mcmc_fUH}
show similarly but for different specifications as the figure titles suggest.

In the remainder of this subsection, we provide additional details to execute the MCMC algorithm.
First, we discuss the tuning parameters for Steps 1, 2, and 3. 
For these steps, we employ the normal densities as a proposal density, i.e., we draw $\tilde{\theta}_\tau \sim \mathcal{N}(\theta_\tau^{(m)},\Omega_\tau^{(m)})$ where $\Omega_\tau^{(m)}$ is the tuning parameter.
For the first ten draws, i.e., 500 iterations, we set $\Omega_\tau^{(m)}:= \Omega_0$, the identity matrix multiplied by $10^{-5}$. 
For the next 990 draws, we set $\Omega_\tau^{(m)}$ to be the sample covariance of $\{\theta_\tau^{(\cdot)}\}$ up to that point  multiplied by 2.38 and divided by the dimension of $\theta_\tau$ with 95\% of probability and set $\Omega_\tau^{(m)}:= \Omega_0$ with 5\% of probability.
This way, the algorithm automatically adjusts the tuning parameters. 
        For more on this adaptation scheme, see \cite{HaarioSaksmanTamminen2001} and  \cite{HaarioSaksmanTamminen2005}, among others.
        Then, we stop adapting the tuning parameter at the 1,000$^{\rm th}$ draw to make sure the sequence after the 1,000$^{\rm th}$ draw to the end is a Markov chain. 
        Note that since the normal proposal densities are symmetric, we do not consider the ratio of proposal densities to compute the acceptance rates; (\ref{ar:1}), (\ref{ar:2}), and (\ref{ar:3}).
         
        Second, the proposal function for Step 4 explicitly considers the support of $u_t$. This improves computational efficiency because, then, the algorithm would not propose a candidate outside the support, which is surely rejected.
       The support of $u_t$ depends on $\sigma_u$; see section 4.
       In particular, we can show  that for all $t$,  $(1-u_t^{(m-1)})/c_u \leq \tilde{\sigma}_u  \leq  \overline{\sigma}_u = 0.9/c_u$.
        So, in Step 4, we draw $\tilde{\sigma}_u$ from a normal distribution that is truncated so that it has the relevant support;
        \[
        \tilde{\sigma}_u  \sim \mathcal{N}\left(\sigma_m^{(m-1)},(0.01)^2\right)
        \times
        \mathbbm{1}\left(\tilde{\sigma}_u\in\left[\underline{\sigma}_u^{(m)}, \overline{\sigma}_u\right]\right)
        \]
        where $\underline{\sigma}_u^{(m)}:=\max_t\left\{\frac{1-u_t^{(m-1)}}{2.5758}\right\}$. Then, $q_{\sigma_u}^m(\cdot|\sigma_u^{(m-1)})$ is accordingly defined to be the truncated normal density, which is used in the acceptance rate (\ref{ar:4}).
        We note that at each iteration $m$, the lower bound $\underline{\sigma}_u^{(m)}$ has to be re-computed because it depends on $\{u_t^{(m-1)}\}_{t=1}^T.$
        
        Third, we introduce the proposal function for Step 5. 
        To improve the efficiency of the algorithm, we draw candidates for $u_t^{(m)}$ from the proposal density with the known support. 
        That is, we draw 
      \[
        \tilde{u}_t \sim \mathcal{N}\left(u_t^{(m-1)},(0.01)^2\right) \times \mathbbm{1}\left(\tilde{u}_t \in \left[\underline{u}(\sigma_u^{(m)}),1\right]\right),
  \]
  where $\underline{u}(\sigma_u) = 1 - c_u \tilde{\sigma}_u$; see equation (4) in the main paper. 
  The truncated normal density defines $q_u^m(\cdot|u_t^{(m-1)})$ accordingly, which we use in the acceptance rate (\ref{ar:6}).

       Fourth, we now describe the initial values we use to start the algorithm. 
        The bid density \eqref{gb1} 
        has a compact support that depends on the \emph{unknown} parameter $\theta$ and the unobserved heterogeneity $u_t$.
        So, it is important to find $(\{\psi_\tau^{(0)},\eta_\tau^{(0)}\}_{\tau=1}^3, u_1^{(0)},\ldots,u_T^{(0)})$ at which the likelihood is nonzero. 
        We initially set $\psi_\tau^{(0)} = (0,\ldots,0) \in\mathbb{R}^k$, with $k=5$ for all $\tau =1,2,3$, i.e., the cost distributions are all $\mathcal{U}[0,1]$ at the initial value, where $\mathcal{U}[a,b]$ denotes the uniform distribution over $[a,b]$.
        We choose $k = 5$ because it is computationally feasible and yet large enough to flexibly model the densities, i.e., the predictive densities are smooth and the high order terms of Legendre polynomials would have little contribution.
        We also set $\eta_\tau^{(0)}=0$ for all $\tau$.
        Then, we find $\{u_t^{(0)}\}$ for all $t$ so that the initial likelihood is not zero. 
        For each $t=1,\ldots,T$, $u_t^{(0)}$ must satisfy the inequality 
        \begin{align}
        \underline{b}(\theta^{(0)},\mathcal{I}_t) \leq \frac{b_{it}}{u_t^{(0)}} \leq 1,\label{con4u0}
        \end{align} 
        where $\underline{b}(\theta^{(0)},\mathcal{I}_t)= \beta_\tau(0|\theta^{(0)},\mathcal{I}_t)$ for all $\tau$, the lower bound of the equilibrium bid under $\theta^{(0)}$ if $u_t = 1$.
        For all $i\in\mathcal{I}_t$, the inequality \eqref{con4u0} gives an interval of admissible values of $u_t^{(0)}$. 
        For any $b_{it}$,  \eqref{con4u0} implies 
        \begin{align}
        b_{it}
        \leq 
        u_t^{(0)} 
        \leq 
        \frac{b_{it}}{\underline{b}(\theta^{(0)},\mathcal{I}_t)}.\label{con4u1}
        \end{align}
        Since the inequality \eqref{con4u1} has to be satisfied by all bidders in $\mathcal{I}_t$, we have  the bound for $u_t^{(0)}$ as below. 
        \begin{align}
        \max_{i\in\mathcal{I}_t}\{b_{it}\} 
        \leq 
        u_t^{(0)} 
        \leq 
        \min_{i\in\mathcal{I}_t} \left\{\frac{b_{it}}{\underline{b}(\theta^{(0)},\mathcal{I}_t)}\right\}.\label{utini}
        \end{align}
        If the interval above is well defined, i.e., the lower bound is smaller than the upper bound,
        \begin{align}
        \max_{i\in\mathcal{I}_t}\{b_{it}\} \leq 
        \min_{i\in\mathcal{I}_t}\left\{\frac{b_{it}}{\underline{b}(\theta^{(0)},\mathcal{I}_t)}\right\},\label{utini1}
        \end{align} 
        then 
        we draw 
        $
        u_t^{(0)}\sim \mathcal{U}\left[\max_{i\in\mathcal{I}_t}\{b_{it}\}, 
        \min_{i\in\mathcal{I}_t}\left\{\frac{b_{it}}{\underline{b}(\theta^{(0)},\mathcal{I}_t)}\right\}\right]
      $.
        If this can be done for all $t=1,\ldots,T$, we start the MCMC algorithm (steps 1 to 5) above to explore  the posterior.
        However, we may not have \eqref{utini1} as we desire for some $t$. 
        Even if \eqref{utini1} is violated just at one $t$, the bid density \eqref{gb1} is zero, and so is the likelihood.
       If this happened, one should try other initial values. 
       For example, the earlier version of the paper, which considers several job categories, tries a new $\eta_\tau^{(0)}$ by increasing it by 0.1 and repeat the procedure here until it finds an initial value with a nonzero likelihood, whenever the inequality condition is not satisfied at $\eta_\tau^{(0)}$.
       For the ``printing papers" category, however,
       \eqref{utini1} is satisfied for all $t$ at the (first) initial values $\theta^{(0)}$ suggested above. 
      We set $\sigma_u^{(0)} = \overline{\sigma}_u$.

Fifth, 
the boundary-value method requires a benchmark bidder in each distinct $\mathcal{I}$, and the method might fail to converge (or takes a long time) depending on the choice of the benchmark bidder, in which case \cite{Fibich_Gavish_2011} suggest using another bidder as a benchmark.
In our computation, we compute $L_1$ distance between two outcome functions, which will become the bidding strategy at convergence, for each bidder in subsequent iterations of the boundary-value algorithm. 
When the maximum of the $L_1$ distances (of all bidders in $\mathcal{I}$) falls below a prespecified tolerance level (we use a tight tolerance, \texttt{tol = 1e-10}), we consider the algorithm converges. 
If the maximum does not fall before the prespecified maximum number of iterations (we use \texttt{MaxIter = 100}), we rerun the boundary-value algorithm with setting another bidder to be the benchmark. 
If it happens for all bidders for any of the configurations in the data, the MCMC algorithm rejects the candidate and consider another candidate in the next MCMC step.

Finally, 
the boundary-value method of \cite{Fibich_Gavish_2011} also requires a starting value whenever it is called in the MCMC algorithm. 
For the first MCMC iteration, we use the bidding strategy associated with the symmetric model where bidders' are risk-neutral and draw costs from $\mathcal{U}(0,1)$. 
For the rest of MCMC iterations, we use the bidding strategies from the previous MCMC iteration as the initial point to start the boundary-value method in the next MCMC iteration. 
This approach reduces computing time because two subsequent MCMC draws tend to be close each other and, therefore, the starting point should be near the solution.

\section{Estimation with Simulated Data}
\label{section:montecarlo}
In this section, we consider a simulation exercise to evaluate the performance of our method.
We generate two datasets of different sizes and use them to separately estimate the parameters. Then we compare the posterior distributions to verify that they are well supported at the true parameter and that the posterior distribution from the larger of the two samples is more condensed around the parameter than the posterior from the smaller sample. The main objective of this exercise is to illustrate that our method correctly captures unobserved heterogeneity in procurements if there is a such variation in the data generating process.

In congruence with
empirical application, we choose a data-generating process with 3 types of bidders, each with a different cost density and CRRA parameter. 
Bidders submit bids in FPPs, where the variability of the bidder configuration is substantial. 
We consider ten different bidder configurations: $A\in\left\{(11), (111), (22), (222), (33), (333), (12), (13), (23),(123)\right\},$ and simulate $\tilde{T}_A\in\{20,100\}$ 
for each configuration $A$, i.e., $T=\tilde{T}_A\times 10$ procurements, from the data-generating process:
\begin{eqnarray*}
    f_1(c)&=0.1\cdot\mathbbm{1}(c\in[0,1]) + 0.9\cdot \textbf{beta}(c;1,4), \hspace{.5cm} \eta_1 =0.7; \\
    f_2(c)&=0.1\cdot\mathbbm{1}(c\in[0,1]) + 0.9\cdot \textbf{beta}(c;1,3), \hspace{.5cm} \eta_1 =0.4; \\
    f_3(c)&=0.1\cdot\mathbbm{1}(c\in[0,1]) + 0.9\cdot \textbf{beta}(c;2,4), \hspace{.5cm} \eta_1 =0.1,
\end{eqnarray*}
where $\textbf{beta}(\cdot;\alpha,\beta)$ is the Beta density of cost with parameters $(\alpha,\beta)$ and $f_u(\cdot|\sigma_u)$ is defined in Equation (4) of the main text with variance $\sigma_u = 0.1$.
That is, we simulate two bid samples with sizes of $T=200$ and $1,000$. For each of them, we apply our method to explore the posterior distribution of the model primitives. 

\begin{figure}[t!]
\caption{Predictive Cost Densities and True Densities, Simulated Sample\label{MCfcost}}
\begin{center}
\begin{tabular}{c}
\includegraphics[width=5in]{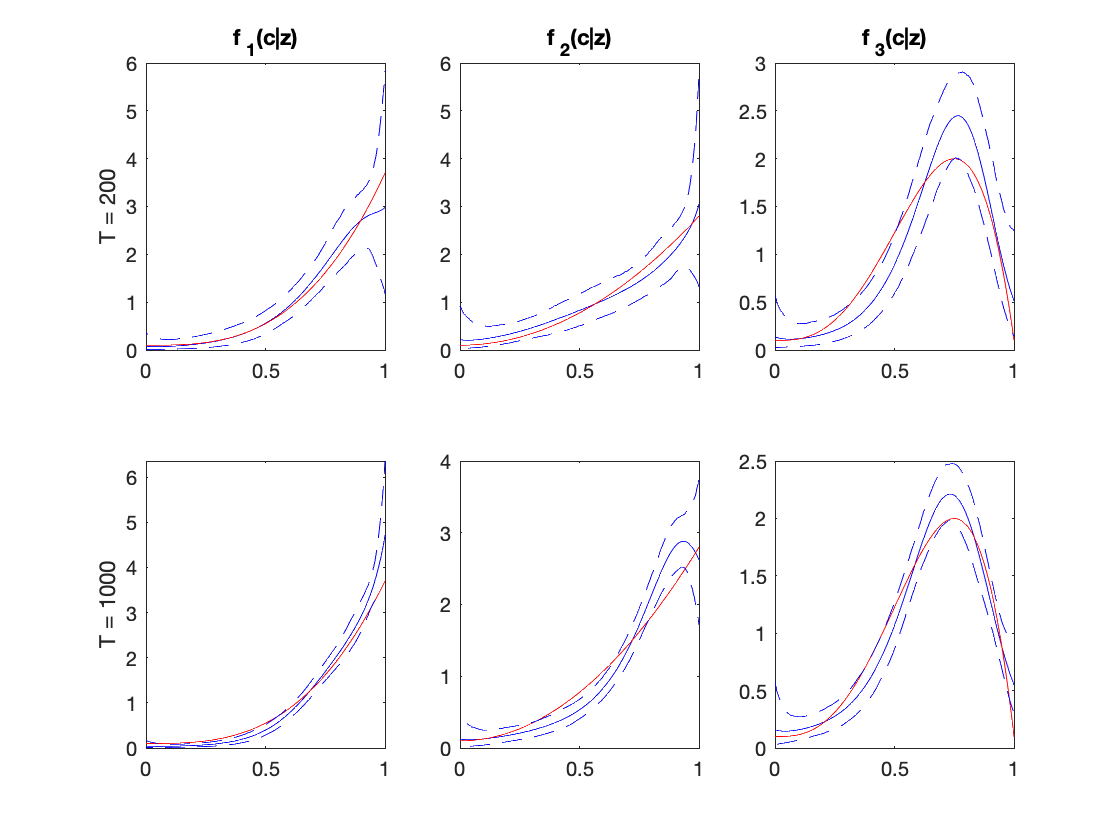}
\end{tabular}
\caption*{\footnotesize 
Each panel shows the predictive cost densities (in solid blue lines) along with the its 2.5 and 97.5 percentiles (in dashed blue lines) and the true density (in red line).
The upper and lower panels are associated with the sample of size $200$ and $1000$ procurements, respectively.
The left, middle, and right panels are for type 1, 2, and 3 bidders, respectively. 
}
\end{center}
\end{figure}

\begin{figure}[t!]
\caption{Posterior of Risk-Aversion Coefficients, Simulated Sample\label{MCetaH}}
\begin{center}
\begin{tabular}{c}
\includegraphics[width=5in]{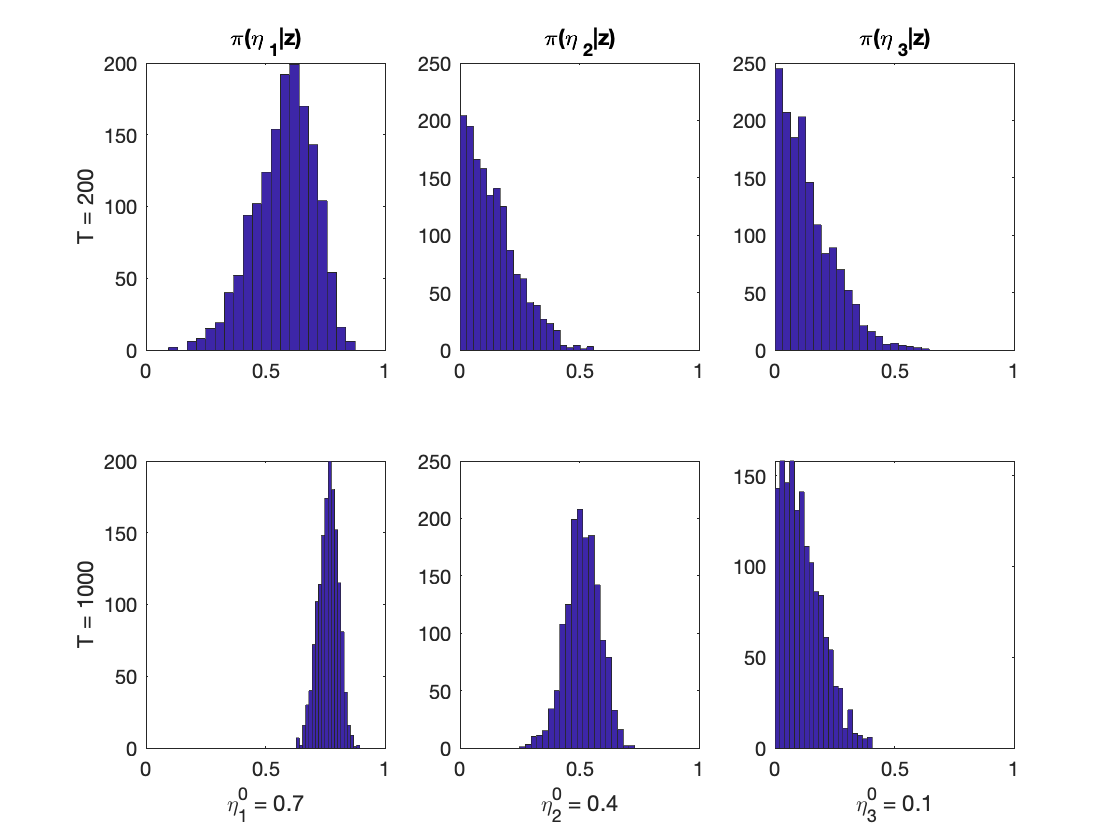}
\end{tabular}
\caption*{\footnotesize 
Each panel displays the histogram of the posterior mean of CRRA parameters, with true values $\{\eta_1^0, \eta_2^0, \eta_3^0\}=(0.7, 0.4, 0.1)$.
The (upper, lower) panels are associated with the sample of size $(200,1000)$ procurements.
The left, middle, and right panels are for type 1, 2, and 3 bidders. 
}
\end{center}
\end{figure}

\begin{figure}[t!]
\begin{center}
\caption{Density of the Unobserved Heterogeneity, Simulated Sample\label{fig:montecarloUH}}
    \includegraphics
[width=0.7\textwidth]   
{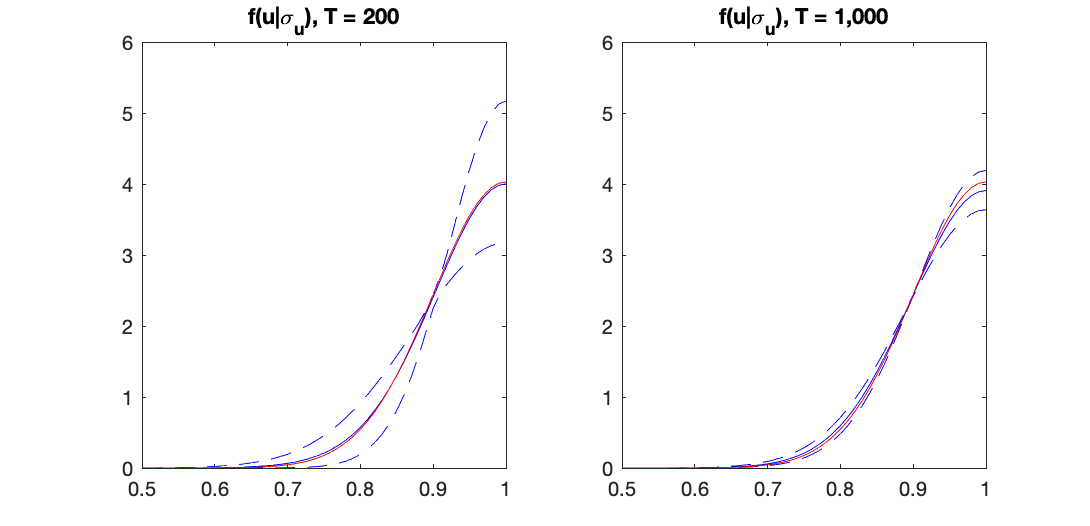}
\end{center}
\caption*{\footnotesize Notes. The figures display the posterior density $f_u(\cdot|\sigma_u)$, for  $T=200$ and $T=1,000$. Each point along the blue plain line denotes pointwise posterior mean of the density, and the blue dashed-lines denote the 2.5 and 97.5 percentiles.  The true density is in the red.}
\end{figure}

The estimation results from this simulation exercise are presented in Figures \ref{MCfcost}, 
\ref{MCetaH}, and \ref{fig:montecarloUH}.  
Figure \ref{MCfcost} displays the posterior predictive densities (plain lines) of the cost for each type along with the point-wise 95\% credible band (dashed lines) and the true cost densities (red lines),
and Figure \ref{MCetaH} shows the histograms of the risk-aversion parameters drawn from the posterior. 
The cost densities and the CRRA parameters are estimated  accurately, and the estimates become more precise and coalesce around the true value as the sample size increases. 

Figure \ref{fig:montecarloUH} displays the pointwise posterior mean of $f_{u}(\cdot|\sigma_u)$ (in plain) with a 95\% credible interval  constructed by pointwise 2.5 and 97.5 percentiles (dashed lines), and the true density of the unobserved heterogeneity (red line).
The posterior means are  accurate, and as the sample size grows, the posterior becomes more coalesced around the true density.

As a robustness check, moreover, we repeat the exercise above but restricting $u_t$ to be one for all procurements, i.e., no unobserved heterogeneity ($\sigma_u =0$), as in our empirical findings of section 5.1. 
We find that our method correctly converges to the true model primitives as before,
including 
the degenerate distribution of $u$, i.e., $\sigma_u \approx 0$.

\section{Kolmogorov-Smirnov Test on MCMC Samples}\label{sec:KStest}
To examine whether multiple MCMC samples are random draws from the same (posterior) distribution, the Bayesian literature often employs a frequentist hypothesis test; see, for example, $\chi^2$-test in chapter 4 of \cite{Geweke_2005_book} and Kolmogorov-Smirnov (KS) test in chapter 12 of \cite{Robert_Casella_2004}. 
If the draws in the samples are independent, we can apply the frequentist methods  without any modifications to test the null hypothesis that the samples are drawn from the identical distribution.
Since MCMC draws are serially correlated, however, one needs to make some adjustment of the samples.
The most common approach in the Bayesian literature is to use \emph{thinned} subsamples. 
For example, \cite{Brooks2003} and \cite{Robert_Casella_2004} employ thinned MCMC samples to conduct the KS test to examine whether two MCMC samples follow the same posterior distribution. 
Recall that a thinned subsample collects every $\texttt{thin}^{\rm th}$ draw in the original MCMC sample, e.g., $\theta^{(1\cdot \texttt{thin})}, \theta^{(2\cdot \texttt{thin})}, \theta^{(3\cdot \texttt{thin})}, \ldots$. That is, every two consecutive draws, $\theta^{(m\cdot \texttt{thin})}$ and $\theta^{((m+1)\cdot \texttt{thin})}$, in the thinned subsample are separated by $\texttt{thin}$ iterations. So, the serial correlation should be small for a large $\texttt{thin}$.
For thinned subsamples, therefore, the $p$-value of a frequentist hypothesis testing would approximate the theoretical $p$-value. We consider the approximation here to be by nature similar to the frequentist convention to approximate the sampling distribution of the test statistics by its asymptotic distribution because both require the sample size to grow to infinity so that one can use a larger $\texttt{thin}$.

Following the literature, we use thinned MCMC samples to conduct the KS test in the paper. As section \ref{section:posteriorcomputation} describes, we collect every $50^{\rm th} (=\texttt{thin})$ draw in all our MCMC samplings and use the second half of them for all statistical inference we conduct, including computation of $p$ values of KS test in section 5; see, for example, Tables 1 and 3. 

For most cases, we apply the KS test to the samples of parameters 
drawn from the posterior distribution, e.g., we draw the CRRA coefficients $(\eta_1,\eta_2,\eta_3)$ from the posterior and apply the KS test to see if they are distributionally different across the bidder types in Table 1. 
However, the comparison of predictive costs in sections 5.1 and 5.3 may require some additional clarification,
because predictive costs are not directly drawn from the posterior.
Recall that the posterior predictive cost density of type $\tau\in\{1,2,3\}$ is given as
\begin{align}
    f_\tau(c|z) = \int_{\mathbb{R}^k} f(c|\psi_\tau) 
    \pi(\theta,\sigma_u,(u_t)_{t=1}^T|z) 
    d\psi_\tau,\label{eq:predfc}
\end{align}
where $\psi_\tau\in\mathbb{R}^k$ is included in $\theta$.
We evaluate \eqref{eq:predfc} by 
$
    M^{-1}\sum_{m=1}^M f(c|\psi_\tau^{(m)}), 
$
which 
almost-surely converges to  \eqref{eq:predfc} as $M$ grows.
Note that Figure 3 in the main paper plots those predictive cost densities for $\tau\in\{1,2,3\}$ and for each different specification of risk aversion.
We supplement the graphical comparisons in Figure 3 by statistical evidence based on KS test results. To do so, for each specification of utility functions, (heterogeneous CRRA, homogeneous CRRA, and No CRRA), we draw a sample of $(c_\tau^{(m)})_{m=1}^M$ from \eqref{eq:predfc}, where $c_\tau^{(m)} \sim f(c|\psi_\tau^{(m)})$ and $(\psi_\tau^{(m)})_{m=1}^M$ are drawn from the posterior, $\pi(\theta,\sigma_u,(u_t)_{t=1}^T|z)$, and we apply the KS test to these samples.

Such statistical evidence might not add much information when the two densities we compare 
look clearly
different from each other 
in the graphs, 
like in the example of the predictive cost densities with heterogeneous risk aversion (top panels in Figure 3 or the dashed lines in the middle and bottom panels) and the densities with no risk aversion (bottom panels).
Nevertheless, the test results should help the researcher conclude whether the distributions under comparison are identical when they look similar, like in the case of the predictive cost densities with heterogeneous risk aversion and the predictive cost densities with a common risk aversion coefficient (middle panels).

Our sensitivity analysis in section 5.3 provides 
additional 
examples where the test results are useful.
We obtain the predictive cost densities for each alternative specification and compare them against the ones under the main specification. The cost densities under alternative specifications are almost identical to the ones under the main specification, especially for type 2 and type 3 bidders; in Figures \ref{fig:sens_fc_1} and \ref{fig:sens_fc_2}.
However, there is some slight difference for cases like type 1 bidder under small prior. For all cases, however, we find no evidence against the hypothesis of identical distributions at any conventional significance level; therefore, we conclude that the predictive cost densities are robust for the specifications we consider.

\section{Algorithm for Policy Simulations}\label{sec:algorithmPolicy}
For the policy analysis in section 5.2, we need to evaluate the predictive procurement cost at a counterfactual reserve price $\rho$,
for which this section develops our algorithm.

\subsection{Case with a Given Bidder Configuration }\label{sec:givenItilde}
For the moment, we consider the case where a one-dimensional reserve price $\rho$ is applied to all the bidders in a given set $\tilde{\mathcal{I}}$ regardless of their types.
Here, we construct a model for the first-price auction that maps to the FPP with $\rho$ using the equivalence relationship similar to the relationship between \eqref{procure1} and \eqref{ok11}.
In particular, we compute 
the expected procurement cost
in the FPP from the seller's expected revenue in the auction, which we evaluate by simulating bid data in the associated auction.

Due to the bid data normalization in the paper, the cost lies in $[\underline{c},\overline{c}] =[0,1]$, i.e., \eqref{conv} simplifies to $v_i = 1 - c_i$. 
Then,  $v_i$ follows $\tilde{F}_i(x) = 1 - F_i(1- x)$, where $F_i(x)$ is the CDF of $c_i$ for all $i \in \tilde{\mathcal{I}}$.
In addition, the reserve price in the data, i.e., the current reserve price, is $\rho_c = \overline{c} = 1$. 
This reserve price in the associated auction is then $\tilde{\rho}_c = 1 - \rho_c = 0$.
Similarly, for any counterfactual reserve price $\rho$ in the procurement, 
the reserve price in the associated action
is $\tilde{\rho} = 1 - \rho$. 

The objective functions \eqref{procure} and \eqref{auction} imply that $\beta_i(c_i) = 1 - \tilde{\beta}_i(v_i)$, where $\{\beta_i\}_{i\in\mathcal{I}}$ and $\{\tilde{\beta}_i\}_{i\in\mathcal{I}}$ are the bidding strategies in the procurement and auction, respectively, with $v_i = 1 - c_i$.
Therefore, the procurement cost is $\min_{i\in\tilde{\mathcal{I}}}\{\beta_i(c_i)\} = 1 - \max_{i\in\tilde{\mathcal{I}}}\{\tilde{\beta}_i(v_i)\}$ at the current reserve price, for realized costs $\{c_i\}_{i\in\tilde{\mathcal{I}}}$.
If $\rho < 1$, the procurement cost is 
\begin{align}
\min_{i\in\tilde{\mathcal{I}}}\{\beta_i(c_i;\rho)\mathbbm{1}(c_i \leq \rho) + \rho_c\cdot\mathbbm{1}(c_i > \rho)\} 
= 1 - \max_{i\in\tilde{\mathcal{I}}}\left\{\tilde{\beta}_i(v_i;\tilde{\rho})\mathbbm{1}(v_i \geq\tilde{\rho}) \right\},\label{nn}
\end{align}
where the equilibrium strategies indicate the dependence on the reserve prices.
To simplify the expressions, we suppress the dependence of the strategies on the bidder configuration and the model primitives -- the parameters for the distribution functions and CRRA coefficients. 
Let us explain the left-hand side of \eqref{nn}.
If bidder $i$'s cost is less than $\rho$, she bids $\beta_i(c_i;\rho)$.
Otherwise, bidder $i$ bids any number above $\rho$ to avoid winning the procurement; this case does not need to explicitly appear in \eqref{nn} for the procurement cost (a losing bid cannot be the procurement cost). 
If some bidders bid below $\rho$, the procurement cost equals the minimum of the bids.
If no bidder bids below $\rho$, the 
buyer
purchases similar goods and services at  $\rho_c$.
The second indicator takes care of that case.
This assumption is consistent with the way by which a reserve price is chosen; see Section 3.
The equality holds because $\rho_c = 1$ and $\beta_i(c_i;\rho)\mathbbm{1}(c_i \leq \rho) = 1 - \tilde{\beta}_i(v_i;\tilde{\rho})\mathbbm{1}(v_i \geq\tilde{\rho})$.

The (expected) procurement cost that we need for policy analysis  is the expectation of the left-hand side of \eqref{nn}, where the expectation is for $\{c_i\}_{i\in\tilde{\mathcal{I}}}$, for which we take the expectation of the right-hand side, i.e., one minus the seller's (expected) revenue.
That is, 
\begin{align}
\Lambda(\rho,\theta,\mathcal{I},\tilde{\mathcal{I}})
= 1 - E\left[\max_{i\in\tilde{\mathcal{I}}}\left\{\tilde{\beta}_i(v_i;\tilde{\rho})\mathbbm{1}(v_i \geq\tilde{\rho}) \right\}\right]\label{eq:LR}
\end{align}
where the expectation on the right-hand side of \eqref{eq:LR} is the seller's (expected) revenue of the associated auction and the third argument $\mathcal{I}$ on the left-hand side will become apparent in the next subsection. Note that we also use notations $\Lambda(\rho,\theta)$ and $\Lambda(\rho,\theta,\mathcal{I})$. The latter is the procurement cost integrating out $\tilde{\mathcal{I}}$ and 
the former integrating out both $\mathcal{I}$ and $\tilde{\mathcal{I}}$.

To compute $\{\tilde{\beta}_i(v_i;\tilde{\rho})\}_{i \in \tilde{\mathcal{I}}}$ with $|\tilde{\mathcal{I}}| \geq 2$, we further extend the algorithm of \cite{Fibich_Gavish_2011}.
For the case with $|\tilde{\mathcal{I}}| \geq 2$,
define the rescaled value
\begin{align}
v_i^{\tilde{\rho}}  := \left(\frac{v_i - \tilde{\rho}}{1 - \tilde{\rho}}\right) \mathbbm{1}(v_i \in (\tilde{\rho},1]) 
\label{eq:vrho}
\end{align}
and let $\tilde{F}_i^{\tilde{\rho}}(x) := \tilde{F}_i((1-\tilde{\rho}) x + \tilde{\rho})$ for $x \in [ 0,1]$ 
be the CDF of $v_i^{\tilde{\rho}}$ and $\tilde{f}_i^{\tilde{\rho}}(\cdot)$ be its PDF.
Then, 
$v_i^{\tilde{\rho}}$ has the unit support, which is required for the algorithm  of \cite{Fibich_Gavish_2011}.
Then, the algorithm solves for the inverse bidding strategies $\{\tilde{\phi}_i^\rho(\cdot)\}_{i\in\tilde{\mathcal{I}}}$ in the form of \eqref{procFOC1}
with $\tilde{F}_i^{\tilde{\rho}}(\cdot)$ and  $\tilde{f}_i^{\tilde{\rho}}(\cdot)$
in place of 
$\tilde{F}_i(\cdot)$ and  $\tilde{f}_i(\cdot)$, giving the bidding strategies 
$\{\tilde{\beta}_i^\rho(\cdot)\}_{i\in\tilde{\mathcal{I}}}$.
Finally, 
we have 
\begin{align}
\tilde{\beta}_i(v_i;\tilde{\rho}) = 
\left\{
\begin{array}{ll}
\tilde{\rho} + (1-\tilde{\rho}) \tilde{\beta}_i^{\tilde{\rho}}
(v_i^{\tilde{\rho}}) & \textrm{ if $v_i \geq \tilde{\rho}$}\\
0  & \textrm{otherwise.}
\end{array}
\right.
\label{eq:strrt}
\end{align}

The discussion so far suggests how to evaluate \eqref{eq:LR} via simulation.
For any reserve price $\rho$ and cost densities $\{F_i\}$ for $i\in\tilde{\mathcal{I}}$,
we construct their auction counterparts $\tilde{\rho}$ and  $\{\tilde{F}_i\}$. Then,
we draw values $\{v_{im}\}_{i\in \tilde{\mathcal{I}}} \sim \prod_{i\in \tilde{\mathcal{I}}} \tilde{F}_i$ independently for all $m \in\{ 1,\ldots,M_3(\tilde{\mathcal{I}})\}$, where $M_3(\tilde{\mathcal{I}})$ is the number of auctions in the simulation for the bidder configuration $\tilde{\mathcal{I}}$. 
Then, by evaluating \eqref{eq:strrt} at the drawn values, we compute a set of simulated equilibrium bids.
Then, the average of the $M_3(\tilde{\mathcal{I}})$ simulated winning (maximum) bids consistently estimates the expectation in \eqref{eq:LR}.
Let $i^*$ be the index of the bidder in $\tilde{\mathcal{I}}$ who has the highest value. 
\begin{align}
\frac{1}{M_3(\tilde{\mathcal{I}})}
\sum_{m=1}^{M_3(\tilde{\mathcal{I}})}
\tilde{\beta}_{i^*}(v_{i^*m};\tilde{\rho})\mathbbm{1}(v_{i^*m} \geq\tilde{\rho})
\stackrel{a.s}{\longrightarrow}
E\left[\max_{i\in\tilde{\mathcal{I}}}\left\{\tilde{\beta}_i(v_i;\tilde{\rho})\mathbbm{1}(v_i \geq\tilde{\rho}) \right\}\right].\label{eq:revsim}
\end{align}
Using the simulated bid data, we also evaluate the efficiency measures.
That is, we evaluate the probability of allocation (at least one bidder wins) by
\begin{align*}
\frac{1}{M_3(\tilde{\mathcal{I}})}
\sum_{m=1}^{M_3(\tilde{\mathcal{I}})}
\mathbbm{1}(v_{i^*m} \geq \tilde{\rho})
\stackrel{a.s}{\longrightarrow}
\Pr(\textrm{at least one bidder wins}),
\end{align*}
and the probability that the efficient bidder wins by
\begin{align*}
\frac{1}{M_3(\tilde{\mathcal{I}})}
\sum_{m=1}^{M_3(\tilde{\mathcal{I}})}&
\mathbbm{1}\left(\tilde{\beta}_{i^*}(v_{i^*m};\tilde{\rho}) 
=
\max_{i\in\tilde{\mathcal{I}}} \{ \tilde{\beta}_{i}(v_{im};\tilde{\rho}) \}
\right)
\mathbbm{1}(v_{i^*m} \geq \tilde{\rho})
\\&
\stackrel{a.s}{\longrightarrow}
\Pr(\textrm{the efficient bidder wins}).
\end{align*}
Note that these probabilities depend on the model primitives, reserve price, and bidder configuration.

\subsection{Endogenous Bidder Configuration and Bidder-Specific Reserve Price}
Up to now, we have considered the case where $\tilde{\mathcal{I}}$ is given and have focused on how we compute the outcomes for the procurement  using the outcomes of the associated auction. 
Section 5.2 allows $\tilde{\mathcal{I}}$ to be endogenously determined when the procurer chooses a binding reserve price.
From now on, we extend the algorithm to compute $\Lambda(\rho,\theta,\mathcal{I})$ by incorporating the random $\tilde{\mathcal{I}}$ and bidder-specific reserve price, i.e., $\rho = (\rho_i)_{i\in\mathcal{I}}$ is now a vector 
(the type-specific case is nested here.).
We do not refer to the associated auction anymore, which is implicit; refer to the previous subsection.

Let $\mathcal{I}$ be the set of bidders that would be exogenously given if the procurement had a non-binding reserve price.
To evaluate the procurement cost $\Lambda(\rho,\theta,\mathcal{I})$,
we simulate $M_3$ sets of the bidder costs, i.e., $\{c_{im}\}_{i\in\mathcal{I}} \sim \prod_{i\in\mathcal{I}} F_i(\cdot)$, independently for all $m \in\{1, \ldots, M_3\}$. 
In each simulated procurement $m$, 
bidder $i$ with $c_{im} > \rho_i$ does not enter the procurement.
So, the binding reserve price $\rho$ and realized costs $\{c_{im}\}_{i\in\mathcal{I}}$ in each $m$ 
revise the bidder configuration. 
To be specific, we denote the revised configuration by
$ 
\tilde{\mathcal{I}}_m = \tilde{\mathcal{I}}(\mathcal{I}, \{\rho_i\}_{i\in\mathcal{I}}, \{c_{im}\}_{i\in\mathcal{I}}) = \{i \in \mathcal{I}: c_{im} \leq \rho_i \}. 
$
Since the costs $\{c_{im}\}_{i\in\mathcal{I}}$ are random, $\tilde{\mathcal{I}}_m$ is random. 
The support $\mathcal{S(I)}$ of $\tilde{\mathcal{I}}_m$ is finite because $|\mathcal{I}|<\infty$. 
Let $Q(\tilde{\mathcal{I}})$ be the probability of $\tilde{\mathcal{I}}\in\mathcal{S(I)}$ for a generic element $\tilde{\mathcal{I}}$ of  $\mathcal{S(I)}$.
We evaluate $Q(\tilde{\mathcal{I}})$ by its consistent estimate
$
M_3(\tilde{\mathcal{I}})/M_3 
\stackrel{p}{\longrightarrow} Q(\tilde{\mathcal{I}}),
$
where 
$M_3(\tilde{\mathcal{I}}) = \sum_{m=1}^M\mathbbm{1}(\tilde{\mathcal{I}} = \tilde{\mathcal{I}}_m)$.
Note that we used $M_3(\tilde{\mathcal{I}})$ in the previous subsection.

Section 5.2 assumes that $\tilde{\mathcal{I}}$ is commonly known among the bidders in $\tilde{\mathcal{I}}$.
For the cases with $|\tilde{\mathcal{I}}| \geq 2$, 
let $\rho^e(\tilde{\mathcal{I}},\rho) := \min\{\rho_i:i \in\tilde{\mathcal{I}}\}$.
Note that $\rho^e(\tilde{\mathcal{I}},\rho)$ is the (scalar-valued) effective reserve price, which is applied to all the bidders in $\tilde{\mathcal{I}}$. 
So, $\rho^e$ plays the role of $\rho$ in section \ref{sec:givenItilde}.
The reason why 
$\rho^e(\tilde{\mathcal{I}},\rho)$ is the effective reserve price is because all bidders in $\tilde{\mathcal{I}}$, by observing $\tilde{\mathcal{I}}$, learn that a bidder facing $\rho^e(\tilde{\mathcal{I}},\rho)$ participated in the procurement, and 
this bidder would never bid above $\rho^e(\tilde{\mathcal{I}},\rho)$; thus, there is no chance of winning if bidding above  $\rho^e(\tilde{\mathcal{I}},\rho)$.
So, $\rho^e(\tilde{\mathcal{I}},\rho)$ is effectively the (scalar valued) reserve price. 
For all $\tilde{\mathcal{I}}\in\mathcal{S(I)}$ such that $|\tilde{\mathcal{I}}| \geq 2$, therefore, the algorithm in section \ref{sec:givenItilde} evaluates the expected procurement cost, $\Lambda(\rho^e(\tilde{\mathcal{I}},\rho),\theta,\mathcal{I},\tilde{\mathcal{I}})$, which we consistently estimate by \eqref{eq:revsim} using the subsample of $M_3(\tilde{\mathcal{I}})$ simulated procurements with $\tilde{\mathcal{I}}$ . 
If $\tilde{\mathcal{I}} = \{i\}$ for $i \in \mathcal{I}$, the procurement cost is $\rho_i$. 
If $\tilde{\mathcal{I}}$ is empty, the procurement cost is $\rho_c$. 
Hence, we compute the expected procurement expected cost for the given $\mathcal{I}$, 
\begin{align}
\Lambda(\rho,\theta,\mathcal{I}) = \sum_{\tilde{\mathcal{I}}\in\mathcal{S(I)}} Q(\tilde{\mathcal{I}}) \Lambda(\rho^e(\tilde{\mathcal{I}},\rho),\theta,\mathcal{I},\tilde{\mathcal{I}}),\label{eq:Limm}
\end{align}
where we have consistent estimates for all the components on the right-hand side.
We observe the distribution of $\mathcal{I}$ in the data $z$. 
Therefore, we compute 
the expected procurement cost
with unknown $\mathcal{I}$  by integrating $\mathcal{I}$ out by its empirical distribution. That is,
\begin{align}
\Lambda(\rho,\theta) &= \sum_{\mathcal{I}} \left[\frac{1}{T}\sum_{t=1}^T\mathbbm{1}(\mathcal{I} = \mathcal{I}_t)\right] \Lambda(\rho,\theta,\mathcal{I})\nonumber
\\&=
\sum_{\mathcal{I}} \left\{\left[\frac{1}{T}\sum_{t=1}^T\mathbbm{1}(\mathcal{I} = \mathcal{I}_t)\right]
\sum_{\tilde{\mathcal{I}}\in\mathcal{S(I)}} Q(\tilde{\mathcal{I}}) \Lambda(\rho^e(\tilde{\mathcal{I}},\rho),\theta,\mathcal{I},\tilde{\mathcal{I}})\right\},\label{eq:Lrho}
\end{align}
where the second equality uses \eqref{eq:Limm}.
We obtain the efficiency measures under $(\rho,\theta)$ integrating $(\mathcal{I},\tilde{\mathcal{I}})$  out, similarly.

\subsection{Additional Note on Policy Simulation}
We have three types, $\rho = (\rho_1,\rho_2,\rho_3)\in\mathbb{R}^3$, and bidder $i$ faces $\rho_{\tau(i)}$ in all procurements she enters.
Section 5.2 considers two cases: common reserve price and type-specific reserve price. 
For the case with a common reserve price, we set $\rho_1 = \rho_2 = \rho_3$ and consider all reserve prices $\rho_\tau \in \{\frac{15}{100},\frac{16}{100},\frac{17}{100},\ldots,\frac{99}{100},1\}$. We evaluate the predictive procurement cost, 
\begin{align}
\frac{1}{M_2}\sum_{m=1}^{M_2}\Lambda(\rho,\theta^{(m)}) \stackrel{a.s}{\longrightarrow} E[\Lambda(\rho,\theta)|z],\label{eq:1m2}
\end{align} 
at all $\rho$, where $\{\theta^{(m)}\}_{m=1}^{M_2}$ are the MCMC draws from the posterior. 
This exercise gives Figure 4 in the main paper. 
We report the reserve price with the smallest \eqref{eq:1m2} in section 5.2 with the predictive outcome variables at that reserve price.
Note that we obtain the predictive efficiencies by integrating $\theta$ out, similarly.

For the case with type-specific reserve prices, 
we solve the policymaker's problem 
as follows. 
Let $\mathcal{A}_\tau^1$ denote the set of equally spaced grid points between $\underline{\rho}_\tau^1 = 0.2$   and $\overline{\rho}_\tau^1 = 1$ with step size of $\texttt{step}^1 = 0.2$.
Then, $\mathcal{A}^1 = \mathcal{A}_1^1\times\mathcal{A}_2^1\times\mathcal{A}_3^1$ collects all possible combinations of $(\rho_1,\rho_2,\rho_3)$.
Here, the superscript 1 indicates that they are used for the initial set of grid points.
The first iteration solves the optimization problem over $\mathcal{A}^1$. 
At each $s^{\rm th}$ iteration in the optimization process, we denote the solution by
$\hat{\rho}^s:=(\hat{\rho}_1^s,\hat{\rho}_2^s,\hat{\rho}_3^s) := \arg\min_{\rho\in\mathcal{A}^s} E[\Lambda(\rho,\theta)|z]$. 
Then, we repeat the search by revising the set of reserve prices. 
To be specific, we define $\mathcal{A}_\tau^{s+1}$ for each type $\tau\in\{1,2,3\}$ as the set of grid points between  
$\underline{\rho}_\tau^{s+1} = \max\{\hat{\rho}_\tau^s-\texttt{step}^s,\underline{\rho}_\tau^s\}$
and 
$\overline{\rho}_\tau^{s+1} = \min\{\hat{\rho}_\tau^s+\texttt{step}^s,\overline{\rho}_\tau^s\}$
with the equal step size $\texttt{step}^{s+1} = \texttt{step}^s/2$ and define $\mathcal{A}^{s+1} = (\mathcal{A}_1^{s+1}\times\mathcal{A}_2^{s+1}\times\mathcal{A}_3^{s+1})\setminus \mathcal{A}^s$.
Then, we solve the optimization problem over $\mathcal{A}^{s+1}$ to find 
$\hat{\rho}^{s+1} := \arg\min_{\rho\in\mathcal{A}^{s+1}} E[\Lambda(\rho,\theta)|z]$. 
We repeat the search revising the set of reserve prices and reducing the step size by half until the step size falls below a certain threshold, $\underline{\texttt{step}}$.
Throughout this procedure, we evaluate the posterior predictive cost \eqref{eq:1m2} by the $M_2=1,000$ MCMC draws from the posterior by computing $\Lambda(\rho,\theta,\mathcal{I},\tilde{\mathcal{I}})$ in \eqref{eq:Lrho} by Monte Carlo with simulation size of $M_3 = 5,000$. We use $\underline{\texttt{step}} = 0.005$.

\section{Sensitivity Analysis}\label{sec:sen}
This section provides additional results from our sensitivity analysis: 
the graphs of the predictive cost densities and the counterfactual analysis of inviting one additional bidder from each type under alternative specifications considered in section 5.3 of the main paper.
 
\subsection{Predictive Cost Densities}
\begin{figure}[t!]
\begin{center}
\caption{Sensitivity of Cost Densities\label{fig:sens_fc_1}}
    \includegraphics
[width=0.9\textwidth]   
{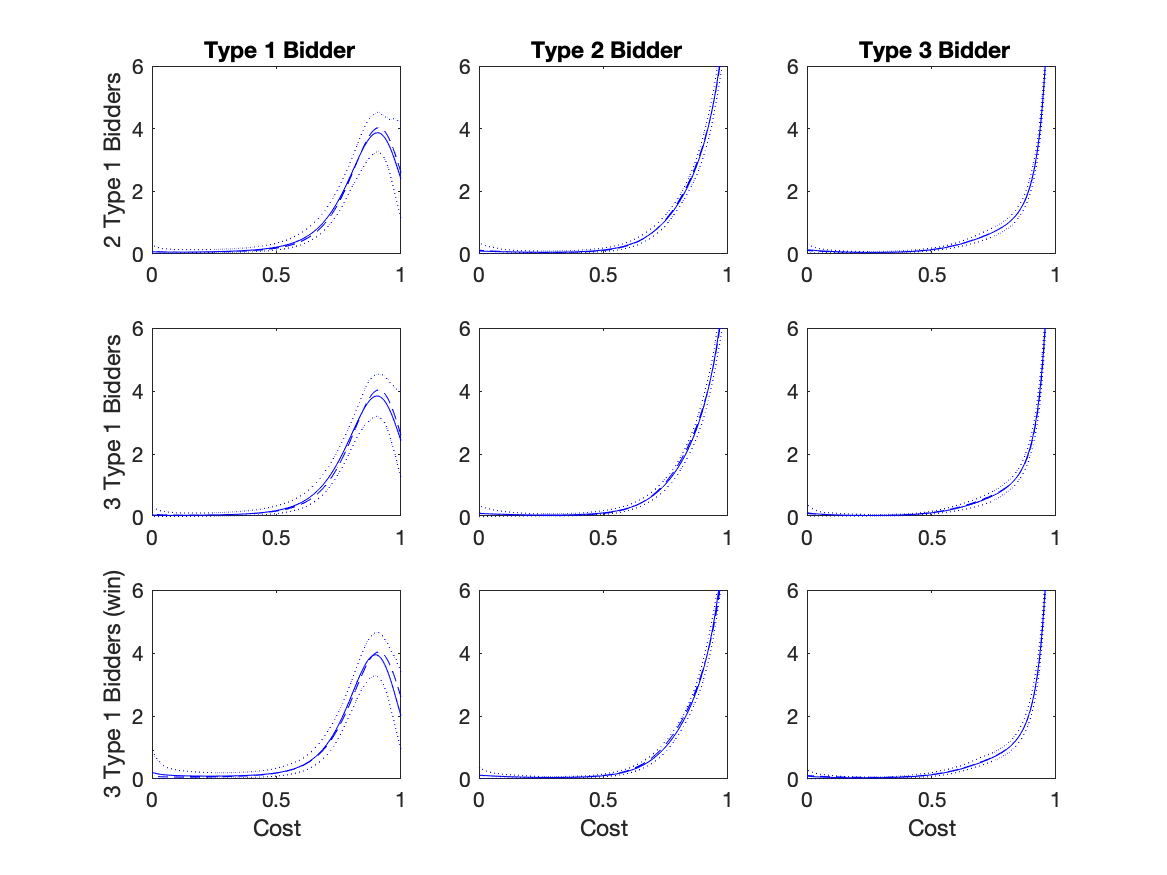}
\caption*{\footnotesize 
The diagrams show the posterior predictive cost density (solid) and 95\% credible band (dotted). 
Each row corresponds to a different definition of bidder types, and each column corresponds to a type of a bidder.
The dashed lines are the predictive densities under the main specification,  pasted from the top panels of Figure 3 in the main paper.
}
\end{center}
\end{figure}

The top panels of Figure \ref{fig:sens_fc_1} show, for each type $\tau \in \{1,2,3\}$, the posterior mean of the cost density at every point $c \in [0,1]$ by a solid line and a 95\% credible band around the predictive density by dotted lines for the specification where the two most frequent entrants are type 1 bidders.  
The middle (bottom) panels summarize the results for the cases where the three most frequent entrants (winners) are type 1 bidders. 
For comparison, the dashed lines are the predictive densities under the main specification, pasted from the top panels of Figure 3 in the main paper.   

\begin{figure}[t!]
\begin{center}
\caption{Sensitivity of Cost Densities, Figure continued\label{fig:sens_fc_2}}
    \includegraphics
[width=0.9\textwidth]   
{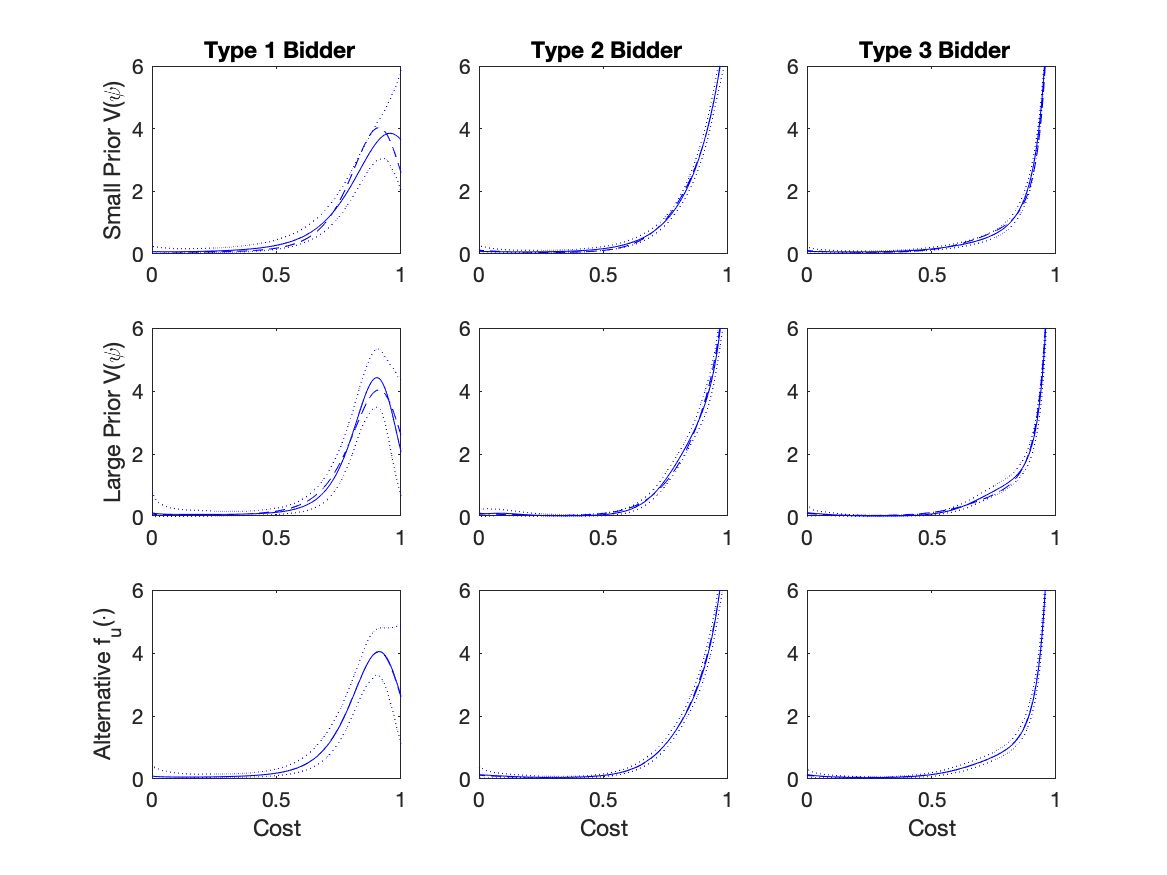}
\caption*{\footnotesize 
This figure is a continuation of Figure \ref{fig:sens_fc_1}. 
The upper and middle blocks correspond to the small and large prior variances of $\psi$, and the lower one to the alternative specification of the density of unobserved heterogeneity.
Each panel shows the posterior predictive cost density (solid) and 95\% credible band (dotted) for each type. 
The dashed lines are the predictive densities pasted from the upper block of Figure \ref{fig:sens_fc_1}. 
}
\end{center}
\end{figure}

Figure \ref{fig:sens_fc_2} continues to show the predictive densities under alternative priors and density specification for the unobserved heterogeneity.
The upper and middle panels are associated with the cases where the prior variance of $\psi$ is small and large, respectively. 
The lower panels are with the alternative specification of the density of the unobserved heterogeneity.

The posterior predictive densities are overall robust. 
In particular, the predictive cost densities of type 2 and 3 bidders are not affected by any of those elements that the researcher selects because the bid samples for those types are pretty large; see sections 3.2 and \ref{sec:data}.
On the other hand, the predictive cost density of the type 1 bidder appears to be less robust as the bid sample of type 1 is small, especially when the alternative prior is directly concerned about the parameters of the cost densities. Note that the influence of prior disappears as the sample size increases, as suggested by the Bernstein von-Mises Theorem; see Chapter 10 of \cite{van_der_Vaart_1998}. 

However, the way the predictive density depends on those elements are as expected.
Recall from section 4 that the prior mean of the cost density is the uniform distribution on $[0,1]$.
When the stronger prior is used (the first row of Figure \ref{fig:sens_fc_2}), 
the predictive cost density of type 1 is flatter than the one under the main specification  (dashed line). 
On the other hand, the predictive cost density under the main specification appears to be flatter than the one under the weaker prior  (the second row of Figure \ref{fig:sens_fc_2}).

Finally, the distribution of the unobserved heterogeneity is essentially degenerate even under the more flexible specification. 
The predictive cost densities are all identical to the ones under the main specification (the third row of Figure \ref{fig:sens_fc_2}). 

\subsection{Inviting One Additional Bidder}
Table \ref{tab:CF} presents the predictive procurement costs and the probability that the efficient bidder wins when the procurer invites one additional bidder from each type. 
Note that the first block (main specification) reprints the results under the main specification, which appear in the last block of Table 2 in the main paper 
for comparison.
The table shows the results under alternative type definitions, prior specifications, and alternative density of the unobserved heterogeneity.
\begin{table}[th!]  
 \begin{center}
 \caption{Counterfactual Analysis; Inviting One Additional Bidder\label{tab:CF}}
  \scalebox{0.85}{\begin{tabular}{ll|rr}
\hline
\hline
 	&  
	& \multicolumn{1}{c}{Predictive}	
	& \multicolumn{1}{c}{Prob. that} 	
	\\
    Alternative
 	& 
 	Inviting
	& \multicolumn{1}{c}{Procurement Cost} 		
	& \multicolumn{1}{c}{Efficient Bidder Wins} 	
	\\
    Specifications
	&
 	One additional
	& 
	\multicolumn{1}{c}{(1)}
	& 
	\multicolumn{1}{c}{(2)}
	\\
\hline
Main Specification & Type 1 Bidder 	 & 0.791 [0.775, 0.804] 	 & 0.988 [0.981, 0.996] \\
 & Type 2 Bidder 	 & 0.794 [0.782, 0.805] 	 & 0.996 [0.994, 0.998] \\
 & Type 3 Bidder 	 & 0.797 [0.784, 0.811] 	 & 0.996 [0.993, 0.997] \\
&&&\\
2 Type 1 Bidders & Type 1 Bidder 	 & 0.789 [0.776, 0.801] 	 & 0.988 [0.982, 0.995] \\
 & Type 2 Bidder 	 & 0.793 [0.781, 0.805] 	 & 0.995 [0.994, 0.997] \\
 & Type 3 Bidder 	 & 0.795 [0.784, 0.807] 	 & 0.995 [0.993, 0.997] \\
&&&\\
3 Type 1 Bidders & Type 1 Bidder 	 & 0.790 [0.776, 0.802] 	 & 0.990 [0.984, 0.996] \\
 & Type 2 Bidder 	 & 0.793 [0.782, 0.803] 	 & 0.996 [0.994, 0.998] \\
 & Type 3 Bidder 	 & 0.797 [0.786, 0.808] 	 & 0.996 [0.993, 0.998] \\
&&&\\
3 Type 1 Bidders (win) & Type 1 Bidder 	 & 0.780 [0.763, 0.795] 	 & 0.993 [0.987, 0.997] \\
 & Type 2 Bidder 	 & 0.794 [0.782, 0.804] 	 & 0.996 [0.994, 0.998] \\
 & Type 3 Bidder 	 & 0.801 [0.786, 0.813] 	 & 0.996 [0.994, 0.998] \\
&&&\\
Small Prior $V(\psi)$ & Type 1 Bidder 	 & 0.781 [0.757, 0.800] 	 & 0.987 [0.981, 0.996] \\
 & Type 2 Bidder 	 & 0.787 [0.772, 0.800] 	 & 0.996 [0.994, 0.998] \\
 & Type 3 Bidder 	 & 0.791 [0.776, 0.804] 	 & 0.996 [0.993, 0.998] \\
&&&\\
Large Prior $V(\psi)$ & Type 1 Bidder 	 & 0.793 [0.774, 0.808] 	 & 0.988 [0.980, 0.995] \\
 & Type 2 Bidder 	 & 0.795 [0.782, 0.806] 	 & 0.996 [0.994, 0.998] \\
 & Type 3 Bidder 	 & 0.801 [0.787, 0.813] 	 & 0.995 [0.993, 0.997] \\
&&&\\
Alternative $f_u(\cdot)$ & Type 1 Bidder 	 & 0.790 [0.771, 0.804] 	 & 0.988 [0.981, 0.995] \\
 & Type 2 Bidder 	 & 0.793 [0.781, 0.803] 	 & 0.996 [0.994, 0.998] \\
 & Type 3 Bidder 	 & 0.797 [0.783, 0.810] 	 & 0.996 [0.994, 0.997] 
\\
\hline 
\end{tabular}}
 \caption*{\footnotesize 
For each specification, 
this table shows predictive outcome variables along with 95\% credible intervals when the procurer invites one additional bidder from each type. 
The outcome variables are the 
procurement
cost and the probability that the bidder with the lowest cost wins. 
 }
 \end{center}
\end{table}  

The predicted outcomes are similar to the ones under the main specification. 
In particular, the results are qualitatively the same.
Inviting one additional bidder would substantially reduce the procurement cost and, therefore, be more beneficial to the procurer than optimally choosing reserve prices. 
Hence, we find that the insight of \cite{bulow_Klemperer_1996} holds for the ``printing papers" category of Russian procurements where bidders are asymmetric in cost density and risk-aversion.
Moreover, inviting an additional type 1 (3) bidder reduces the most (least) procurement cost, but the difference is not large. Finally, inviting a type 1 bidder would reduce the most efficiency, but only slightly more than inviting other bidders.

\end{document}